\documentstyle[natbib209,aasj,mnfixes,epsfig]{mn-nat}

\newcommand{\mue}{{\mu_{\rm e}}}
\newcommand{\Xn}{{X_{\rm n}}}
\newcommand{\mb}{m_{\rm b}}

\newcommand{\grcc}{{\rm g\, cm^{-3}}}
\newcommand{\ergcc}{{\rm erg\, cm^{-3}}}
\newcommand{\Gammarho}{\Gamma_\rho}
\newcommand{\Gammamue}{\Gamma_{\mue}}
\newcommand{\Rec}{R_{\rm ec}}
\newcommand{\Mdot}{\dot{M}}

\newcommand{\rhond}{\rho_{\rm nd}}

\newcommand{\zd}{z_{\rm d}}
\newcommand{\dzd}{\Delta\zd}
\newcommand{\del}{\nabla}

\newcommand{\enuc}{\epsilon_{\rm nuc}}
\newcommand{\ebrem}{\epsilon_\nu}
\newcommand{\Enuc}{E_{\rm nuc}}
\newcommand{\Ef}{E_{\rm F}}
\newcommand{\kB}{k_{\rm B}}

\newcommand{\rbot}{r_{\rm bot}}
\newcommand{\rtop}{r_{\rm top}}

\newcommand{\fcomp}{f_{\rm comp}}
\newcommand{\fnuc}{f_{\rm nuc}}

\def\dddot#1{%
        \stackrel{\textstyle.\hspace{0.6pt}\!{.}\!\hspace{0.6pt}.}{#1}}

\begin{document}

\title[Crustal Quadrupole Moments]{Deformations of Accreting Neutron
Star Crusts and Gravitational Wave Emission}

\author[Ushomirsky, Cutler, \& Bildsten]
{Greg Ushomirsky$^1$%
%
, Curt Cutler$^2$, and Lars Bildsten$^3$ \\
$^1$Department of Physics and Department of Astronomy, 601 Campbell
Hall, University of California, Berkeley, CA 94720; \\
gregus@tapir.caltech.edu \\
$^2$Max-Planck-Institut fuer Gravitationsphysik,
Albert-Einstein-Institut, Am Muehlenberg 1, D-14476 Golm bei Potsdam,
Germany; \\
cutler@aei-potsdam.mpg.de \\
$^3$Institute for Theoretical Physics and Department of Physics, Kohn
Hall, University of California, Santa Barbara, CA 93106;\\
bildsten@itp.ucsb.edu}

\maketitle

\begin{abstract}

 Motivated by the remarkably narrow range of measured spin frequencies
of $\sim 20$ accreting (and weakly magnetic) neutron stars in the
Galaxy, \citet{Bildsten98:GWs} conjectured that their spin-up had been
halted by the emission of gravitational waves. If so, then the
brightest persistent X-ray source on the sky, Scorpius X-1, should be
detected by gravitational wave interferometers within ten years.
\citet{Bildsten98:GWs} pointed out that small nonaxisymmetric
temperature variations in the accreted crust will lead to ``wavy''
electron capture layers, and the resulting horizontal density
variations near e$^-$~capture layers create a mass quadrupole moment.
Neglecting the elastic response of the crust, \citet{Bildsten98:GWs}
estimated that even e$^-$~capture layers in the thin outer crust can
develop the quadrupole necessary to balance accretion torque with
gravitational waves, $Q_{22} \sim 10^{37}- 10^{38}$~g~cm$^{-2}$ for
accretion rates $\dot{M}\sim10^{-10}-2\times10^{-8}\
M_\odot$~yr$^{-1}$.

We present a full calculation of the crust's elastic adjustment to the
density perturbations induced by the temperature-sensitive
e$^-$~capture reactions. We find that, due to the tendency of the
denser material to sink rather than spread sideways, neglecting the
elastic response of the crust overestimates, by a factor of $20-50$,
the $Q_{22}$ that results from a wavy capture layer in the thin outer
crust. However, we find that this basic picture, when applied to
capture layers in the deep inner crust, can still generate $Q_{22}$ in
the necessary range, as long as there are $\lesssim5\%$ lateral
temperature variations at densities in excess of $10^{12} \ {\rm g \
cm^{-3}}$, and as long as the crustal breaking strain is high
enough. By calculating the thermal flow throughout the core and the
crust, we find that temperature gradients this large are easily
maintained by asymmetric heat sources or lateral composition gradients
in the crust.  If the composition or heating asymmetries are
independent of the accretion rate, then for $\dot{M}\lesssim
5\times10^{-9}\ M_\odot$~yr$^{-1}$ the induced quadrupole moments have
approximately the same scaling, $\propto\dot{M}^{1/2}$, as that
necessary to balance the accretion torque at the same spin frequency
for all $\dot{M}$.  Temperature gradients in the deep crust lead to a
modulation in the thermal emission from the surface of the star that
is correlated with $Q_{22}$.  In addition, a $\sim 0.5\%$ lateral
variation in the nuclear charge-to-mass ratio in the crust will also
result in a $Q_{22}$ sufficient to halt spin-up from accretion even in
the absence of a lateral temperature gradient.  

We also derive a general relation between the stresses and strains in
the crust and the maximum quadrupole moment they can generate.  We
show under quite general conditions that maintaining a $Q_{22}$ of the
magnitude necessary to balance the accretion torque requires
dimensionless strain $\sigma\sim10^{-2}$ at near-Eddington accretion
rates, of order the breaking strain of conventional materials.  This
leads us to speculate that accreting neutron stars reach the same
equilibrium spin because they all are driven to the maximum $Q_{22}$
that the crust can sustain.
\end{abstract} 


\section{Introduction}
\label{sec:introduction}

  Recent discoveries by the {\it Rossi X-Ray Timing Explorer} indicate
that most of the rapidly accreting ($\dot M \gtrsim 10^{-11} M_\odot \
{\rm yr}^{-1}$) and weakly magnetic ($B\ll 10^{11} \ {\rm G}$) neutron
stars in our Galaxy are rotating in a narrow range of frequencies
around $\nu_s \approx 300 \ {\rm Hz}$ \citep{vanderKlis98}.  From both
evolutionary considerations and their galactic distribution, we know
that the neutron stars in these low-mass X-ray binaries (LMXBs) are
relatively old systems (see \citealt{bhatt91}) that have accreted
enough angular momentum to reach rotation rates closer to the breakup
frequency of $\approx 1.5 \ {\rm kHz}$.  Hence, some mechanism must be
found to halt the spin-up. One possible explanation is that the
neutron stars have reached the magnetic spin equilibrium (where the
spin frequency matches the Keplerian frequency at the magnetosphere)
at nearly identical spin frequencies. This requires that the neutron
star's dipolar magnetic field strength correlates very well with $\dot
M$ \citep*{white97,Miller98}.  In an alternate scenario,
\citet{Bildsten98:GWs} suggested that these stars are rotating fast
enough so that appreciable angular momentum is radiated away as
gravitational waves (GW), allowing an equilibrium where the angular
momentum added by accretion is lost to gravitational
radiation. Equilibria via gravitational wave emission from rotational
instabilities at much more rapid rotation rates had been postulated
earlier by \citet{Papaloizou78:gravity_waves} and \citet{Wagoner84}.

 The angular momentum loss rate from quadrupolar GW emission scales as
$\nu_s^5$, so that gravitational radiation effectively provides a
``wall'' beyond which accretion can no longer spin the star
up. \citet{Bildsten98:GWs} estimated that ``wavy'' electron capture
layers in the neutron star's crust, caused by a large-scale
temperature asymmetry misaligned from the spin axis, can provide the
quadrupole needed ($Q_{22} \sim 2\times 10^{38}$~g~cm$^2$ at the
Eddington accretion rate, $\dot{M}_{\rm Edd}\equiv2\times10^{-8}
M_\odot$~yr$^{-1}$) to reach this limiting situation at $\nu_s\approx
300$ Hz. Another possibility is GW emission from a continuously
excited r-mode (i.e., a Rossby wave) in the neutron star core
\citep{Bildsten98:GWs, Andersson99:accreting_rmode}, though
\citet{Levin99} has shown that such steady-state r-mode can
potentially be thermally unstable.  More recently, \citet{bu99:rmodes}
have shown that {\it steady-state} equilibrium between the accretion
torque and r-mode gravitational wave emission is incompatible with
observations of the quiescent luminosities of neutron star transients.
Finally, \citet{BU99} have shown that the extra dissipation due to a
viscous boundary layer between the crust and the core is many orders
of magnitude stronger that the viscous mechanisms previously
considered, making it unlikely that the r-modes are excited in the
cores of accreting neutron stars. We will not consider the r-mode
hypothesis further here, but will instead concentrate on a
self-consistent calculation of the mass quadrupole moment generated in
the crust.

\subsection{ Equilibrium Quadrupolar Gravitational Wave Emission} 
\label{sec:grav}

 We start by calculating the quadrupole moment that
the neutron star (NS) must have so that
the spin-up torque from accretion, $N_a$, is balanced by emission of 
quadrupolar gravitational radiation. Consider a NS
that is perturbed from sphericity by a
density perturbation $\delta \rho \equiv {\it Re}\{\delta \rho_{lm}(r)
Y_{lm}(\theta, \phi)\}$. Let $Q_{lm}$ be the perturbation's multipole
moment, defined by
\begin{equation}\label{qlm}
Q_{lm} \equiv \int{\delta \rho_{lm}(r) r^{l+2} dr}.
\end{equation}
We concentrate on perturbations $Q_{22}$ with $l=m=2$; these
perturbations radiate at a frequency $\nu_{gw} = 2\nu_s$.  The
resulting rate of loss of angular momentum $N_{gw}$ is
\begin{equation}\label{ii}
N_{gw}={\dot E_{gw}\over \Omega} = \frac{G}{c^5}\frac{1}{5\Omega}\bigl<\dddot I_{ab} \dddot I^{ab}\bigr>,
\end{equation}
\noindent 
where $I_{ab} \equiv \int{\delta\rho\,r_a r_b d^3V}$, $\Omega =
2\pi\nu_s$, and ``$\bigl< \ldots \bigr>$'' means ``time-averaged over
one period.''  A little algebra shows that
$\, \bigl<\dddot I_{ab} \dddot I^{ab}\bigr> = 
(256\pi/15) \Omega^6 Q^2_{22}$, 
so equation (\ref{ii}) becomes
\begin{equation}\label{djz}
N_{gw} = \frac{256\pi}{75} \frac{G\Omega^5 Q^2_{22}}{c^5}.
\end{equation}
For simplicity, we assume that the accreted angular momentum is that
of particles arriving from the inner edge of the accretion disk
(placed at the NS radius), so that $N_a=\dot M (GMR)^{1/2}$, where
$M$ and $R$ are the mass and radius of the NS. The required quadrupole
moment $Q_{\rm eq}$ such that gravitational wave emission is in
equilibrium with the accretion torque, $N_a$, is then
\begin{eqnarray}\label{eq:qneed} 
Q_{\rm eq}&=&3.5\times 10^{37} {\rm g \ cm^2}M_{1.4}^{1/4}R_6^{1/4} 
\\ \nonumber
&&\times \left(\dot M\over 10^{-9}
M_\odot {\rm \ yr^{-1}}\right)^{1/2}\Bigg(\frac{300 \ {\rm Hz}}
{\nu_s}\Bigg)^{5/2},
\end{eqnarray}
where $M_{1.4}=M/1.4 M_\odot$ and $R_6=R/10 \ {\rm km}$.  The range of
$\dot M$'s typically encountered in the low-mass X-ray binaries is
$10^{-10}-2\times 10^{-8} \ M_\odot {\rm yr^{-1}}$, requiring
$Q_{22}\approx 10^{37}-10^{38}$~g~cm$^2$ for $\nu_s=300 \ {\rm Hz}$
\citep{Bildsten98:GWs}.  

This paper is devoted to learning whether a quadrupole this large can
be generated from deformed capture layers in the crust
\citep{Bildsten98:GWs}, and to finding the magnitude and distribution
of the elastic strain required to sustain it. We answer the latter
question in some detail in \S~\ref{sec:max_intro}, where we show that
$Q_{22}$ is related to the typical strain, $\bar\sigma$, via
\begin{equation}
Q_{22}\approx1.2 \times 10^{38}{\rm g\ cm}^2
        \left(\frac{\bar\sigma}{10^{-2}}\right)
        \frac{R_{6}^{6.3}}{M_{1.4}^{1.2}}.
\end{equation} 
A more complete version of this relation that includes the dependences
on the density at the crust-core transition and the composition of the
crust is given in equation~(\ref{eq:qmax}).  If we presume that the
crust is pushed to some yielding strain $\bar\sigma_{\rm max}$ by the
physical effects calculated here, we find
\begin{equation}
\nu_{s,eq}\approx 295 \ {\rm Hz}
        \left(10^{-2}\over\bar\sigma_{\rm max}\right)^{2/5}
        \frac{M_{1.4}^{0.6}}{R_6^{2.4}}
        \left(\frac{\dot M}{10^{-8} \ {\rm M_\odot \ yr^{-1}}}\right)^{1/5}, 
\end{equation}
a rather suggestive relation that points to a possible answer as to
why so many LMXBs are in a narrow spin frequency range, namely that
accretion always drives the crustal strain to the breaking point.

Before we launch into the detailed discussion of how the crust is
stressed and how it responds, we calculate the strength of the GW signal from
such an equilibrium radiator.  Consider a neutron star at a distance
$d$ with an energy flux in gravitational waves, $\dot E_{gw}$, but
with an unknown spin orientation to the observer.  We define the source's
``angle-averaged'' field strength $h_a$ (at Earth) by
\begin{equation}
h_a^2 \equiv \frac{1}{4\pi}\int{(h_+^2 + h_\times^2) d\Omega} 
\end{equation} 
where the integral is over source orientations%
\footnote{ 
A common measure of GW source strength is the characteristic amplitude
$h_c$.  It is related to $h_a$ by $h_c\approx1.15h_a$.}.
The standard formula for the effective stress-energy of gravitational
waves yields
\begin{equation}
N_{gw} = \frac{c^3\Omega d^2 h_a^2}{G},
\end{equation}
which, when we write it in terms of $Q_{22} $ (using 
equation \ref{djz}) gives
\begin{equation}
h_a ={16\over 5}\left(\pi\over 3\right)^{1/2} {GQ_{22}\Omega^2\over
dc^4}. 
\end{equation} 
When the angular momentum loss by gravitational radiation 
balances angular momentum gain by accretion, 
\begin{eqnarray}
h_a = 3.5 \times 10^{-27} {R_6^{3/4}\over M_{1.4}^{1/4}}
\left(300 \ {\rm Hz}\over\nu_s\right)^{1/2} \\ \nonumber
\times\left(F_x\over 10^{-8} \ 
{\rm erg \ cm^{-2} \ s^{-1}}\right)^{1/2},
\end{eqnarray}
\citep{Wagoner84,Bildsten98:GWs}.  Here we have replaced $\dot M$ and
$d$ with the observed X-ray flux, $F_x=G\dot{M}M/4\pi R d^2$.  The
gravitational wave strength for a neutron star accreting at the
Eddington limit at the galactic center is then $h_a\approx 5\times
10^{-27}$. Prior accurate knowledge of the position on the sky and
orbital periods of many of these X-ray binaries will allow for deep
searches with the suite of laser-interferometric gravitational wave
detectors currently under construction (LIGO, VIRGO, GEO-600, and
TAMA-300; see \citealt{Brady98} and \citealt{BradyCreighton99}). The
nearby source Scorpius X-1 is the obvious first target. Its X-ray flux
is $F_x\approx 2\times 10^{-7} \ {\rm erg \ cm^{-2} \ s^{-1}}$ and the
spin period is still unknown, but likely in the range of 300 Hz
\citep{vanderKlis98} giving $h_a\approx 1.6\times
10^{-26}$. \citet{bradshaw99} recently determined that Sco X-1 is at a
distance $d=2.8\pm 0.3 \ {\rm kpc}$, giving a luminosity close to the
Eddington value for cosmic abundances, $2\times 10^{38} \ {\rm erg \
s^{-1}}$, implying $\dot M\approx 2\times 10^{-8} \ M_\odot {\rm \
yr^{-1}}$ and a required quadrupole $Q_{22} \approx 1.6\times
10^{38}$~g~cm$^2$~$(300 \ {\rm Hz}/\nu_s)^{5/2}$ for equilibrium GW
emission.

\subsection{Origin of the Crustal Quadrupole Moment}
\label{sec:origin-of-crustal-Q}

\begin{figure}
\begin{center}
\epsfig{file=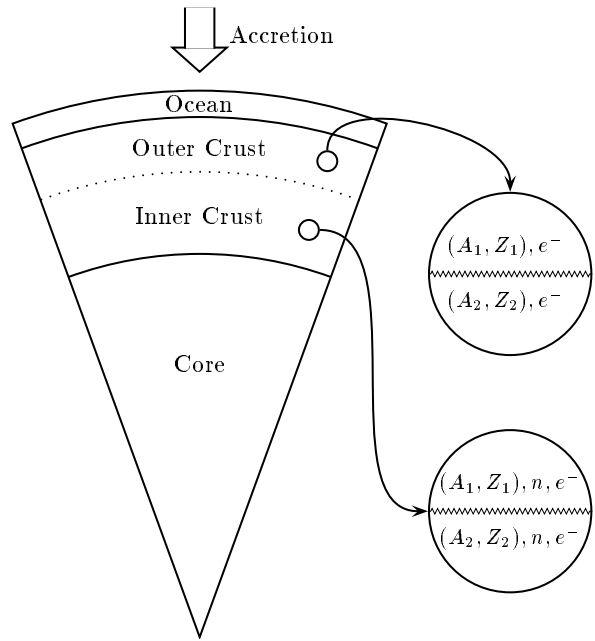,width=8.5cm}
\end{center}
\caption{\label{fig:crust-diagram} 
The schematic structure of the neutron star, displaying the inner and
outer crust lying on the liquid core. The ocean on top of the crust
ends when the material is dense enough to crystallize. The outer crust
consists of nuclei in a lattice and a background of degenerate,
relativistic electrons, whereas the inner crust (at densities
$\gtrsim(4-6)\times 10^{11} \ {\rm g \ cm^{-3}}$) has free neutrons in
addition to the nuclear lattice. }
\end{figure}

 The crust of a neutron star is a thin ($\approx1$~km) layer of
crystalline ``ordinary'' (albeit neutron-rich) matter that overlies
the liquid core composed of neutrons, protons, and electrons (see
Figure~\ref{fig:crust-diagram}).  The composition of the crust, i.e.,
the mass number $A$ and the charge $Z$ of the nuclei, varies with
depth.  As an accreted nucleus gets buried under an increasingly thick
layer of more recently accreted material, it undergoes a series of
nuclear reactions, including electron captures, neutron emissions, and
pycnonuclear reactions \citep{Sato79,HZ90,Blaes90}.  The crust
therefore consists of layers of different nuclear composition, as
indicated schematically in Figure~\ref{fig:crust-diagram}.

Essentially all the pressure in the outer crust, and an appreciable fraction of
the pressure in the inner crust, is supplied by degenerate relativistic
electrons.  The pressure must be continuous across the boundaries of
the compositional layers.  However, electron capture reactions reduce
the number of electrons per nucleon, and hence
require density jumps between the compositional layers depicted in
Figure~\ref{fig:crust-diagram}. In the outer crust, these density
jumps are as large as $\approx 10\%$, while in the inner crust the
density contrast is smaller, $\lesssim1\%$.  

\begin{figure}
\begin{center}
\epsfig{file=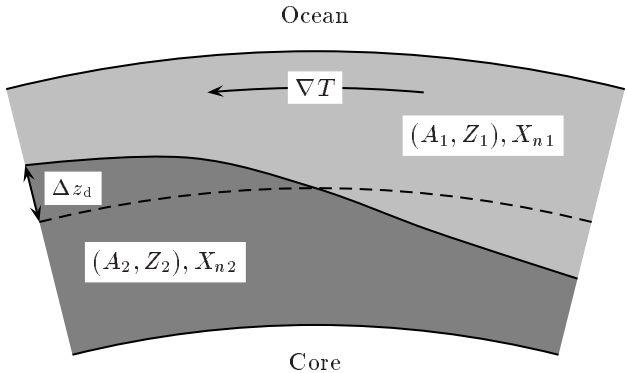,width=8.5cm}
\end{center}
\caption{\label{fig:sink-diagram} 
A cartoon description of how a transverse temperature gradient in the
crust will lead to a varying altitude for the electron captures.  The
dashed line denotes the unperturbed location of the e$^-$ capture
boundary between the layer of material composed of nuclei $(A_1,Z_1)$,
a mass fraction $X_{n1}$ of free neutrons, and free electrons, and a
layer with nuclei $(A_2,Z_2)$, $X_{n2}$ fraction of free neutrons, and
free electrons.  When a lateral temperature gradient $\nabla T$ is
introduced, the capture boundary shifts to a new location shown by the
solid line.  Note that this cartoon presumes that the crust is
infinitely rigid and therefore does not adjust elastically to the
shift $\dzd$ of the capture boundary.}
\end{figure}

\citet{BC98} showed that at the typical crustal temperatures of
accreting neutron stars, $\gtrsim2\times 10^8 {\rm K}$, the electron
capture rates are sensitive to the local temperature (see discussion
in \S~\ref{sec:el-cap-rates}). In this case, regions of the crust that
are hotter undergo electron-capture transitions at a lower density
(and thus larger radius) than the colder regions.  This effect is
illustrated schematically in Figure~\ref{fig:sink-diagram}, which
shows a patch of the crust near a capture layer where nucleus
$(A_1,Z_1)$ is transformed into $(A_2,Z_2)$.  If there is no lateral
temperature gradient, then the capture layer is spherically symmetric,
as indicated by the dashed line.  The lateral temperature gradient
(shown by the arrow) induces some extra captures on the left (hotter)
side, and reverses some captures on the right (cooler) side.  If the
lateral temperature contrast is $\delta T$, then e$^-$ captures on the
hot side of the star happen at the Fermi energy that is roughly
$\Upsilon\kB\delta T$ lower, and captures on the cold side proceed at
the Fermi energy about $\Upsilon\kB\delta T$ higher than in the
unperturbed case \citep{Bildsten98:GWs}. Typical values of $\Upsilon$
are $10-20$ (see the top panel in Figure~\ref{fig:dzd-vs-deltaT}).
Since the Fermi energy increases with depth, captures occur a distance
$\dzd$ higher in the crust on the hot side, and a distance $\dzd$
lower on the cold side. The resulting capture layer is indicated by
the solid line in Figure~\ref{fig:sink-diagram}.

Therefore, a large-scale temperature asymmetry deforms the capture
layers. Such a temperature gradient, if misaligned from the spin axis,
will give rise to a nonaxisymmetric density variation and a nonzero
quadrupole moment $Q_{22}$ {\it even if the composition of the crust
has no lateral variation.}  Of course, a lateral composition gradient
in the NS crust can exist without a temperature gradient, and would
also create a quadrupole moment.

On dimensional grounds, the quadrupole moment generated by a
temperature-dependent capture boundary is $Q_{22}\sim Q_{\rm fid}
\equiv \Delta\rho\dzd R^4$, where $\Delta\rho$ is the density jump at
the electron capture interface.  This fiducial value, $Q_{\rm fid}$,
is the quadrupole moment that would result if crustal matter just
moved horizontally to regain horizontal pressure balance.  Using this
estimate, \citet{Bildsten98:GWs} argued that a single wavy capture
boundary in the outer crust could generate $Q_{22}$ sufficient to
buffer the spinup due to accretion, provided that temperature
variations of $\sim20\%$ are present in the crust.

  However an important piece of physics is missing from this fiducial
estimate: the shear modulus of the crust $\mu$.  This must be
important to the estimate, since if $\mu$ vanishes, the crust becomes
a liquid and cannot support a non-zero $Q_{22}$. In addition, a NS
crust is thought to have a small shear modulus relative to the
pressure, $\mu/p\sim10^{-2}-10^{-3}$ typically.  Much of this paper is
concerned with improving upon the above fiducial estimate. Treating
the crust as an elastic solid, we assume the existence of a horizontal
temperature or composition gradient, and then solve for the
displacement field that brings the crust into equilibrium, with the
gravitational, pressure, and shear-stress forces all in balance.  From
the density perturbation $\delta\rho$ we calculate $Q_{22}$ for a
single electron capture layer and find that it is typically $5-50$
times smaller than the fiducial estimate, depending on the capture
layer depth (see \S~\ref{sec:quadrupole-vs-depth}). Hence, in order to
generate a $Q_{22}$ large enough for substantial GW emission, we must
include capture layers in the more massive neutron-rich part of the
crust.

Before launching into detailed calulations, we now describe why it is
plausible to assume that lateral temperature and composition gradients
are present in the crusts of accreting neutron star. We present a
detailed model of thermal gradients in the crust in
\S~\ref{sec:thermal-pert}.

\subsection{Possible Causes of Temperature Asymmetries} 
\label{sec:asymm}

 In LMXBs, accretion will replace the primordial crust after about
$\sim 5\times 10^7\ {\rm yr}\ (\dot{M}/10^{-9}M_\odot\ {\rm yr^{-1}})$.
This is much shorter than the lifetime of such systems.  While it is
plausible that the primordial crust is spherically symmetric in
composition, we suggest that the accreted crust need not be.

The accreted crust is composed of the compressed products of nuclear
burning of the accreted hydrogen and helium in the NS's upper
atmosphere.  The nuclear mix entering the top of the crust depends
sensitively on the burning conditions and is still not well known.
\citet{Schatz99} showed that the products of {\it steady-state}
burning in the upper atmosphere are a complicated mix of elements far
beyond the iron peak. The exact composition (and the average $A$)
depends on the local accretion rate, which could have a significant
non-axisymmetric piece in the presence of a weak magnetic field.

 However, except in the highest accretion rate LMXBs, nearly all of
the nuclear burning occurs in Type I X-ray bursts, sudden consumption
of fuel accumulated for hours to days prior to ignition. The
rotational modulation observed during type I X-ray bursts in several
LMXBs (first detected in 4U~1728-34 by
\citealt{strohmayer96:_millis_x_ray_variab_accret}; see
\citealt{klis99:_millis} for a review) provides conclusive evidence
that bursts themselves are not axisymmetric. Until the origin of this
symmetry breaking is clearly understood, it is plausible to postulate
that these burst asymmetries get imprinted into the crustal
composition or result from them. Finally, it is possible that there is
a feedback mechanism that causes composition asymmetries to grow:
lateral composition variations lead to temperature asymmetries (as
shown below), which, in turn, affect the burning conditions of the
elements entering the crust.

 Clearly, it is almost impossible to compute the magnitude of the
composition asymmetry from first principles, and we shall not attempt
to do so here. Instead, we postulate that such asymmetry exists at
some level and explore its consequences.  In particular, we show that
a composition asymmetry in the crust will modulate the heat flux
through it.  Thus, one of our predictions is a relation between the
modulation in the persistent thermal emission of accreting neutron
stars and their quadrupole moments.  In other words, the lateral
temperature gradient in the deep crust that generates a $Q_{22}$
sufficient to halt accretional spin-up also gives rise to a certain
thermal flux modulation, which we quantify in
\S~\ref{sec:resulting-temperature-variations} and
\S~\ref{sec:quadrupole-vs-mdot}. 

The consequences of the lateral composition asymmetry of the crust are
two-fold.  First, different elements have different charge-to-mass
ratio, and hence different thermal conductivity and neutrino
emissivity, both of which scale as $Z^2/A$.  This lateral variation of
the transport properties modulates the heat flux in the NS crust,
leading to lateral temperature variations $\delta T$.  Secondly, the
nuclear reactions in the deep crust release energy and heat it
locally. The heat release is again dependent on the particular
element, and hence varies laterally if the crust is compositionally
asymmetric, also giving rise
to a temperature gradient.  We quantify both of these thermal effects
in \S~\ref{sec:thermal-pert}.  In addition to these thermal effects, a
composition gradient generates a quadrupole moment directly, because
elements with different $Z/A$ have different characteristic electron
pressures, and the resulting transverse pressure gradient elastically
deforms the crust.

\subsection{Outline of the Paper} 

The purpose of this paper is three-fold.  In the first part, we
calculate the temperature asymmetry in the crust that can arise from
asymmetric conductivities and/or heating.  As stressed in
\S~\ref{sec:asymm}, at this point it is just a (testable!) conjecture
that there are large-scale temperature asymmetries misaligned with the
spin of the neutron star. We show how these can arise due to
composition gradients, despite the large thermal conductivity of the
core.  In \S~\ref{sec:crust-structure}, we discuss the structure of
the neutron star crust; in particular, we extend the work of
\citet{Bildsten98:GWs} and \citet{BC98} on the structure of electron
capture layers by considering regions where there is a large density
of free neutrons. In \S~\ref{sec:thermal-pert}, we calculate the
magnitude of the temperature variations induced by the composition
variations, both through their effect on the conductivity and on the
local heating rate in the crust.  We find that $10\%$ lateral
variations in the heating rate or conductivity result in $\lesssim5\%$
temperature asymmetry in the deep crust at accretion rates of order
the Eddington rate, and $\lesssim1\%$ lateral temperature variations
at $10^{-2}\dot{M}_{\rm Edd}$.

In the second part of our paper (\S~\ref{sec:crust-deformation}
and~\ref{sec:quadrupole-scalings}), we calculate the elastic
adjustment of the crust induced by a deformed electron capture layer
or a smooth composition gradient, and determine the resulting mass
quadrupole moment, $Q_{22}$.  If we consider the $Q_{22}$ generated by
a single deformed capture layer in the outer crust (which contains
only $\sim 10^{-5} M_{\odot}$), we find that $Q_{22}$ is smaller than
the fiducial estimate of \citet{Bildsten98:GWs} by a factor $20-50$.
But deformed capture boundaries in the deep, inner crust {\it can}\
generate sufficient $Q_{22}$ to halt the spin-up of neutron stars at
300~Hz, provided there are $\sim 1\%$ lateral temperature variations,
and provided the induced strains do not crack the crust. Quadrupole
moments due to multiple capture layers add linearly, and hence the
required temperature asymmetry is even smaller. Moreover, a smooth
$0.5\%$ composition gradient results in a similar quadrupole moment
even in the absence of a lateral temperature gradient.  Our solutions
exhibit typical strains $\sigma \gtrsim 10^{-2}$ at near-Eddington
accretion rates, and $\sigma\sim10^{-3}$ at $10^{-2}\dot{M}_{\rm Edd}$
(see \S~\ref{sec:maxQ}).  The former is larger than the breaking
strain for terrestrial rocks under atmospheric pressure, but is
perhaps possible for highly compressed neutron star crusts. The level
of crustal strain required to sustain the large quadrupole moments is
perhaps the most problematic feature of our model.

The third part of this paper is an investigation of the relation
between $Q_{22}$ and crustal shear stresses.  Specifically, we derive
a relation, Eq.~(\ref{eq:qform}), that expresses $Q_{22}$ as an
integral over shear stress terms in the crust.  Eq.~(\ref{eq:qform})
holds independent of any detailed model of how those stresses are
generated, and it immediately gives us an upper limit on $Q_{22}$ for
a given crustal breaking strain. In \S~\ref{sec:self-gravity}, we
estimate the correction to the results due to the neglect of the
gravitational potential perturbation (the Cowling approximation).

Finally, we close in \S~\ref{sec:conclusions} with a summary of our
efforts and a discussion of what is still missing from the theoretical
picture.

\section{Structure of the Accreted Neutron Star Crust} 
\label{sec:crust-structure}

\citet{Bildsten98:GWs} confined his discussion to the outer crust
(before neutron drip at $\rho<\rhond\approx (4-6)\times 10^{11} \
{\rm g \ cm^{-3}}$, \citealp{HZ90}), which is held up by relativistic
degenerate electrons.  However, as we found in the course of our work,
the capture layers that produce the
largest quadrupole moments are at densities much greater than neutron
drip, and so we need to model the entire crust. Our modeling of the
thermal structure of the crust and core mostly follows \citet{Brown99}.

The nuclear mix entering the top of the crust is not well known.  This
mix depends sensitively on the conditions of hydrogen and helium
burning in the upper atmosphere. For steady burning, \citet{Schatz99}
showed that the products are a complicated mix of elements far beyond
the iron peak. The more relevant case of
time-dependent nucleosynthesis in X-ray bursts is still unresolved
\citep{Rembges97,thiel97,Rembges98}.  What eventually needs to be done
is a calculation of the nuclear evolution of these complicated mixes
throughout the deep crust. Lacking these inputs, we take for our model
of the crustal composition the tabulation given by
\citet{HZ90,HZ90b}. These authors start by assuming that pure iron 
enters the top of the crust, and that at each pressure the crust is 
composed of a single nuclear species, with the 
transitions between species being abrupt.  
Starting with $^{56}$Fe, their calculations produce the sequence of
the most energetically favorable nuclei (and the range of densities
and pressures for each one) under the constraint that only electron
captures, neutron emissions, and pycnonuclear reactions are allowed
(i.e. the crust is too cold for a thermal reshuffling of nucleons to
occur).

Taking the composition to be pure instead of mixed has negligible
effect on the hydrostatic composition of the crust. However the
composition does have a large effect on the crust's thermal
conductivity, $K$; we use an estimate of $K$ that is appropriate for a
lattice of mixed composition \citep{Schatz99}.  Also, while
\citet{HZ90,HZ90b} approximate the capture transitions as infinitely
sharp, we resolve the actual, finite-thickness capture layers by
integrating the capture rate equation, following \citet{BC98}.  We
extend their work by including the presence of free neutrons, which
allows us to resolve capture layers at $\rho>\rhond$.

\subsection{Temperature Sensitivity of Electron Capture Rates}
\label{sec:el-cap-rates}

In accreting neutron stars, the nuclear transformations of the crustal
material are driven by increasing compression from the weight of the
overlying matter. The location and thickness of the reaction layers
are determined by the competition between the corresponding reaction
rate (at a given $\rho$ and $T$) and the local compression timescale,
i.e., the rate at which the local conditions are changing.  The
compression timescale at a given depth is $t_{\rm comp}=p/\dot{m}g$,
where $\dot{m}=\dot M/4\pi r^2$ is the local accretion rate,
$g=GM_r/r^2$ is the local gravitational acceleration, and $p/g$ is
approximately the column density, $\int_r^R \rho\, dr$.  This is just
the time it takes for the pressure on a fluid element to double due to
the extra hydrostatic pressure of new material added at the top.

 Consider the transformation of a region of the crust where the
predominant nucleus has charge $Ze$ and mass $A\mb$. (For simplicity,
consider a region where $\rho < \rhond$.)  As it is compressed, the
electron Fermi energy, $\Ef$, rises to the point where an electron
capture on the nucleus is energetically allowed.  The reaction
transforms an element $(A,Z)$ into $(A,Z-1)$. In practice, the mass
difference between $(A,Z-1)$ and $(A,Z-2)$ is always greater than that
between $(A,Z)$ and $(A,Z-1)$, so the subsequent reaction
$(A,Z-1)+e^{-}\rightarrow (A,Z-2)+\nu_{e}$ is very fast, and typically
proceeds immediately \citep{HZ90,Blaes90}. The second capture releases
of order 1~MeV, making the process effectively irreversible.  The two
successive electron captures can often be treated as one reaction,
with the rate-limiting step being the first capture.  At densities
greater than neutron drip, the captures are accompanied by neutron
emission (typically $6$ neutrons are emitted), and sometimes also by a
pycnonuclear reaction.

What is the exact place where the reaction becomes fast enough to
compete with compression?  At $T=0$, in order for electron captures to
proceed, the fluid element must be compressed until the electron Fermi
energy $\Ef$ exceeds the threshold energy $Q$ \citep[roughly the mass
difference between the $e^{-}$ capturer and the product,][]{Blaes90}.
Hence, at low $T$ the electron capture rate is not very
temperature-sensitive, and the location of the capture layers does not
depend on the local temperature.  The thickness of the electron
capture layers in this case is set by the need to have enough phase
space (determined by $\Ef-Q$) so that the electron captures proceed at
a rate comparable to that of the compression \citep{Blaes90}.

However, \citet{BC98} showed that when the temperature is in excess of
$2\times 10^8$~K (conditions typical for crusts of neutron stars in
LMXBs), there are enough electrons on the thermal tail of the
Fermi-Dirac distribution so that captures can proceed even when
$\Ef<Q$.  The capture rate $\Rec$ is approximately given by
\begin{equation}\label{eq:Rec}
\Rec=\left(\frac{\ln 2}{ft}\right)\frac{2Q^2 (\kB T)^3}{(m_e c^2)^5}
        \exp\left(\frac{\Ef-Q}{\kB T}\right)
\end{equation}
(equations [2] and [5] of \citealt{BC98}).  The $ft$ value is for the
first e$^-$ capture transition and depends on the degree of
forbiddenness of the reaction.  Typical $ft$ values range anywhere
from $10^3$ to $10^8$ seconds for the transitions that are relevant.
In this paper we use $ft=10^4\ {\rm sec}$ for all reactions. As we
show below, the dependence of the location of the capture layer on
$ft$ is only logarithmic, so even an error of a few orders of
magnitude is not important to our work. On the other hand, the
sensitivity to the local temperature is exponential, which is why even
modest lateral temperature gradients can generate sizeable quadrupole
moments.

\begin{figure}
\begin{center}
\epsfig{file=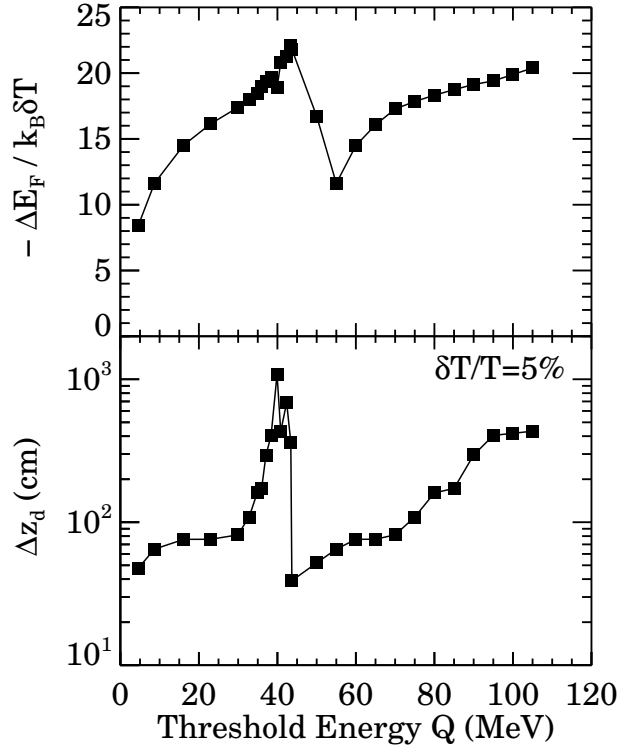}
\end{center}
\caption{%
\label{fig:dzd-vs-deltaT} 
Top: Sensitivity $\Upsilon=-\Delta\Ef/\kB\delta T$ of the location of
e$^-$ capture layers to local temperature perturbations as a function
of the capture layer threshold $Q$.
Bottom: Vertical shift of a capture layer $\dzd$ for a fiducial
temperature perturbation $\delta T/T=0.05$, disallowing elastic
readjustment of the crust, plotted as function of the capture layer
threshold energy $Q$.  These plots are generated using the formalism
described in Appendix~\ref{sec:source-terms} (see
Eqs.~[\ref{eq:dzd-vs-deltaT-precise}] and~[\ref{eq:beta-vs-deltaT}]).
The complicated dependence of $\dzd$ and $\Upsilon$ on the capture
layer depth stems from the differences in the number of electrons
captured and number of neutrons emitted in each capture layer.}
\end{figure}

How does the location of the capture layer change with the local
temperature?  Most electron captures happen when the lifetime of an
element to electron capture, $t_{\rm ec}=1/\Rec$ becomes comparable to
the local compression timescale $t_{\rm comp}$
\citep{BC98,Bildsten98:GWs}. Prethreshold captures then proceed at
$\Ef\approx Q-\Upsilon\kB T$ \citep{Bildsten98:GWs}, where
\begin{equation}\label{eq:beta-Efermi}
\Upsilon=\ln\left[\frac{p_c(\Ef)}{\dot{m}g}\frac{\ln2}{ft}
	\frac{2Q^2 (\kB T)^3}{(m_e c^2)^5}\right]
	\sim 10-20,
\end{equation}
depending on the capture layer pressure $p_c$ and other
parameters. The procedure for solving this equation is described in
Appendix~\ref{sec:source-terms}.  The function $\Upsilon$ is plotted
in the top panel of Figure~\ref{fig:dzd-vs-deltaT}. 

We define $\Delta z_d$ to be the distance that the capture layer {\it
would} shift vertically if the crust were absolutely rigid; i.e., if
there were no elastic readjustment. Then $\Delta z_d$ is approximately
given by
\begin{equation}\label{eq:dzd-vs-deltaT}
\frac{\dzd}{h} \approx \Upsilon \frac{d\ln p}{d\ln\Ef}\frac{\kB T}{Q}
\frac{\delta T}{T},
\end{equation}
where $h$ is the local scale height.  At densities lower than neutron
drip, $p\propto\Ef^4$, while for $\rho>\rhond$ the dependence of $p$
on $\Ef$ is even steeper, as electrons supply only a fraction of the
total pressure. This relation is formulated more precisely in
Appendix~\ref{sec:source-terms}, and $\dzd$ for a fiducial $5\%$
temperature perturbation $\delta T/T$ is shown in the bottom panel of
Figure~\ref{fig:dzd-vs-deltaT}.

\subsection{Capture Layers at Densities Higher than Neutron Drip}
\label{sec:Capture-Layers}

\begin{figure}
\begin{center}
\epsfig{file=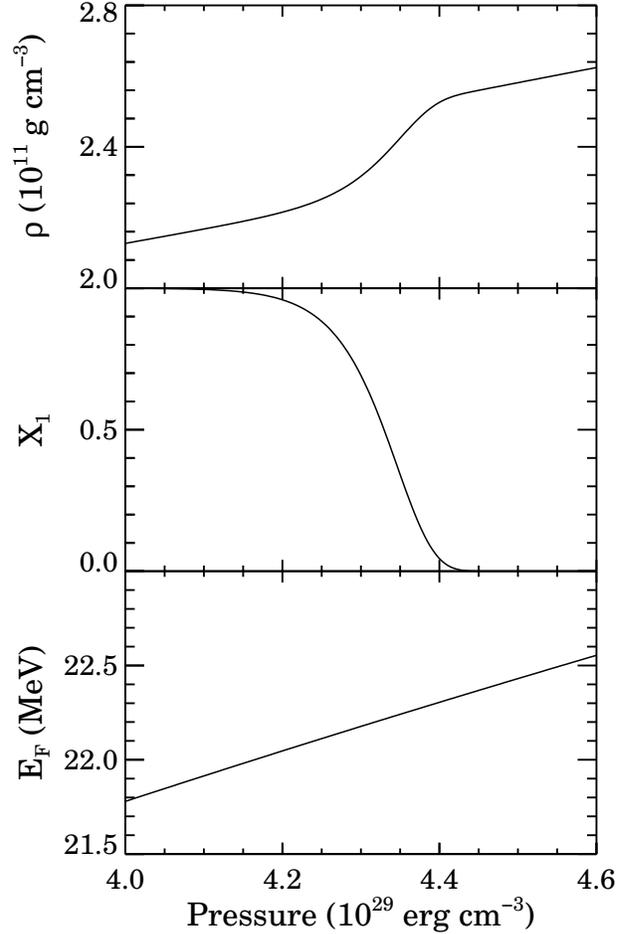}
\end{center}
\caption{\label{fig:cap-layer1} 
Structure of an electron capture layer in the outer crust, at
threshold energy $Q=23$~MeV.  The reaction is
$^{56}$Ca$+2e^{-}\rightarrow^{56}$Ar$+2\nu_e$.  The upper panel shows
the run of density with pressure, with the density jump
$\Delta\rho/\rho\approx10\%$ clearly evident.  The middle panel shows the
abundance $X_1$ of $^{56}$Ca. The bottom panel shows the electron
Fermi energy $\Ef$.  Note that the slope of $\Ef$ does not change in
the capture region and is always $d\ln\Ef/d\ln p=1/4$.}
\end{figure}

\begin{figure}
\begin{center}
\epsfig{file=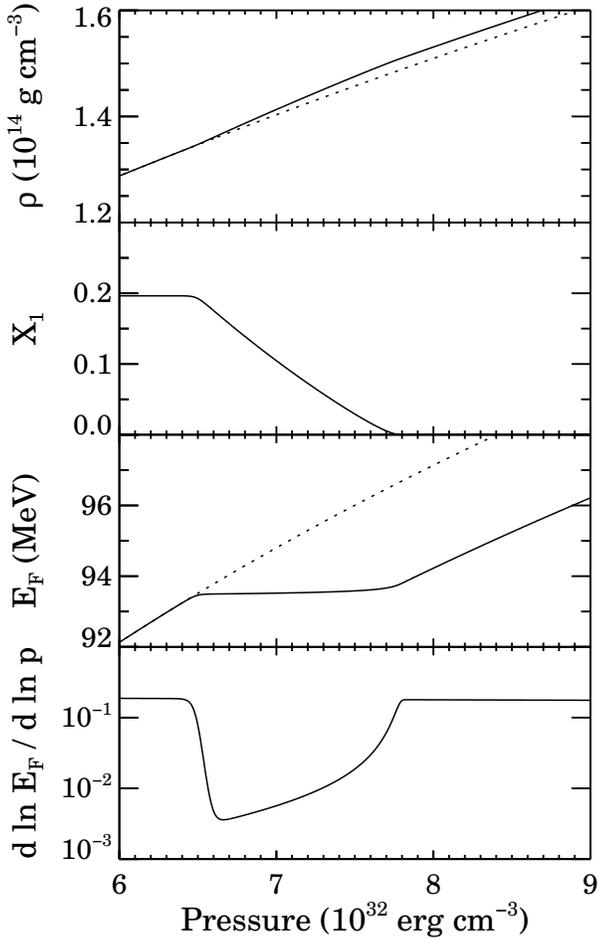}
\end{center}
\caption{\label{fig:cap-layer2} 
Top three panels are same as Fig.~\ref{fig:cap-layer1}, but for a
capture layer in the inner crust, with threshold $Q=95$~MeV.  For
comparison, dotted lines show what the run of $\rho$ (top panel) and
$\Ef$ (third panel) would be if the capture layer was not present.
The fourth (bottom) panel shows the slope $d\ln\Ef/d\ln p$, computed
using Eq.~(\ref{eq:dlnEf-dlnp}).}
\end{figure}

An exact calculation of the crust's composition would require a
reaction network large enough to allow for the possibility of several
elements undergoing captures at the same time. This is beyond the
scope of our paper.  For simplicity we assume that between the capture
layers, the crust is composed of a single species of nucleus $(A,Z)$,
electrons, and neutrons (once $\rho>\rhond$).  Capture layers,
however, contain a mix of elements: a mass fraction $X_1$ of element
$(A_1, Z_1)$ that is abundant above the capture layer (see
Figure~\ref{fig:sink-diagram}), a mass fraction $\Xn$ of free
neutrons, a mass fraction $X_2 = 1-X_1-\Xn$ of element $(A_2, Z_2)$
(the end product of electron captures onto $(A_1,Z_1)$), and electron
density $n_e=\rho/\mue\mb$.  With this simplification, we only need to
integrate one rate equation at a time.

We consider a capture layer, where elements $(A_1,Z_1)$ are
transformed into $(A_2, Z_2)$, and, at the top of the layer, where the
mass fraction of $(A_2,Z_2)$ is zero, the neutron mass fraction is
$X_{\rm n1}$.  Suppose that the reaction is a simple one, i.e. it
consists of capturing $Z_1-Z_2$ electrons, and emitting $A_1-A_2$
neutrons.  Then, at the bottom of the layer, where the reaction is
complete and the mass fraction of $(A_1,Z_1)$ is zero, the mass
fraction of free neutrons $X_{\rm n2}$ is such that $A_1/(1-X_{\rm
n1})=A_2/(1-X_{\rm n2})$.  Similarly, if electron captures and neutron
emissions in the layer are accompanied by a pycnonuclear reaction (i.e
fusing of two nuclei $(A,Z)$ into a single $(2A,2Z)$ nucleus), then
simple bookkeeping shows that $A_1/(1-X_{\rm n1})=(1/2) A_2/(1-X_{\rm
n2})$.

Now consider some point in the capture layer where both reactants
$(A_1,Z_1)$ and products $(A_2, Z_2)$ are present.  In practice,
neutron emissions are triggered by electron captures, and the
timescale for neutron emission is always much shorter than the
electron capture timescale \citep{Sato79,HZ90}.  We therefore assume
that these steps proceed simultaneously, so the proportion of neutrons 
per nucleus of each type stays fixed; e.g.,   
if only $(Z_1-Z_2)/2$ electrons have been captured, 
then exactly $(A_1-A_2)/2$ neutrons have been emitted.
This gives 
\begin{equation}\label{eq:Xn}
\Xn=\frac{X_1 X_{\rm n1}+(1-X_{\rm n1}-X_1)X_{\rm n2}}{1-X_{\rm n1}} \, .
\end{equation}
The electron mean molecular
weight is 
\begin{equation}\label{eq:mue}
\frac{1}{\mue}=\frac{X_1 Z_1}{A_1}+\frac{(1-\Xn-X_1)Z_2}{A_2} \, .
\end{equation}
The change in mass fraction $X_1$ is computed from the continuity
equation for species $(A_1,Z_1)$,
\begin{equation}\label{eq:partialx}
\frac{\partial (nX_1)}{\partial t}+\del\cdot\left(n X_1\vec{v}\right)=
        - n X_1\Rec(p,X_1,T),
\end{equation}
where $n$ is the baryon number density, $\Rec$ is the electron capture
rate (\ref{eq:Rec}), and $\vec{v}=-\hat{r}\dot{M}/4\pi r^2\rho$ is the
very slow downward motion of the fluid element. We assume that the
accretion flow is predominantly radial and time independent, so
Eq.~(\ref{eq:partialx}) becomes
\begin{equation}\label{eq:rate-spherical}
\frac{d\ln X_1}{dr}=\left(\frac{4\pi r^2}{\Mdot}\right)\rho\Rec(p,X_1,T),
\end{equation}
making clear the competition between the capture rate and the
compression timescale, $t_{\rm comp}$,  defined earlier. 

We use the sequence of reactions shown in Tables 1 and 2 of
\citet{HZ90} to describe the crust's composition. Their sequence
starts with $^{56}$Fe with $\Xn=0$ for densities $\rho<1.5\times 10^9\
{\rm gr\ cm^{-3}}$ and ends with $^{88}$Ti with $\Xn=0.8$ at densities
$\rho>1.61\times10^{13}\ {\rm gr\ cm^{-3}}$.  For the threshold energy
$Q$, we use the value of the electron chemical potential at the
(abrupt) transitions between nuclei in the calculation of
\citet{HZ90b}, as shown in their Table~2.  \citet{HZ90} stop their
calculations at $\rho>10^{13}\ {\rm gr\ cm^{-3}}$, where their EOS
state for accreted matter approaches the standard \citet{bbp} equation
of state for cold catalyzed matter, and so the authors stop their
tabulation.  At these densities the equation of state is dominated by
neutron pressure, and the exact $A$ and $Z$ of the nuclei have little
effect on the EOS.  But an important difference emphasized by
\citet{haensel97} is that $Z\gtrsim50$ at the bottom of a cold
catalyzed crust, while the accreted model has $Z\sim20$.  While this
has little effect on the EOS, the difference in $Z$ does affect the
shear modulus \citep{Sato79,haensel97}.

The crust extends to densities $\rho\approx 2\times 10^{14}\ \grcc$,
and there are many more capture layers in the deep crust, each of
which contributes to the NS quadrupole moment.  Because the tables in
\citet{HZ90} only extend to $\rho = 1.61\times10^{13}\ {\rm gr\
cm^{-3}}$, in all our calculations we insert an extra, ad hoc, movable
capture layer in the bottom part of the crust, and study the
quadrupole moment induced by this layer as function of its position.
We take the crust to be made up of $(A,Z)=(88,22)$ above the capture
layer and $(A,Z)=(82,20)$ below. We selected these values because in
most of the capture reactions listed by \citet{HZ90}, two electrons
are captured and six neutrons are released.

Figures~\ref{fig:cap-layer1} and~\ref{fig:cap-layer2} show the
structure of the capture layers at densities below and above
$\rhond$.  The top panel shows the run of density with pressure
(downward direction is to the right).  At $\rho<\rhond$
(Figure~\ref{fig:cap-layer1}), the pressure is supplied entirely by
degenerate electrons.  In this particular capture layer, $Z$ changes
from 20 to 18, so the density jump is
$\Delta\rho/\rho=\Delta\mue/\mue\approx2/Z=10\%$.  The density change
in a capture layer at $\rho>\rhond$ (Figure~\ref{fig:cap-layer2})
is much smaller, as electrons provide a much smaller fraction of the
pressure. 

For $\rho>\rhond$, the capture layers become much thicker, both in
pressure coordinates and in physical coordinates. We understand this
as follows.  Since the reaction rate~(\ref{eq:Rec}) is exponentially
sensitive to $\Ef-Q$, the transition is always sharp in $\Ef$
coordinates (see bottom panel of Figures~\ref{fig:cap-layer1}
and~\ref{fig:cap-layer2}).  However, in pressure coordinates, the
width of the capture layer is set by $\Delta p/p=(d\ln\Ef/d\ln
p)^{-1}(\Delta\Ef/\Ef)$, where $d\ln\Ef/d\ln p$ is given by
equation~(\ref{eq:dlnEf-dlnp}) in Appendix~\ref{sec:source-terms}.
Exponential sensitivity of the reaction rate ensures that
$\Delta\Ef/\Ef$ remains approximately constant.  At $\rho<\rhond$,
$d\ln\Ef/d\ln p=1/4$. But for $\rho>\rhond$, it becomes smaller, and
within the capture layer itself, extremely small, and the capture
layer becomes correspondingly thick (see bottom panel of
Figure~\ref{fig:cap-layer2}). In fact, around neutron drip, the
increased width of capture layers, coupled with the increasing number
of capture layers per unit depth, makes the capture layers overlap.

Because we only integrate one capture reaction at a time, our code
cannot deal with overlapping layers, and so we artificially disregard
several reactions indicated in \citet{HZ90}.  This should not lead to
any serious problems, as our equation of state at these depths is
insensitive to the exact $(A,Z)$.  Second, as shown in
\S~\ref{equations}, the quadrupole moments due to different capture
layers add linearly, so overlapping capture layers can in principle be
dealt with using superposition.

\begin{figure}
\begin{center}
\epsfig{file=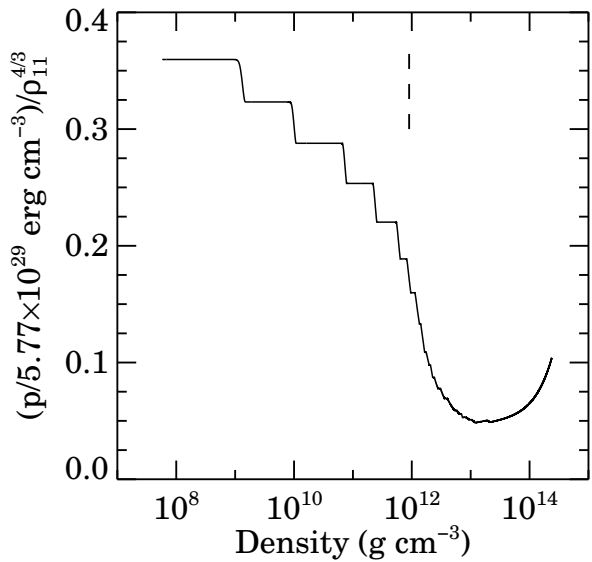}
\end{center}
\caption{\label{fig:EOS}
Equation of state used in this work. For densities below neutron drip,
the EOS is just that of relativistic electrons. Neutron drip is
indicated by a dashed vertical line. We have plotted the EOS in a
manner such that the y-axis is $\mue^{-4/3}$ for $\rho<\rhond$. The
large steps at $\rho<\rhond$ are from the electron captures, whereas
at densities above neutron drip, the density contrasts weaken.}
\end{figure}

\subsection{Hydrostatic Structure of the Accreted Crust}
\label{sec:Hydr-Struct}

 With the composition set, we now discuss the equation of state. The
electron pressure is that of fully degenerate relativistic $T=0$
particles, $p_e=5.77\times10^{29}\ \ergcc\ (\rho_{11}/\mue)^{4/3}$,
where $\rho_{11}=\rho/10^{11}\ \grcc$. Free neutron pressure is given
by the $p_n(n_n)$ fit of \citet{Negele73}, where $n_n=\Xn\rho/\mb$. We
neglect the ion pressure and the non-ideal and thermal electron
effects on the equation of state. The total pressure is then
$p(\rho)=p_e(\rho)+p_n(\rho)$. The resulting $p-\rho$ relation is
shown in Figure~\ref{fig:EOS}, where we have plotted it in such a way
as to exhibit the changing balance between electron pressure and
neutron pressure.  It agrees quite well with the more sophisticated
treatments of \citet{HZ90b} and \citet{Brown99}.

In order to construct the crust, we pick a starting radius,
pressure, and mass,  and integrate the Newtonian equations of hydrostatic
balance and mass conservation, 
\begin{equation}\label{eq:hydro-mass}
\frac{dp}{dr}=-\rho g, \hskip 30 pt \frac{dM_r}{dr}=4\pi r^2\rho,
\end{equation}
(where $M_r$ is the mass enclosed inside $r$, and $g=GM_r/r^2$)
together with the rate equation~(\ref{eq:rate-spherical}) for the
appropriate species, down towards the core of the star.\footnote{
We initially used a Runge-Kutta integrator with adaptive stepsize
control.  However, the stepsize adjustment algorithm tended to take
steps that proved too large when we used them to solve 
the thermal structure and elastic perturbation equations. 
Namely, the very uneven mesh
generated by the integrator led to large roundoff errors and
convergence difficulties. Our practical solution was to limit the
maximum step size to a small fraction ($10^{-6}$) of the radius. This
generates a mesh that is mostly even and has extra resolution near
capture layers as necessary.}
Following \citet{Brown99}, we decouple the calculation of the thermal
structure from the hydrostatic calculation.  This is justified since
the equation of state is nearly independent of temperature.  Only the electron
capture rate is temperature-sensitive, and so the absolute locations
of the capture layers that we find could be in error by a few meters.
However, since the effect we are studying depends only on relative
motion of the layers, this inaccuracy is of no concern.  We start the
integration well above the crust, at $p\sim10^{21}\ \ergcc$, in order
to later apply the thermal boundary condition.  The crust begins 
where the ratio of Coulomb energy to thermal energy,
\begin{equation} \label{eq:gamma-coul}
\Gamma_{\rm Coul}\equiv \frac{Z^2e^2}{akT}
\end{equation}
(where $a=(3/4\pi n)^{1/3}$ is the internuclear spacing) exceeds
$170$.  We stop the integration when we have reached the fiducial
density $\rho\approx 2\times 10^{14}\grcc$ at the crust-core boundary
\citep[for a review and recent results see][]{Pethick95}.  Our (purely
Newtonian) fiducial crust is $1.1$~km thick and has a mass of $0.06
M_\odot$, with a mass of $10^{-4}M_\odot$ in the outer
crust. \citet{Brown99} points out that at high enough accretion rates,
the combination of the low nuclear charge $Z$ characteristic of the
accreted crust and high temperatures may melt the crust (i.e., make
$\Gamma_{\rm Coul}\le 170$) around neutron drip.  We take this
possibility into account by rerunning our $Q_{22}$ calculations with
the top of the crust at neutron drip.

\subsection{Steady-State Thermal Structure of the Crust}
\label{sec:Steady-State-Thermal}

\begin{figure*}[t]
\begin{center}
\epsfig{file=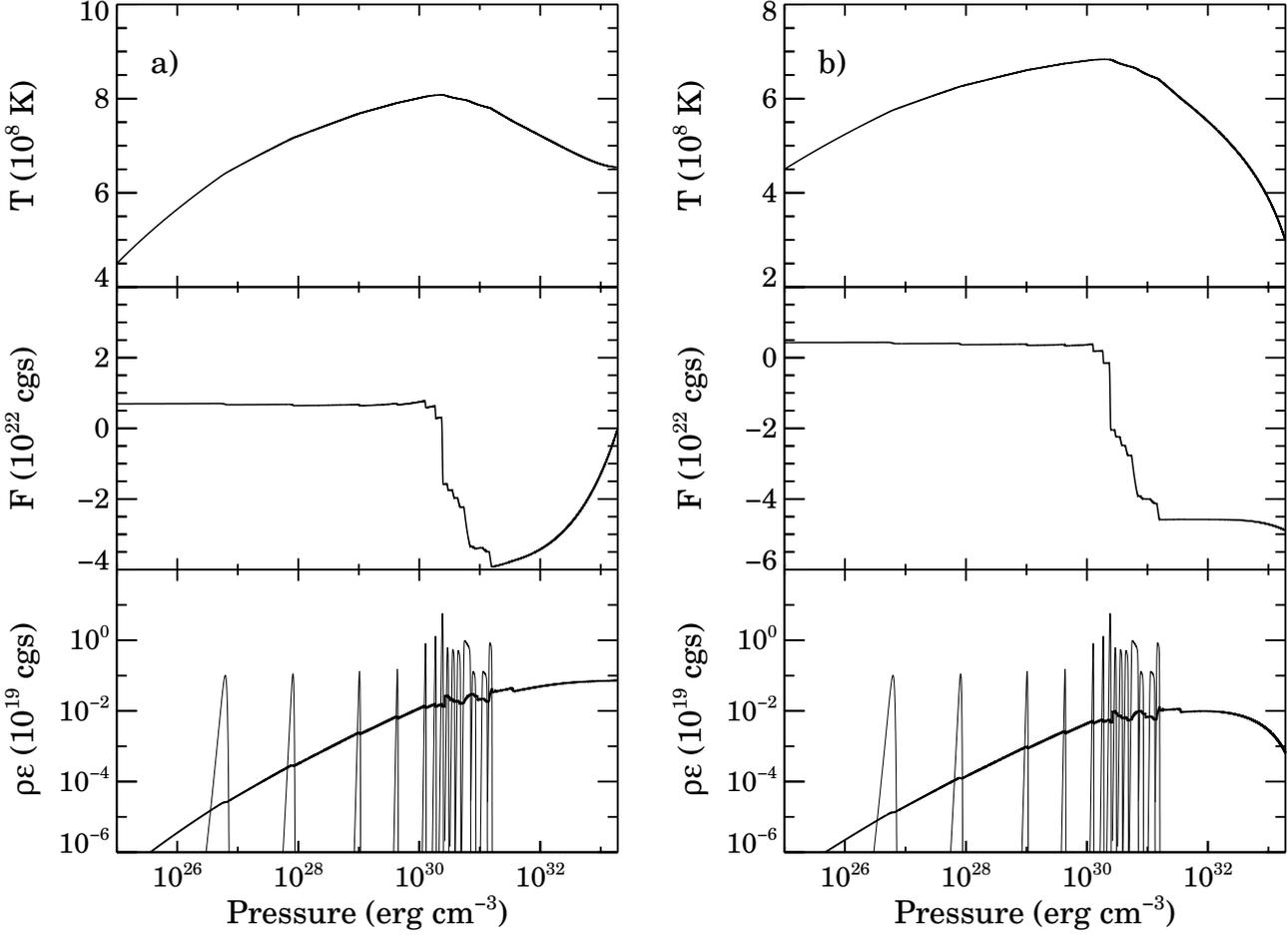}
\end{center}
\caption{\label{fig:thermal-struct} 
Left (a): Thermal structure of Model~S ($0.5\dot{M}_{\rm Edd}$
accretion rate, superfluid gap $\Delta=1\ {\rm MeV}$).  Top panel
shows the temperature of the crust. Middle panel shows the heat flux
$F_r$ in the crust ($F_r<0$ means the heat is flowing toward the
core).  Bottom panel shows the local energy release $\rho\enuc$ (in
erg~cm$^{-3}$~s$^{-1}$) due to reactions in the crust (sharp spikes)
and the local rate of neutrino cooling $\rho\ebrem$ (also in
erg~cm$^{-3}$~s$^{-1}$).
Right (b): Same as (a), but for Model~N ($0.5\dot{M}_{\rm Edd}$
accretion rate, superfluid gap $\Delta=0\ {\rm MeV}$, i.e. normal
fluid in the core).}
\end{figure*}

With the hydrostatic structure in hand, we compute the steady state
thermal profile of the crust by solving the heat equation (without GR
correction terms)
\begin{equation}\label{eq:heat}
\del\cdot\vec{F}=\rho\epsilon=\rho\left(\enuc-\ebrem\right),
\end{equation}
where the thermal flux $\vec{F}$ obeys
\begin{equation}\label{eq:flux}
\vec{F}=-K\del T \, .
\end{equation}
Here $\enuc$ is the local energy deposited by nuclear reactions,
$\ebrem$ is the local energy loss due to neutrino emission, and $K$ is
the thermal conductivity. We neglect the compressional heating, which
is negligible compared to the nuclear energy release \citep{Brown99}.
In our calculation we mostly follow \citet{Brown99} and
\citet{Brown98} as far as the microphysics is concerned, except for
our treatment of nuclear energy release due to reactions in the
crust. In particular, for $\ebrem$ we adopt a formula based on liquid
phase electron $\nu\bar{\nu}$ bremsstrahlung (\citealt{Haensel96},
Eq.~[8]) because we expect the crust to be quite impure and hence
neutrino emission due to electron-ion and electron-impurity scattering
will dominate over phonon scattering.  For the same reason we use the
electron-ion scattering conductivity (\citet{Schatz99}, Appendix A,
generalization of \citet{YU80} results) when computing $K$, rather
than phonon-mediated scattering.

\citet{Brown99} and \citet{Brown98} approximated the heat deposition
due to nuclear reactions as being uniform in the region around neutron
drip.  In this paper, since we are resolving individual capture
layers, we also resolve the heat release from them (though we find
that this more accurate treatment does not lead to significant
differences).  Let $\Enuc$ be the energy (ergs/nucleon) deposited in
the transition layer by a single reaction (usually a pair of e$^{-}$
captures, accompanied by neutron emission at $\rho > \rhond$).
The energy generation rate in the transition layer is then
\begin{equation}\label{eq:enuc}
\rho\enuc=\frac{1}{X_{\rm t}}\frac{dX}{dr} \left(\frac{\Enuc}{4\pi
r^2}\right) \left(\frac{\Mdot}{\mb}\right).
\end{equation}
where $X_{\rm t}$ is the mass fraction of the source nucleus at the
top of the layer.  The total energy (erg~s$^{-1}$) deposited in the
transition layer is $\Enuc\Mdot/\mb$.

  Equation (\ref{eq:heat}) requires two boundary conditions. The
boundary condition at the bottom of the crust is set by the ability of
the core to radiate the heat flux from the crust as neutrinos.  
For the bottom boundary condition we consider two cases: a normal core
and a superfluid core. For the normal core
we use modified Urca neutrino emissivity \citep[][{equation
[11.5.24]}]{shapiro83}. If
the core is in a superfluid state, neutrino emissivity is
suppressed%
\footnote{We neglect neutrino emission via $e$-$e$ bremsstrahlung,
which is not exponentially suppressed by nucleon superfluidity and
dominates neutrino emission from superfluid NS cores at
$T\lesssim10^8$~K \citep{kaminker99:ee-brem}}
by a factor $\exp(-\Delta/\kB T)$, where $\Delta$ is the
superfluid gap energy. In general, the gap energy will vary with the 
density in the core. However, we approximate superfluid effects by just
including an overall exponentional factor in the modified Urca
formula,
\begin{equation}\label{eq:Lcore}
L_{\rm core}=5.3\times10^{39} {\rm \ erg\ s^{-1}}
\frac{M}{M_\odot}\left(\frac{\rho_{\rm nuc}}{\rho}\right)^{1/3}
T_9^8 \exp\left(-\frac{\Delta}{\kB T}\right),
\end{equation}
where $\rho_{\rm nuc}$ is the nuclear density. If the core is composed
solely of normal particles then $\Delta=0$, while the typical values
of gap energy for superfluid cores is $\Delta\approx 1{\rm MeV}$.  The
boundary condition at the bottom of the crust is just that all heat
going into the core comes out as neutrinos, $F+L_{\rm core}/4\pi
r^2=0$, where $F$ is the radial heat flux; $\vec{F}=F\hat{r}$.

When nuclear burning in the upper atmosphere is steady, the
outer boundary condition is set by the temperature at the
hydrogen/helium burning layer, roughly 
\begin{equation}\label{eq:Tburn}
T_{\rm burn}\approx 5.3\times10^8\ {\rm K}
	\left(\frac{\dot{m}}{\dot{m}_{\rm Edd}}\right)^{2/7},
\end{equation}
where $\dot{m}$ is the local accretion rate \citep{Schatz99}.  At
sub-Eddington accretion rates ($\dot{m}\lesssim(0.1-1)\dot{m}_{\rm
Edd}$, the exact boundary is not known) the burning in the upper
atmosphere is not stable, leading to type I X-ray bursts (see
\citealt{Bildsten-review98} for a recent review).  In that case the
outer boundary condition is more complicated.  However,
\citet{Brown99} found that for high accretion rates, the local heating
in the deep crust makes the temperature there very insensitive to the
outer boundary temperature.  That is no longer true for low accretion
rates, where the temperature in the inner crust is much more sensitive
to the outer boundary temperature.  While the thermal time at the
burning layer is quite short, the thermal time in the majority of the
crust is on the order of years, much longer than either the burst
duration ($\sim10$~s) or the burst recurrence time (hours to days).
Hence, we expect that in the time-averaged sense, the outer boundary
temperature approaches that given by the steady calculation.  Thus we
adopt a very simple boundary condition by setting $T=T_{\rm burn}$ at
$p=10^{21}\ \ergcc$ (the approximate location of the hydrogen/helium
burning layer).

Because superfluidity of the core changes the flux profile in the
crust, we ran our calculations for two models.  Both have the same
accretion rate $\dot{M}=0.5\dot{M}_{\rm Edd}$, use identical
microphysics, and share the same hydrostatic structure.  However,
model~S (Figure~\ref{fig:thermal-struct}a) has a superfluid core, with
gap energy $\Delta=1\ {\rm MeV}$, while model~N
(Figure~\ref{fig:thermal-struct}b) has a normal core
(i.e. $\Delta=0$). These models are very similar to the ones obtained
by \citet{Brown99}, and we refer the reader to that paper for an
in-depth review and discussion of their overall thermal properties.

\section{Temperature Perturbations due to Physical Asymmetries}
\label{sec:thermal-pert}

 As discussed in \S~\ref{sec:asymm}, there are several possible causes
of spin-misaligned lateral temperature variations in the crust.
Rather than simply assume a given temperature contrast $\delta T$
(from which we can calculate $Q_{22}$), in this section we calculate
just how large either the composition variations or nuclear heating
variations must be to imprint a given $\delta T$.  This tells
us how large either of these effects must be to generate a sufficient
$Q_{22}$ for gravitational radiation to balance the accretion torque.

\subsection{Possible Sources for the  Temperature Variations}
\label{sec:possible-dT-sources}

Lateral differences in the crustal composition will have two
effects. The first is lateral variations in the amount of energy
deposited in the crust by nuclear reactions (since nuclear
transmutations of different elements deposit different amounts of
energy).  We denote lateral variations of this type as
$\fnuc=\delta\Enuc/\Enuc$, so that a nonzero $\fnuc$ means that more
energy is released on one side of the crust than the other%
\footnote{Since the rate at which the energy released in the crustal
nuclear reactions depends on the local accretion rate (see
Eq.~[\ref{eq:enuc}]), a similar effect would occur if the crust had
laterally uniform composition entering at the top, but different local
compression rates.}.  
Second, the charge-to-mass ratio $Z^2/A$ is likely to differ if
burning proceeds to different $A$ on different sides of the star.  The
conductivity in the crust scales as $K\propto(Z^2/A)^{-1}\rho T^{n_k}$
\citep{Schatz99}, while neutrino emissivity is
$\ebrem\propto(Z^2/A)\rho T^{n_e}$ \citep{Haensel96}.  For the
microphysics employed here, $n_k\approx 1$ and $n_e\approx 6$, so that
the conductivity and neutrino emissivity will vary laterally with the
composition variation. We denote the Eulerian perturbation in the
charge-to-mass ratio as $\fcomp=\delta(Z^2/A)/(Z^2/A)$.  Regions with
$\fcomp>0$ are more opaque and radiate neutrinos more efficiently. The
hydrostatic structure is hardly affected however, as the dependence of
the EOS on $A$ and $Z$ is very weak.

  Both $\fnuc$ and $\fcomp$ will lead to a calculable lateral
temperature variation, $\delta T$. An important issue to clarify is
the role of the core, which has high thermal conductivity and hence is
nearly isothermal. Although the core tends to smooth out temperature
variations (i.e., decrease $\delta T$), we show that it does not force
them to zero in the crust.  Because the core's thermal conductivity is
much higher than the crust's, $\delta T$ in the core is much smaller
than in the crust.  Hence, we approximate the core as perfectly
conducting and isothermal. The crust is internally heated by nuclear
reactions near neutron drip and therefore is not at the same
temperature as the core.  The crustal thermal equilibrium is set
mostly by the heat flux in the radial direction, implying that the
radial temperature gradient on the side with positive $\fnuc$ or
$\fcomp$ must be steeper in order to connect to the same core
temperature. We now calculate this $\delta T$.

\subsection{The Thermal Perturbation Equations and Boundary Conditions}
\label{sec:thermal-bcs}

While the ultimate effect of $\fnuc$ and $\fcomp$ is to shift the
location of capture layers, to first order we can calculate $\delta T$
by considering the effects of $\fnuc$ and $\fcomp$ on a spherically
symmetric background.  After computing $\delta T$, we will find in
\S~\ref{sec:crust-deformation} how the crust hydrostatically
readjusts.  To first order, we then set $\delta\rho=0$, and the
conductivity perturbation equation is simply
\begin{equation}\label{eq:delta-K}
\frac{\delta K}{K} = -\fcomp + n_k \frac{\delta T}{T}, 
\end{equation}
while the neutrino emissivity perturbation is $\delta\ebrem/\ebrem =
\fcomp + n_e \delta T/T$. 

The nuclear energy generation rate perturbation requires special
attention.  Nuclear reactions are generally quite
temperature-sensitive.  However, despite the temperature perturbation,
the total energy release of a complete capture layer depends only on
the local accretion rate and the total $\Enuc$ of the element.  A
temperature perturbation shifts the capture layer and hence leads to a
local change in $\enuc$ on scales smaller than the distance over which
the layers shift.  However, it cannot change the total amount of
energy released.  Since we neglect the shifts of capture layers at
this stage, we simply have $\delta\enuc/\enuc=\fnuc$. In some sense,
we average the energy generation rate over the scale of the entire
capture layer. This approximation simplifies the calculation
considerably. The Eulerian perturbation to $\epsilon \equiv
\enuc-\ebrem$ (the local nuclear heating rate minus the neutrino
cooling rate) is therefore
\begin{equation}\label{eq:delta-etot}
\frac{\delta\epsilon}{\epsilon}=
        \fnuc\frac{\enuc}{\epsilon}
        -\left(\fcomp+n_e\frac{\delta T}{T}\right)
        \frac{\ebrem}{\epsilon}.
\end{equation}

We assume that the angular dependence of all perturbed quantities
is $\propto Y_{lm}$, i.e. $\delta T (r,\theta,\phi)=\delta T(r)
Y_{lm}(\theta,\phi)$. Perturbing the heat equation
(\ref{eq:heat}) and the flux equation (\ref{eq:flux}) and keeping only
first-order terms gives
\begin{equation}\label{eq:delta-F}
\delta F^a=\left(\frac{\delta K}{K}F
        -K\frac{d\delta T}{dr}\right) Y_{lm} \hat{r}^a
        -\frac{K\delta T}{r} \del^a Y_{lm}
\end{equation}
and
\begin{eqnarray}
\label{eq:del2T}
\frac{1}{r^2}\frac{d}{dr}\left(r^2 K\frac{d\delta T}{dr}\right)
-l(l+1)\frac{K\delta T}{r^2}
	&=&\rho\epsilon
	\left(\frac{\delta K}{K}-\frac{\delta\epsilon}{\epsilon}\right)
	\nonumber \\
	&+& F\frac{d}{dr}\left(\frac{\delta K}{K}\right).
\end{eqnarray}
Substituting for $\delta K/K$ and $\delta\epsilon/\epsilon$, and using
Eq.~(\ref{eq:flux}), Eq.~(\ref{eq:del2T}) becomes 
\begin{eqnarray}\label{eq:delta-T-problem}
&&\frac{1}{r^2}\frac{d}{dr}\left(r^2 K\frac{d\delta T}{dr}\right)
-n_k\frac{F}{KT}K\frac{d\delta T}{dr}
-l(l+1)\frac{K\delta T}{r^2} \nonumber \\
&&-\left\{n_k\frac{F^2}{KT} 
+\rho\left(n_k(\enuc-\ebrem)+n_e\ebrem\right)
\right\}\frac{\delta T}{T}  \nonumber \\
&&=\rho\bigg\{\fcomp(2\ebrem-\enuc)-\fnuc\enuc\bigg\},
\end{eqnarray}
where we have neglected $dn_k/dr$, and for simplicity have taken
the composition perturbation to be radially uniform, so $d\fcomp/dr =0$.  

The thermal perturbation problem, Eq.~(\ref{eq:delta-T-problem}),
requires two boundary conditions. At the top of the crust, the exact
boundary condition can be obtained by matching to a flux-temperature
relation in the ocean, where the thermal profile is set by
compressional heating and that portion of the nuclear energy released
in the deep crust that flows upward through the ocean, rather than
down into the core \citep{BC95,Brown98}.  But since the ocean's
thermal conductivity is much higher than the crust's, we simplify our
calculation by adopting the boundary condition, $\delta T({\rm
top})=0$. The variation in the {\it flux} coming out of the crust is
{\it not} zero; in fact it is potentially observable.

Now consider the boundary condition at the crust-core interface.  The
thermal conductivity in the core is at least several orders of
magnitude higher than in the crust. Hence, any extra flux into the
core can be carried with a very small temperature perturbation, as we
now show. First consider a superfluid core. The boundary condition for
the spherically symmetric calculation is $F_{\rm core}=0$, since
neutrino emission is suppressed and the core cannot radiate away any
significant heat flowing into it
(\S~\ref{sec:crust-structure}). However, we are now lifting the
restriction of spherical symmetry, so heat can flow into the core on
one side and out the other.  The size of $\delta T$ in the core is
then related to the magnitude of the radial flux perturbation by
$\delta F_{\rm r, core}\sim (K_{\rm core}T/R)(\delta T/T)|_{\rm
core}$. The magnitude of the transverse flux $\delta F_{\perp,\rm
core}$ is of the same order, contrary to the situation in the crust,
where the transverse heat flux is much smaller.
The radial flux perturbation in the crust is $\delta F_{\rm r,
crust}\sim (K_{\rm crust}T/\Delta R)(\delta T/T)|_{\rm crust}$, where
$(\delta T/T)|_{\rm crust}$ is the typical magnitude of the
temperature perturbation in the crust, and $\Delta R$ is the thickness
of the crust.  Continuity of the radial flux at the crust-core
boundary then gives
\begin{equation}\label{eq:deltaT-core}
\left.\frac{\delta T}{T}\right|_{\rm core}\sim
        \frac{K_{\rm crust}}{K_{\rm core}}
        \frac{R}{\Delta R}
        \left.\frac{\delta T}{T}\right|_{\rm crust}
        \ll\left.\frac{\delta T}{T}\right|_{\rm crust} \, .
\end{equation}
When the NS core is not a superfluid, it can emit neutrinos and the
equilibrium model has a nonzero flux $F_{\rm core}$ going into the
core (see Figure~\ref{fig:thermal-struct}). The radial flux
perturbation in the core is then determined by the competition of two
terms in Eq.~(\ref{eq:delta-F}), $(\delta K/K)F_{\rm core}$ and
$Kd\delta T/dr\sim (K_{\rm core}T/R)(\delta T/T)$. The core is nearly
isothermal, so $K_{\rm core}T/R\gg F_{\rm core}$ (i.e., the proper
estimate of $F_{\rm core}$ is $K_{\rm core}\Delta T/R$, where $\Delta
T\ll T$ is the difference in temperature between, say, the center and
the crust-core boundary).  Therefore, the second term, $Kd\delta
T/dr$, dominates, and, just as in the case of a superfluid core, we
have $\delta F_{\rm r, core}\sim(K_{\rm core}T/R)(\delta T/T)|_{\rm
core}$.  Thus we again arrive at Eq. (\ref{eq:deltaT-core}), i.e., the
typical magnitude of $\delta T$ in the crust is several orders of
magnitude larger than $\delta T$ in the core.

Hence for both the normal and superfluid core, to good accuracy we can
take $\delta T = 0$ as our boundary condition at the crust-core
boundary. With this boundary condition, we do not need to model the
core, as we effectively assume that it is perfectly
conducting. However, perturbed flux {\it is} flowing through the core.

\begin{figure*}[t]
\begin{center}
\epsfig{file=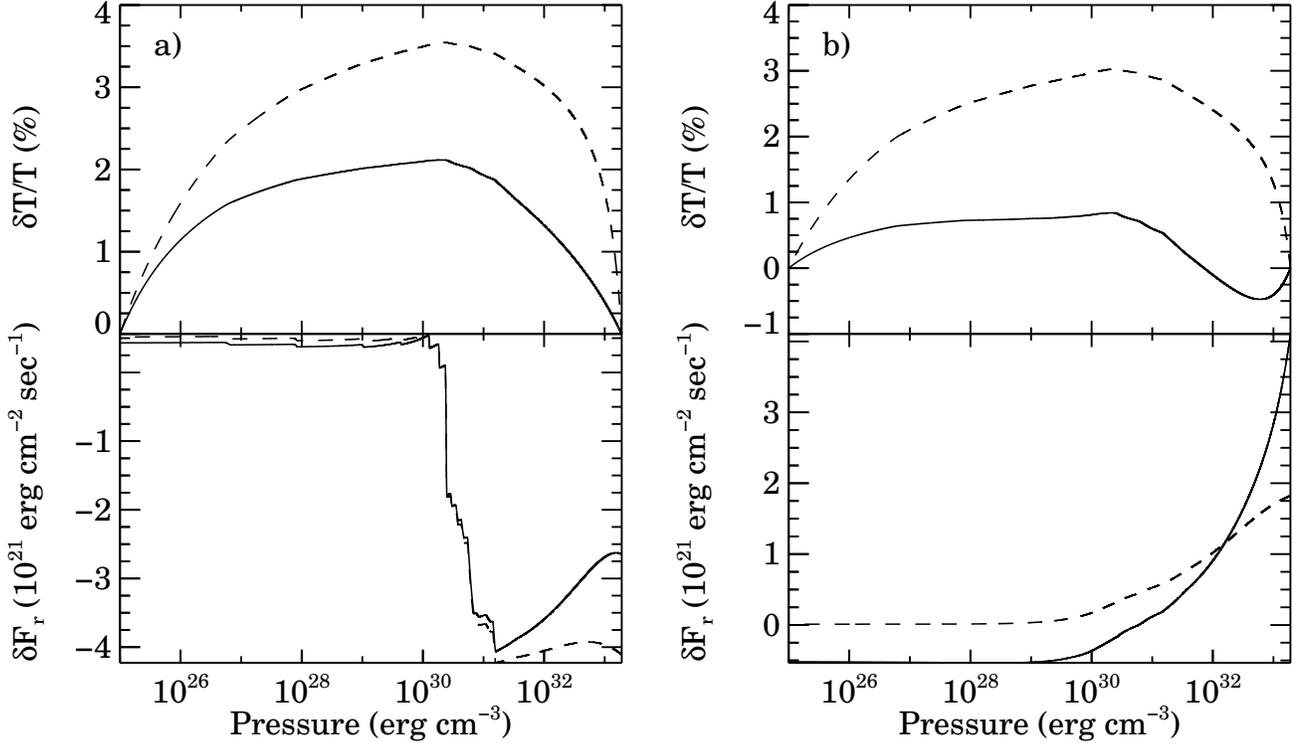}
\end{center}
\caption{\label{fig:thermal-pert}
Left (a): Temperature and flux perturbations in the crust resulting from
laterally varying energy release, $\fnuc=0.1$, but unperturbed
conductivity, $\fcomp=0$.  Angular dependence of the perturbation is
$l=2$, i.e., a quadrupole. Top panel shows the temperature perturbation
$\delta T/T$, while the bottom panel displays the radial flux
perturbation $\delta F_r$.  Solid lines correspond to Model~S
(superfluid core, $\dot{M}=0.5\dot{M}_{\rm Edd}$), while dashed lines
represent Model~N (normal core, same accretion rate).  Transverse heat
flux $\delta F_\perp$ is not shown, as it is much smaller than $\delta
F_r$ in the crust. 
Right (b): same as (a), but the energy release is
not perturbed, $\fnuc=0$, while the charge to mass ratio $Z^2/A$ is
varied laterally, $\fcomp=0.1$.}
\end{figure*}

\subsection{The Resulting Temperature Variations}
\label{sec:resulting-temperature-variations} 

Solutions of the perturbation problem (\ref{eq:delta-T-problem}) are
shown in Figure~\ref{fig:thermal-pert} for the quadrupole ($l=m=2$)
case%
\footnote{The thermal perturbation equations do not
depend on $m$, so our solutions are valid for any $l=2$ perturbation.},
for accretion rate $\dot{M}=0.5\dot{M}_{\rm Edd}$.  In
Figure~\ref{fig:thermal-pert}a we presume that the nuclear heating
varies laterally by $10\%$, i.e. $\fnuc=0.1$, but take $\fcomp=0$, so
conductivity and neutrino emissivity are unperturbed.  The resulting
temperature perturbations are shown in the top panel, with the solid
line representing the superfluid-core case (gap energy $\Delta=1\ {\rm
MeV}$), while the dashed line corresponds to the case of a normal core
($\Delta=0$).\footnote{We have likely overestimated the neutrino
emission suppression in a superfluid core (i.e. our model~S).  As
\citet{Brown99} points out, the neutrino emissivity depends very
sensitively on temperature so that even a moderate number of normal
particles in the core can radiate away an appreciable fraction of the
flux from nuclear reactions in the deep crust.  Hence it is likely
that $\delta T/T$ of model~N is more representative.}
Figure~\ref{fig:thermal-pert}b shows the solutions for the opposite
case, where $\fnuc=0$ but $\fcomp=0.1$.  Bottom panels of the figures
show the perturbation in the radial heat flux $\delta F_r$.  The
transverse flux {\it in the crust} is negligible in all cases, and is
not shown here.

We see that the typical $\delta T/T$ for $\Mdot=0.5\Mdot_{\rm Edd}$
near the bottom of the crust is roughly $(0.1-0.3)f$, where $f$ is
either $\fcomp$ or $\fnuc$.  Equivalently, the temperature
perturbation $\delta T$ is a few~$\times 10^6$~K for $\fcomp$ or
$\fnuc$ of $10\%$, with the maximum $\delta T$ attained around neutron
drip.  The temperature perturbations tend to be larger in models with
a normal core than with a superfluid core.  One might have expected
that the amplitude of $\delta T/T$ would be of order $\fnuc$ or
$\fcomp$.  This is not the case because the crust can radiate away
some of the ``extra'' flux it needs to carry by emitting neutrinos,
thus reducing the lateral temperature gradient. However, this enhanced
neutrino emission does not completely eliminate the lateral
temperature gradient.

\begin{figure}
\begin{center}
\epsfig{file=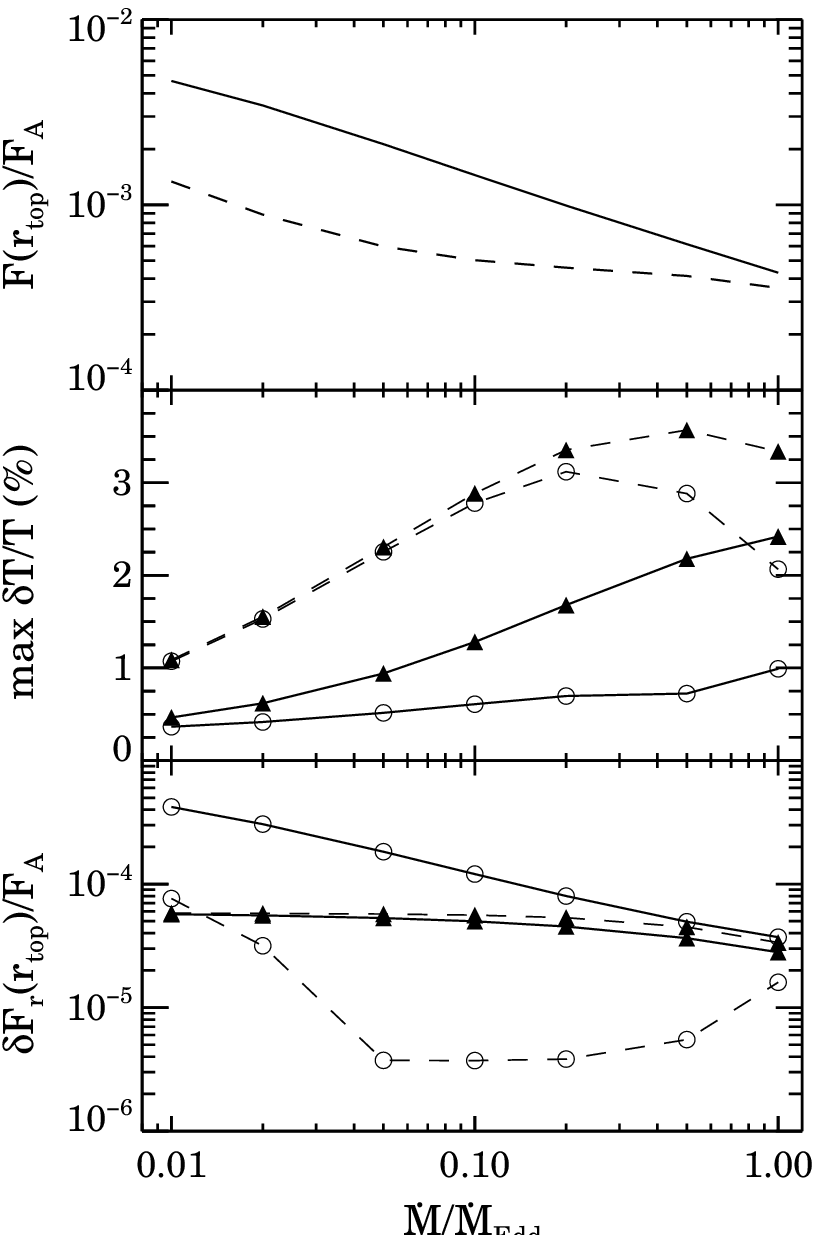}
\end{center}
\caption{\label{fig:flux-dT-pert} 
Top panel: The unperturbed flux emerging from the top of the crust due
to the (spherically symmetric) direct heat input by the nuclear
reactions around neutron drip.  The flux $F(\rtop)$ has been scaled by
the accretion flux $F_{\rm
A}=1.4\times10^{25}$~erg~cm$^2$~s$^{-1}(\dot{M}/\dot{M}_{\rm Edd})$,
so $F/F_{\rm A}=10^{-3}$ corresponds to a luminosity of $200$~keV per
accreted baryon emerging from the top of the crust.  Solid line
denotes the model with a superfluid core ($\Delta=1$~MeV), dashed line
denotes the model with a normal core ($\Delta=0$).  Middle panel: the
maximum magnitude of the crustal temperature perturbation $\delta T/T$
induced by asymmetric heat sources ($\fnuc=0.1$, lines marked with
filled triangles) and opacity/neutrino emissivity variations
($\fcomp=0.1$, lines marked with open circles).  Solid lines are the
results for the model with a superfluid core, dashed lines correspond
to the model with a normal core.  These results scale linearly with
$\fnuc$ and $\fcomp$.  Bottom panel: the perturbation to the flux
emerging from the top of the crust due to the asymmetry in nuclear
energy release ($\fnuc=0.1$) and $Z^2/A$ ($\fcomp=0.1$).  The flux
perturbation $\delta F_r$ has been scaled by the accretion flux
$F_{\rm A}$. The legend is same as in the middle panel.}
\end{figure}

In Figure~\ref{fig:flux-dT-pert}, we survey the dependence of the
temperature perturbations on the accretion rate.  Most of the heat
released by the nuclear reactions near neutron drip (roughly
$1$~MeV/accreted baryon) flows towards the core and is lost as
neutrino emission either from the crust or from the core.  However, a
fraction of the heat flows towards the surface of the star.  This flux
$F(\rtop)$ is plotted in the top panel of
Figure~\ref{fig:flux-dT-pert}, scaled by the accretion flux
$F_{A}\approx (1/4\pi R^2)(GM\dot{M}/R)$, or $\approx200$~MeV$/4\pi
R^2$/accreted baryon.  The solid line denotes a model with a normal
core, while the dashed line corresponds to the model with a superfluid
core. In the superfluid case a smaller fraction of the heat escapes to
the surface because the crust is hotter and hence crustal
bremsstrahlung is more effective at radiating the heat deposited by
nuclear reactions on the spot.

The middle panel of Figure~\ref{fig:flux-dT-pert} shows the magnitude
of the temperature perturbations that result when either the local
heating rate is perturbed by $10\%$ ($\fnuc=0.1$, lines marked with
filled triangles) or the charge-to-mass ratio is altered, leading to
lateral variations in the conductivity and neutrino emissivity
($\fcomp=0.1$, lines marked with open circles).  The typical magnitude
of $\delta T/T$ is a few percent for the models with a normal core
(dashed lines) and is somewhat smaller for models with a superfluid
core (solid lines).   We must emphasize
that our temperature perturbation calculation is linear, and hence a
$20\%$ asymmetry will result in temperature perturbations that are a
factor of 2 larger, so long as $\delta T/T \ll 1$.

These temperature perturbations displace the capture layers and create
a quadrupole moment. But in addition, they result in lateral
variations in the flux leaving the top of the crust.  The amplitude of
this flux perturbation, $\delta F_r(\rtop)$ is plotted in the bottom
panel of Figure~\ref{fig:flux-dT-pert}, scaled by the accretion flux
$F_A$. The typical magnitude of the flux modulation is $\delta
F_r/F_A\approx 10^{-4} (f/0.1)$.  These hot and cold spots, when moving
in and out of the view of the observer due to the rotation of the
neutron star, can generate a modulation in the persistent emission.
Observational implications of this effect are discussed in
\S~\ref{sec:quadrupole-scalings}.

The existence of lateral temperature variations depends on the
continual heating of the crust. When accretion halts, thermal
diffusion with slowly equalize the temperature laterally.  How long
will this process take?

First, let us compare the transverse flux perturbation in the crust
$\delta F_\perp$ to the radial one, $\delta F_r$, as one would expect
that lateral heat transport could be responsible for equalizing the
temperature. From~(\ref{eq:flux}), the background flux in the crust is
$F\sim KT/\Delta R$, while from~(\ref{eq:delta-F}) the transverse flux
perturbation $\delta F_\perp=K\delta T/R\sim F(\Delta R/R)(\delta
T/T)$.  The perturbed radial flux is the sum of two terms, both of
size $\sim F(\delta T/T)$. So for low angular order $l$, the perturbed
transverse flux is smaller than the radial piece by a factor $\sim
\Delta R/R$. Hence, the transverse heat flux in the crust does not
wash out the temperature perturbations.  The heat is transported
predominantly radially through the crust, and both radially and
transversely through the core.  Hence, if accretions stops, the
temperature perturbation will be washed out in a thermal time at the
bottom of the crust,
\begin{equation}\label{eq:thermal-time}
t_{\rm th}\approx 6\ {\rm yr}\left(\frac{p}{10^{32}\
\ergcc}\right)^{3/4} 
\end{equation}
\citep{brown98:transients}.  The transverse temperature asymmetry will
therefore persist so long as (1) accretion persists on timescales
longer than a few years, and (2) the crust has compositional
asymmetries. 

What does this mean for our quadrupole generation mechanism?  In
steadily accreting sources, we would expect the quadrupole moment not
to vary in time, and hence their spin will be set by the competition
of accretion with gravitational radiation torque.  However, as
suggested by \citet{Bildsten98:GWs}, transiently accreting sources
with long recurrence times may be able to spin up to shorter periods,
since, because of the short thermal time, the temperature variations
in the deep crust and their quadrupole moments may be lower than what
one would expect from just the time-averaged accretion rate.

\section{The Elastic Deformation of the Crust}
\label{sec:crust-deformation}

In this section we derive and solve the perturbation equations that
describe the elastic response of the crust to lateral composition
gradients.  Though our primary interest is {\it rotating\/} NSs with
$\nu_s\approx300$~Hz, we consider deformations of nonrotating,
spherically symmetric background models.  For static deformations
(i.e., no Coriolis force), centrifugal terms modify our results only
by order $(\nu_s/\nu_b)^2\sim5\%$, where $\nu_b\approx1.5$~kHz is the NS
breakup frequency.  A given level of temperature or composition
asymmetry gives rise to ``mountains'' of a certain size, where that
size is only slightly modified by rotation.  However, because the
deformed star rotates, it emits gravitational waves.

We treat two cases: one where the lateral composition
gradient is due to a wavy e$^-$ capture boundary (and the gradient is
therefore confined to a region near the layer), and another where the
lateral composition gradient is uniform throughout the crust. Real
neutron stars probably exhibit both types of composition gradients at
some level; for small distortions, their effects should add
linearly. These lateral asymmetries, treated as perturbations on a
homogeneous background, cause pressure imbalances that source a
crustal displacement field $\xi^a$.  Treating the crust as an elastic
solid, we solve for the $\xi^a$ that brings the crust back to
equilibrium, with the gravitational, pressure, and shear-stress forces
all in balance.  We then compute the density perturbation $\delta\rho$
and the resulting $Q_{22}$. An important underlying assumption is that
the stresses are small enough that the NS crust {\it can} be in static
equilibrium.  More precisely, we assume that the crust is deforming
slowly enough that the elastic part of the viscoelastic stress tensor
dominates over the viscous part. If the stresses are much larger than
the crust's yield stress, then this cannot be true.

The process by which deformation is built up in a NS is undoubtedly
complex, involving the viscoelastic response of the crust to
temperature and composition gradients that are built up over time,
during which the primordial crust is replaced by accreted matter.  A
full understanding of crustal deformation would involve solving the
time-evolution equations for all relevant aspects of the crust, from
the moment accretion starts. This would involve plastic flow or
breaking whenever the crustal yield strain was reached. Such a
``movie'' of the crust's history is well beyond our current
abilities. Instead, our solution of the perturbed hydro-elastic
equations essentially amounts to taking a ``snapshot'' of the
crust. We find that this approach gives a lot of detailed information,
but necessarily we must put in some things ``by hand.''

One important thing we put in by hand is the reference state of the
crust: for simplicity, we take it to be spherically symmetric. Namely,
we assume that if one could somehow ``undo'' the ``extra'' e$^-$
captures that cause the capture layer to be wavy, then the neutron
star would ``bounce back'' to a configuration with zero quadrupole
moment.  Since the neutron star crust may have undergone a long
history of plastic flow and cracking, it is not clear how closely the
real crust matches this picture. Fortunately, it is easy to see how
this assumption affects our results. Assume that, in the absence of
any lateral composition gradients currently driving the neutron star
away from spherical symmetry, the star would adopt a shape with
multipole moments $Q_{lm}^{\rm hist}$.  The superscript ``hist''
refers to the fact that for ``historical reasons,'' the crust has
evolved to an equilibrium shape that is non-spherical. (Of course,
almost all the solid objects we use in daily life have nonspherical
equilibrium shape: we manufacture them that way.)  Let $Q_{lm}^{\rm
pert}$ be the piece due to the current lateral composition gradient,
assuming the reference shape is spherical.  Both deviations will be
small, and to first order they add linearly:
\begin{equation}\label{eq:Qtot}
Q_{lm}^{\rm total} = Q_{lm}^{\rm hist} + Q_{lm}^{\rm pert}  \, .
\end{equation}
We guess that the dominant historical forces that
shaped $Q_{22}^{\rm hist}$ (e.g., through viscoelastic flow) are the
same ones that currently shape $Q_{22}^{\rm pert}$.  Correspondingly,
we guess that the two pieces tend to add coherently, rather than
to cancel. For example, consider a spherical shell of steel, and apply
a large inward force at two opposite points on the equator. At first,
$Q_{lm}^{\rm total}$ is just $Q_{lm}^{\rm pert}$, the distortion due
to the existing force. But over time the steel also relaxes somewhat,
and the effect is obviously to increase the total deformation.  One
might then wonder how large $Q_{22}^{\rm total}$ could grow over time,
for a fixed $Q_{22}^{\rm pert}$. It is partly to address that question
that we derive, in \S~\ref{sec:max_intro}, an upper limit limit on
$Q_{22}^{\rm total}$, independent of the relative contributions of
$Q_{lm}^{\rm hist}$ and $Q_{lm}^{\rm pert}$.

In the rest of \S~\ref{sec:crust-deformation} we omit the superscript
``pert'' from $Q_{22}^{\rm pert}$, but hopefully it is now clear that
the neutron star's total $Q_{22}$ does have another piece,
$Q_{22}^{\rm hist}$.

Now let us turn to the source of $Q_{22}^{\rm pert}$. As stated above,
we consider two types: composition gradients due to wavy capture
boundaries, and composition gradients that are radially uniform.  Our
model of the wavy capture boundary is as follows. We posit the
existence of some Eulerian temperature perturbation $\delta T \equiv
{\it Re}\{\delta T(r) Y_{lm}(\theta, \phi)\}$, superimposed on a
background that would otherwise have spherically symmetric
composition.  One possible source of such $\delta T$ is the asymmetric
heat flow due to laterally varying nuclear heating or conductivity
($Z^2/A$), as computed in \S~\ref{sec:thermal-pert}. Regions where
$\delta T(r)$ is positive (negative) have their e$^-$ capture layers
shifted to lower (higher) density, as discussed in
\S~\ref{sec:origin-of-crustal-Q} and
\S~\ref{sec:el-cap-rates}. Essentially one ends up with a crustal EOS
that is angle-dependent near the capture layer, and which requires
shear stresses to be in equilibrium.

The other source we consider is a (radially) uniform lateral
composition gradient, $\Delta \mue/\mue \equiv {\it Re} \{C\
Y_{lm}(\theta, \phi)\}$, for some constant $C$. The resulting $Q_{22}$
will scale linearly with $C$; our fiducial choice is
$C=5\times10^{-3}$.  The effect is again to give angular dependence to
the crust's EOS, but this dependence is now small and uniform rather
than large and confined to the e$^-$~capture regions.  It should be
clear that we do not think $\Delta \mue/\mue (r)$ will really be a
constant, but we just consider it in an averaged sense. 
Our use of Lagrangian $\Delta \mue$ here instead of Eulerian $\delta
\mue$ reflects the fact that we are specifying the composition
gradient that {\it would} exist (due to accretion and/or burning
asymmetries, say) if the crust did not elastically adjust.

Having explained the background solution and the sources, we now
derive the crustal perturbation equations.

\subsection{The Crustal Perturbation Equations}
\label{equations}

 Our calculation is based on the following assumptions and
approximations.  We use Newtonian gravity throughout. We model the
crust as an elastic solid with shear modulus $\mu$ and two-parameter
equation of state: $p = p(\rho,\mue)$, where $\rho$ is the density and
$\mue$ is the electron mean molecular weight. We neglect temperature
in the equation of state, except insofar as it affects $\mue$ (see
Appendix~\ref{sec:source-terms} for details).  We neglect the slight
overall downward flow of matter due to accretion; this is justified
since the ram pressure at relevant depths is completely negligible
compared to gravity or shear forces.  We presume that the crust
responds elastically, i.e., we do not allow the stresses to relax
plastically or via cracking. The stress-energy tensor of the solid is
then
\begin{equation}\label{eq:tab}
\tau_{ab} = -p(\rho,\mue) g_{ab} + \mu\biggl(\nabla_a\xi_b + \nabla_b\xi_a -
\frac{2}{3} g_{ab} \nabla^c\xi_c\biggr),
\end{equation}
where $g^{ab}$ is the flat 3-metric and $\nabla_a$ is its associated
derivative operator.  
The equation of static balance is $\nabla^a \tau_{ab} = \rho
\nabla_a\Phi$ where $\Phi$ is the gravitational potential. 

We carry out first-order perturbation theory with a spherically
symmetric background model (constructed in
\S~\ref{sec:crust-structure}) that has zero shear stress. Thus we
treat $\xi^a$ as a first order quantity. We neglect the perturbation
of the gravitational potential $\Phi$ (the Cowling approximation; but
see \S~\ref{sec:self-gravity} where we relax this approximation). We
use $\delta$ to represent Eulerian perturbations and $\Delta$ for
Lagrangian perturbations. For scalar quantities $\Lambda$, they are
related by $\Delta \Lambda = \delta \Lambda + \xi^a \nabla_a\Lambda$,
where $\xi^a$ is the displacement vector of the elastic solid from its
original state. Since the background model is spherically symmetric,
{\it both} the Eulerian and Lagrangian variations of scalars are
proportional to $Y_{lm}$, and $\Delta \Lambda(r) = \delta \Lambda(r) +
\xi_r d\Lambda/dr$.

The Lagrangian pressure perturbation depends on which effect causes
the crustal EOS to have non-trivial angular dependence.  In the case
of a smooth composition gradient, $\Delta\mue/\mue$, we write
\begin{equation}\label{eq:delta-mue-source-dp}
\Delta p = \left.\frac{\partial p}{\partial\rho}\right|_{\mue}\Delta\rho+
	\left.\frac{\partial p}{\partial\mue}\right|_{\rho}\Delta\mue,
\end{equation}
since $p=p(\rho,\mue)$. For the $\delta T$ source term, we write
\begin{eqnarray}\label{eq:delta-T-source-dp}
\Delta p &=& \left.\frac{\partial p}{\partial\rho}\right|_{T}\Delta\rho +
	\left.\frac{\partial p}{\partial\mue}\right|_{\rho}\Delta T
\\ \nonumber
&=&\left.\frac{\partial p}{\partial\rho}\right|_{T}\Delta\rho +
\left.\frac{\partial p}{\partial\mue}\right|_{\rho}
	\left(\delta T + \xi_r \frac{dT}{dr}\right),
\end{eqnarray}
where in the second equality we used the relation between Eulerian and
Lagrangian perturbations.  Evaluation of the source terms and
coefficients in equations~(\ref{eq:delta-mue-source-dp})
and~(\ref{eq:delta-T-source-dp}) is discussed in detail in
Appendix~\ref{sec:source-terms}.  The only transformation we make here
is to decompose the source terms into spherical harmonics, i.e., we
write $\Delta\mue = \Delta\mue(r) Y_{lm}(\theta,\phi)$ and $\delta T =
\delta T(r) Y_{lm}(\theta,\phi)$.  In either case, the perturbation
``sources'' a displacement vector $\xi^a$ of the form
\begin{equation}\label{eq:xi-components}
\xi^a \equiv \xi_r(r) Y_{lm} \hat r^a + \xi_\perp \beta^{-1}r
\nabla^aY_{lm}, 
\end{equation}
where $\hat r^a$ is the unit radial vector and $\beta \equiv \sqrt{l(l+1)}$.

The perturbation equations for the crust are then\footnote{ These
equations can also be derived by explicitly writing out the components
of the stress tensor and its divergence in spherical coordinates. We
found the tabulation of the components of the stress tensor in
\citet{Takeuchi72} very useful.}
\begin{equation}\label{eq:dtab}
\nabla^a \delta \tau_{ab} = \delta \rho \, g\, \hat r_b. 
\end{equation}
Here $g(r)\equiv GM_r/r^2$ is the local gravitational acceleration of
the background model, and $\delta \tau_{ab}$ is given by%
\footnote{
When not explicitly specified, all {\it physical\/} quantities are
obtained by taking real parts of the corresponding complex expressions.}
\begin{eqnarray}\label{eq:dtab-terms}
\delta \tau_{ab} &=& g_{ab} Y_{lm} \delta \tau_{rr} + e_{ab} Y_{lm}\bigl[2\mu(\xi_r/r - \partial_r\xi_r)\bigr] \\ \nonumber
&+& f_{ab} \delta \tau_{r\perp} + \Lambda_{ab} 2\mu \beta
\xi_{\perp}/r, 
\end{eqnarray}
where
\begin{mathletters}\label{eq:dt-rr-and-rperp}
\begin{eqnarray}
\delta \tau_{rr} &=& -\delta p + \mu \bigl[\frac{4}{3}\partial_r
\xi_r - \frac{4}{3}\xi_r/r + \frac{2}{3}\beta\xi_{\perp}/r\bigr], 
	\label{eq:dtrr} \\  
\delta \tau_{r\perp} &=& \mu\biggl[r\partial_r(\xi_{\perp}/r) + 
	\beta \xi_{\perp}/r\biggr],
	\label{eq:dt-rperp}
\end{eqnarray}
\end{mathletters}
and
\begin{mathletters}\label{eq:dtab-term-abbreviations}
\begin{eqnarray}
e_{ab} &\equiv& g_{ab} - \hat r_a\hat r_b, \\
f_{ab} &\equiv& \beta^{-1} r \bigl(\hat r_a \nabla_b Y_{lm} + 
\hat r_b \nabla_a Y_{lm}\bigr), \\
\Lambda_{ab} &\equiv& \beta^{-2}r^2\nabla_a \nabla_b Y_{lm} +
f_{ab}.
\end{eqnarray}
\end{mathletters}
Note that there is no $\delta\mu$ term in Eq.~(\ref{eq:dtab}), since
in Eq.~(\ref{eq:tab}) the shear modulus $\mu$ is multiplied by terms
involving $\xi_a$, which vanish in the background model.  The Eulerian
density perturbation $\delta \rho$ in Eq.~(\ref{eq:dtab}) follows from
the perturbed continuity equation,
\begin{equation}\label{eq:delta-rho}
\delta\rho =  -\nabla^a(\rho \xi_a)=
-\left\{
	\frac{1}{r^2}\frac{\partial}{\partial r}\left(r^2\rho\xi_r\right)
	-\rho\beta\frac{\xi_\perp}{r}
\right\}Y_{lm},
\end{equation}
and the Eulerian pressure perturbation $\delta p = \Delta p + \rho g
\xi_r $ in Eq.~(\ref{eq:dtrr}) follows from either
Eq.~(\ref{eq:delta-mue-source-dp}) or
Eq.~(\ref{eq:delta-T-source-dp}), depending on the type of 
perturbation imposed upon the crust.

The radial and perpendicular pieces of Eq.~(\ref{eq:dtab}) yield two
second-order equations for the variables $\xi_r$ and
$\xi_\perp$. Following \citet{McDermott88}, we rewrite these as four
first-order equations, using variables
\begin{eqnarray}
z_1\equiv\frac{\xi_r}{r},&\qquad& z_2 \equiv  \frac{\Delta\tau_{rr}}{p}
 =  \frac{\delta\tau_{rr}}{p} - z_1 \frac{d\ln p}{d\ln r} \\ \nonumber
z_3\equiv\frac{\xi_\perp}{\beta r},&\qquad&
z_4 \equiv \frac{\Delta \tau_{r\perp}}{\beta p} =
 \frac{\delta\tau_{r\perp}}{\beta p}  \, . 
\end{eqnarray}
The resulting equations are
\begin{mathletters}\label{eq:perts}
\begin{eqnarray}
\frac{dz_1}{d\ln r} &=& -\left(1+2\frac{\alpha_2}{\alpha_3}-
\mathbf{\frac{\alpha_4}{\alpha_3}}\right)z_1
        +\frac{1}{\alpha_3}z_2 \\ \nonumber
        &+&l(l+1)\frac{\alpha_2}{\alpha_3}z_3
        +\mathbf{\frac{1}{\alpha_3} \Delta S}, \\
\frac{dz_2}{d\ln r} &=& \left(\tilde U\tilde V-4\tilde V
                +12\Gamma\frac{\alpha_1}{\alpha_3}
-\mathbf{4\frac{\alpha_1 \alpha_4}{\alpha_3}}\right)z_1 \\ \nonumber
        &+&\left(\tilde V-4\frac{\alpha_1}{\alpha_3}\right)z_2 
        + l(l+1)\left(\tilde
                V-6\Gamma\frac{\alpha_1}{\alpha_3}\right)z_3
	\\ \nonumber
        &+& l(l+1)z_4
        -{\mathbf 4\frac{\alpha_1}{\alpha_3}\Delta S}, \\
\frac{dz_3}{d\ln r} &=& -z_1+\frac{1}{\alpha_1}z_4, \\
\frac{dz_4}{d\ln r} &=& 
        \left(\tilde V-6\Gamma\frac{\alpha_1}{\alpha_3}
+\mathbf{2\frac{\alpha_1\alpha_4}{\alpha_3}}\right)z_1
        -\frac{\alpha_2}{\alpha_3}z_2  \\ \nonumber
        &+&\frac{2}{\alpha_3}\left\{\left[2l(l+1)-1\right]\alpha_1\alpha_2
                +2\left[l(l+1)-1\right]\alpha_1^2\right\}z_3 \\ \nonumber
        &+&\left(\tilde V-3\right)z_4
        +{\mathbf 2\frac{\alpha_1}{\alpha_3}\Delta S},
\end{eqnarray}
\end{mathletters}
where 
\begin{eqnarray}\label{eq:defs}
\tilde U \equiv \frac{d\ln g}{d\ln r} + 2 , \ \ \ \ \ \ \ \tilde V
\equiv \frac{\rho \,g\,r}{p} = -\frac{d\ln p}{d\ln r}, \\ \nonumber
\alpha_1 \equiv \mu/p , \ \ \ \ \ \ \ 
\alpha_2 \equiv \Gamma - \frac{2}{3}\frac{\mu}{p} , 
\ \ \ \ \alpha_3 \equiv \Gamma +\frac{4}{3}\frac{\mu}{p}.
\end{eqnarray}
The terms $\Gamma$, $\alpha_4$, and the source term $\Delta S$ depend
on the type of perturbation.  In the case where the
perturbations are due to a lateral composition gradient $\Delta\mue$,
the derivatives in Eq.~(\ref{eq:delta-mue-source-dp}) are carried out
at constant composition $\mue$, and we have
\begin{equation}\label{eq:mue-source-term}
\Gamma=\left.\frac{\partial\ln p}{\partial\ln\rho}\right|_{\mue}, \ \
\alpha_4 = 0, \ \   
\Delta S = \left.\frac{\partial\ln p
        }{\partial\ln\mue}\right|_\rho\frac{\Delta\mue}{\mue}.
\end{equation}
On the other hand, in case of the $\delta T$ source term, the
perturbations in Eq.~(\ref{eq:delta-T-source-dp}) are at constant
temperature, so
\begin{equation}\label{eq:thermal-source-term}
\Gamma=\left.\frac{\partial\ln p}{\partial\ln\rho}\right|_{T}, \ \ 
\alpha_4 = \left.\frac{\partial\ln p}{\partial\ln T}\right|_\rho 
\frac{d\ln T}{d\ln r}, \ \ 
\Delta S = 
        \left.\frac{\partial\ln p}{\partial\ln T}\right|_\rho
        \frac{\delta T}{T} \; .
\end{equation}
We describe how to compute the various thermodynamic derivatives in
the above equations in Appendix~\ref{sec:source-terms}.

Except for the terms involving $\alpha_4$ and $\Delta S$, which we
have highlighted by writing them in boldface, Eqs.~(\ref{eq:perts})
are the same as the zero-frequency limit of Eqs.~(13a-d) in
\citet{McDermott88}.  However, in the case of adiabatic pulsations
considered by \citet{McDermott88}, one writes $\Delta p = (\partial
p/\partial\rho)|_s\Delta\rho$, where the partial derivative is taken at
constant entropy.  In this case, there are no source terms $\Delta S$,
no terms involving the temperature gradient $dT/dr$, and $\Gamma$ is
just the adiabatic index $\Gamma_1\equiv (d\ln p/d\ln\rho)_{s}$.  The
$\Delta S$ source term in Eq.~(\ref{eq:perts}) arises from the fact
that our perturbations (Eqs.~[\ref{eq:delta-mue-source-dp}]
or~[\ref{eq:delta-T-source-dp}]) involve an explicit change in
composition as well as a change in density.  In addition, our $\Gamma$
term is defined differently from \citet{McDermott88}, and depends on
the type of perturbation.  Consequently our terms $\alpha_2$ and
$\alpha_3$, defined in Eq.~(\ref{eq:defs}), are also different from
\citet{McDermott88}.

Because Eqs.~(\ref{eq:perts}) are just linear equations with an
inhomogeneous forcing term $\Delta S$ proportional to $\Delta \mue$ or
$\delta T$, it is clear that the resulting quadrupole moment is linear
in $\Delta \mue$ or $\delta T$, respectively.  For the $\delta T$
perturbations, the coefficient $(\partial\ln p/\partial\ln T)|_\rho$
in $\Delta S$ vanishes except near capture boundaries, and so the
total $Q_{22}$ for the NS is just the sum of the $Q_{22}$'s generated
by each capture layer individually. In \S~\ref{sec:nature-solutions}
and~\ref{sec:quadrupole-scalings} we are therefore justified in
looking at solutions for a single capture layer, since multiple
capture layers can be dealt with by superposition.

\subsection{Boundary Conditions and Solution Methods}
\label{sec:boundary-conditions}

The solid crust lies between a liquid ocean and a liquid core.  In the
liquid, $\alpha_1=0$ and $\alpha_2=\alpha_3=\Gamma$.  By integrating
Eqs.~(\ref{eq:perts}) across the crust-ocean boundary at $r=\rtop$,
and the crust-core boundary at $r=\rbot$, it is easy to see that
$z_1$, $z_2$, and $z_4$ (i.e., the radial displacement $\xi_r$, and
tractions $\Delta\tau_{rr}$ and $\Delta\tau_{r\perp}$) must be
continuous when going from the crust to a liquid.  On the other hand,
it follows from Eq.~(\ref{eq:perts}c) that $z_3$ (the transverse
displacement $\xi_\perp$) is allowed to have an arbitrary
discontinuity at $\rtop$ and $\rbot$. These connection conditions,
however, require the knowledge of $\xi_r$, $\Delta\tau_{rr}$, and
$\Delta\tau_{r\perp}$ in the liquid parts of the neutron star. We now
show how to use physical considerations to remove this requirement.

For static solutions with $l\neq0$, the Eulerian pressure perturbation
$\delta p(r) Y_{lm}$ must be zero everywhere in the liquid parts of
the star, since a non-zero $\delta p$ would lead to a horizontal
pressure gradient in the fluid $\delta p(r) \nabla_a Y_{lm}$, which
would (in the absence of counterbalancing shear stresses or
perturbations of the gravitational field) cause the fluid to flow.
Hence, at a liquid-crust boundary, we have $\Delta\tau_{rr}({\rm
crust})=-\Delta p({\rm liquid})=\rho g\xi_r$. Therefore, at $\rtop$
and $\rbot$, the Eulerian $\delta\tau_{rr}=0$.  Using this, along with
the constraint that $\delta \tau_{r\perp}$ must vanish in the fluid
(inviscid fluid cannot support shear stresses) one arrives at the
following boundary conditions (in agreement with \citet{McDermott88}):
\begin{equation}\label{eq:bc1}
z_2 = \tilde V z_1 , \ \ \ \  z_4 = 0.
\end{equation}
Note that the above boundary conditions do not require the knowledge
of displacements and tractions in the liquid parts of the star.

Equation~(\ref{eq:bc1}) assumes that there is no density discontinuity
at the solid-liquid boundary. However, at the interface between the
crust and the ocean there may be a discontinuity
$(\rho_s-\rho_l)/\rho\sim5\times10^{-5}$ (where $\rho_s$ and $\rho_l$
are densities of solid and liquid at the interface) arising from the
latent heat of melting of the crust.  There may also be a density jump
of a few percent associated with a phase transition between the inner
crust and the core of the neutron star \citep{PethickRavenhall95,
Pethick95}. When there is a density discontinuity, the boundary
condition at the edge of the solid is modified to
\begin{equation}\label{eq:bc2}
z_2 = \tilde{V} \frac{\rho_l}{\rho_s} z_1 , \ \ \ \  z_4 = 0.
\end{equation}
Of course, Eq.~{(\ref{eq:bc1}) is just a special case of
Eq.~{(\ref{eq:bc2}), where $\rho_s = \rho_l$ at the interface.

Thus we have a system of 4 ODE's with 4 boundary conditions (2 at each
boundary), with non-zero source terms in the interior region.  The
system can be solved in any number of standard ways.  For example, one
can choose the values of $z_1$ and $z_3$ at the top of the crust,
compute $z_2$ and $z_4$ from the boundary conditions, then integrate
to the bottom of the crust (using a Runge-Kutta type algorithm) and
calculate the residuals of the two boundary conditions there.  Then
one can use a 2-dimensional Newton's method to find appropriate
starting values $z_1$ and $z_3$.  In practice, we found this method to
be {\it unstable}.  On the other hand, we find that a Henyey-type
relaxation algorithm \citep[as described in][]{NumericalRecipes}
converges very reliably. However, roundoff errors can cause trouble if
the relaxation mesh is not fine enough and uniform, as already noted.

Given the solutions of Eqs.~(\ref{eq:perts}), the perturbed multipole
moments $Q_{lm}$ are computed in terms of the $z_i$ variables by using
Eq.~(\ref{qlm}),
\begin{equation}\label{eq:qlm-delta-rho}
Q_{lm}= l\int_{\rbot}^{\rtop} 
	\bigl(z_1+(l+1)z_3\bigr) r^{l+2}\rho dr 
	-\left.\bigl[r^{l+3}\rho z_1\bigr]\right|_{\rbot}^{\rtop},
\end{equation}
where $\rbot$ and $\rtop$ are the bottom and top of the crust,
respectively.  If there is a density discontinuity at the crust-core
or the crust-ocean interface, then $\rho$ in the boundary term of
(\ref{eq:qlm-delta-rho}) is replaced by $\rho_l$. Since the
calculation of the quadrupole moments involves subtracting
two large numbers to get a small number, it is useful to have an
independent accuracy check.  An alternate expression for the
quadrupole moment is derived by finding $\delta\rho$ from
Eq.~(\ref{eq:dtab}), integrating it using Eq.~(\ref{qlm}), using the
boundary conditions and carrying out substantial algebra to show 
\begin{eqnarray}\label{eq:qlm-stress}
Q_{lm} =&-& \int_{\rbot}^{\rtop}
  \frac{1}{\tilde{V}}
  \bigg\{ 
	l(l+1)z_4 - 
    2\alpha_1\left(2\frac{dz_1}{d\ln r} + l(l+1)z_3\right)
	\nonumber \\ 
	&+& \left(l+4-\tilde U\right)
	\left(z_2 - \tilde V z_1\right)
	\bigg\}
	r^{l+2}\rho dr.
\end{eqnarray}
Unlike Eq.~(\ref{eq:qlm-delta-rho}), this expression for the multipole
moment remains unchanged even when there are density jumps at
solid-liquid interfaces.  In practice, we refine our solutions to
Eqs.~(\ref{eq:perts}) until $Q_{lm}$ computed from
Eqs.~(\ref{eq:qlm-delta-rho}) and~(\ref{eq:qlm-stress}) differ
fractionally by less than $10^{-6}$.

Since the crust is geometrically thin, one may wonder if accurate
results may be obtained by adopting a plane-parallel approximation,
i.e., neglecting non-sphericity of the crust.  In this case, the
radial momentum balance equation is 
\begin{equation}\label{eq:plane-parallel-mom-balance}
i k \delta\tau_{r\perp}+\frac{d}{dr}\delta\tau_{rr} = \delta\rho g,
\end{equation}
where the lateral ($x$) dependence of all perturbed quantities is
$e^{ikx}$ and $k\sim\sqrt{l(l+1)}/R$ is the transverse wave number.
Integrating the above equation from $\rbot$ to $\rtop$ and using the
boundary condition $\delta\tau_{rr}=0$, we get a plane-parallel
estimate for the quadrupole moment, 
\begin{equation}\label{eq:plane-parallel-quadrupole}
Q_{22}^{\rm pp}=R^4\int
\frac{ik\delta\tau_{r\perp}}{g}dr 
=l(l+1)R^3 \int\frac{1}{\tilde V}z_4\rho r dr.
\end{equation}
Compare the above expression to Eq.~(\ref{eq:qlm-stress}).  In the
plane-parallel case, only the $\delta\tau_{r\perp}$ component of the
shear stress contributes to the quadrupole moment.  As it turns out,
the $\delta\tau_{rr}$ component of the shear stress is much larger
(see \S~\ref{sec:nature-solutions}). In other words, the crust holds
up the majority of the quadrupole moment by stretching like a spring
with a weight attached to it, rather than bending like a crossbar. In
the plane-parallel case, this vertical motion of the crust is fully
taken into account. However, because of the artificial symmetry of
Eq.~(\ref{eq:plane-parallel-mom-balance}), the dominant stress term
does not enter in the plane parallel expression for the quadrupole
moment, Eq.~(\ref{eq:plane-parallel-quadrupole}).  Quadrupole moments
computed using the plane-parallel approximation
(Eq.~[\ref{eq:plane-parallel-quadrupole}]) underestimate the values
computed using full spherical symmetry (i.e., from
Eqs.~[\ref{eq:qlm-delta-rho}] or~[\ref{eq:qlm-stress}]) by large
factors ($\sim10$).

\subsection{The Nature of the Solutions}
\label{sec:nature-solutions}

\begin{figure}
\begin{center}
\epsfig{file=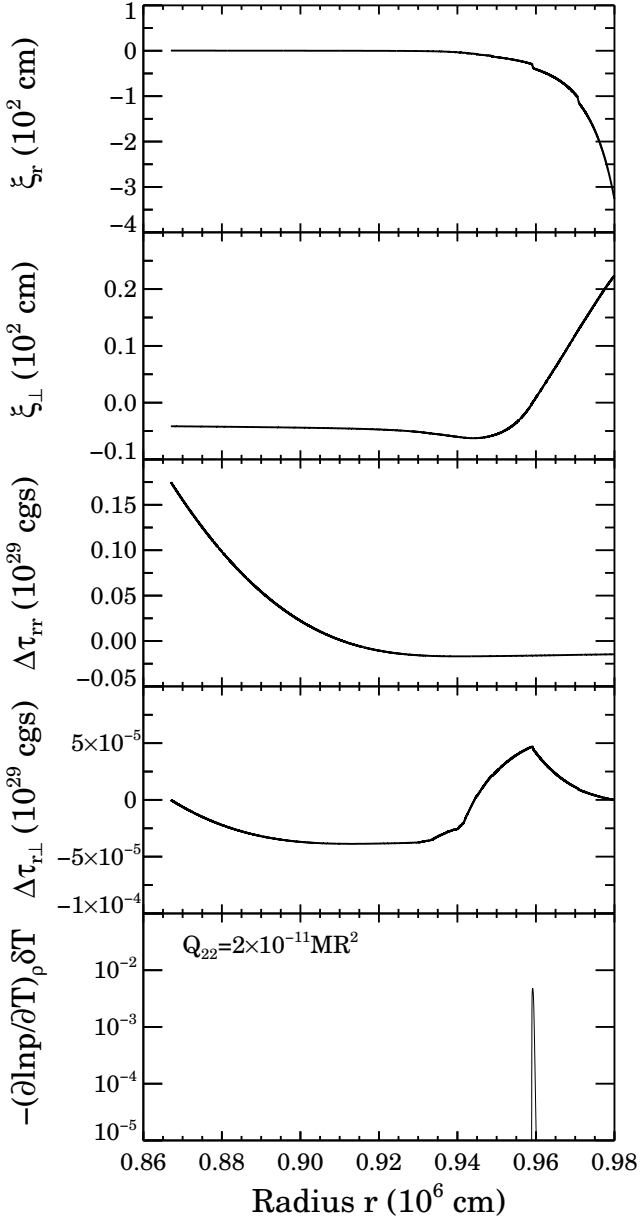}
\end{center}
\caption{%
\label{fig:nthermal-shallow-displ}\label{fig:shallow-displ}
Response of the crust to a temperature perturbation due to an
$\fnuc=0.1$ asymmetry of the nuclear heating rate in the crust.  The
background model has a normal core (gap energy $\Delta=0$) and the
accretion rate is $0.5\dot{M}_{\rm Edd}$.  Only one shallow capture
layer ($Q=23$~MeV) is activated.  Top two panels show vertical and
horizontal Lagrangian displacements, $\xi_r$ and $\xi_\perp$, in
meters, the two middle panels show the stresses, $\Delta\tau_{rr}$ and
$\Delta\tau_{r\perp}$, in $10^{29} \ergcc$. The bottom panel shows the
negative of the source term, $-(\partial\ln p/\partial T)_{\rho}
\delta T$ (see Eq.~[\ref{eq:thermal-source-term}] and
Appendix~\ref{sec:source-terms}).}
\end{figure}

\begin{figure}
\begin{center}
\epsfig{file=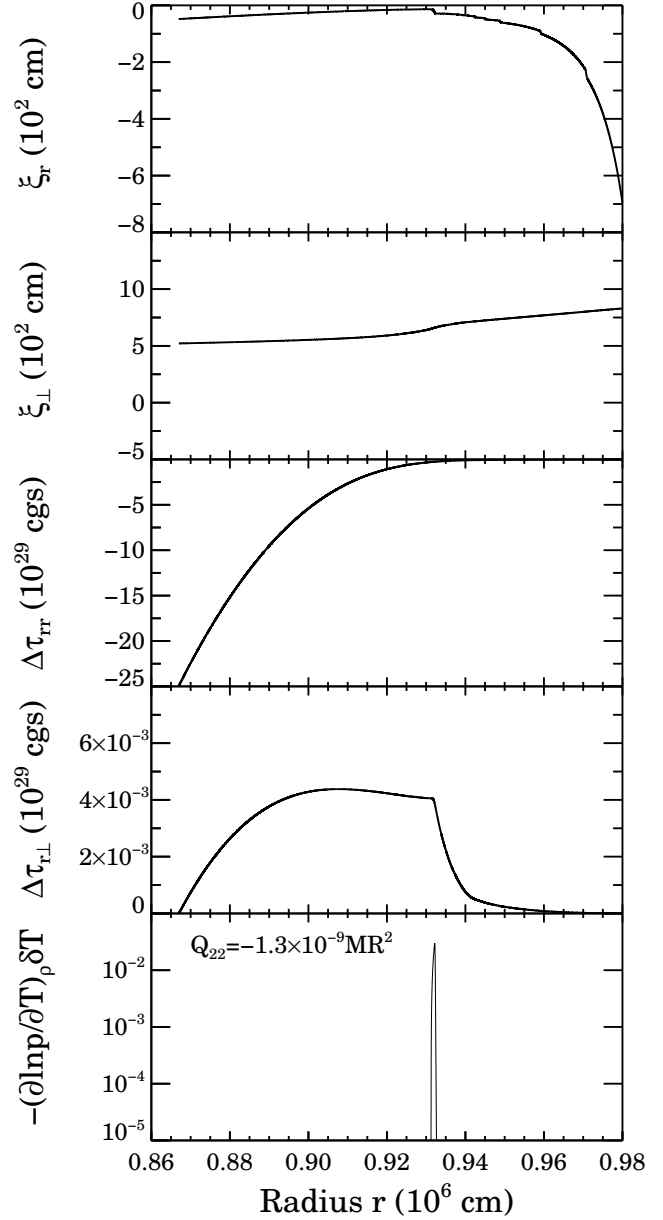}
\end{center}
\caption{%
\label{fig:nthermal-medium-displ}\label{fig:medium-displ}
Same as Figure~\ref{fig:nthermal-shallow-displ}, but for a displacement
of a capture layer at intermediate depth, $Q=42.4$~MeV.}
\end{figure}

\begin{figure}
\begin{center}
\epsfig{file=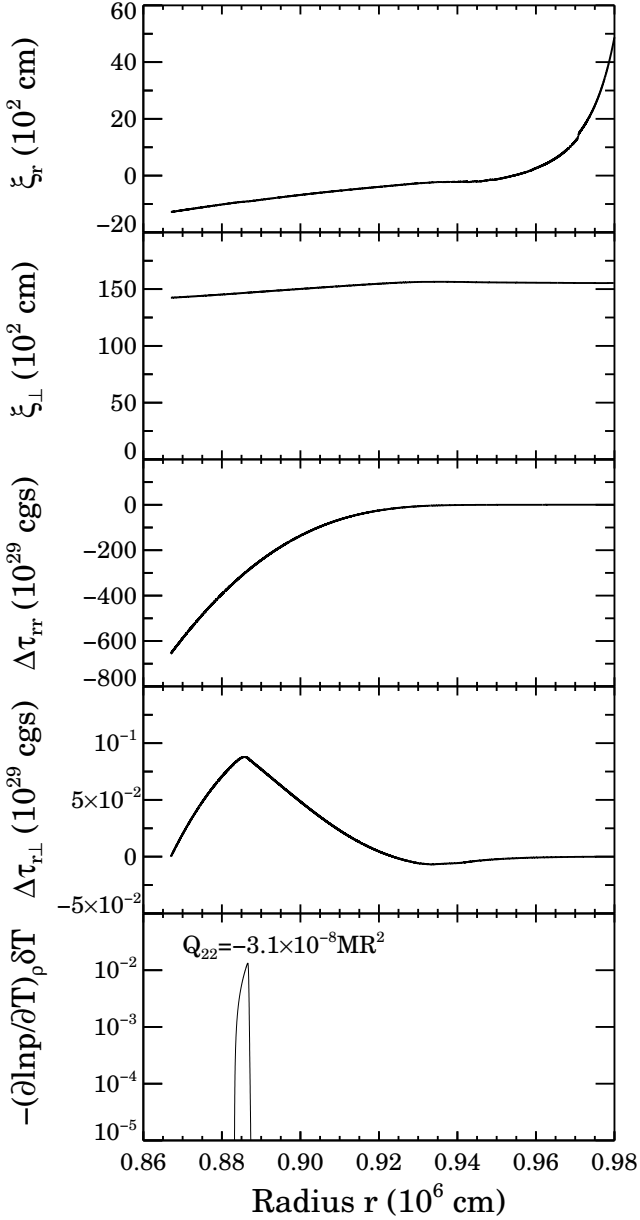}
\end{center}
\caption{%
\label{fig:nthermal-deep-displ}\label{fig:deep-displ}
Same as Figure~\ref{fig:nthermal-shallow-displ}, but for a deep
capture layer, $Q=95$~MeV.}
\end{figure}

\begin{figure}
\begin{center}
\epsfig{file=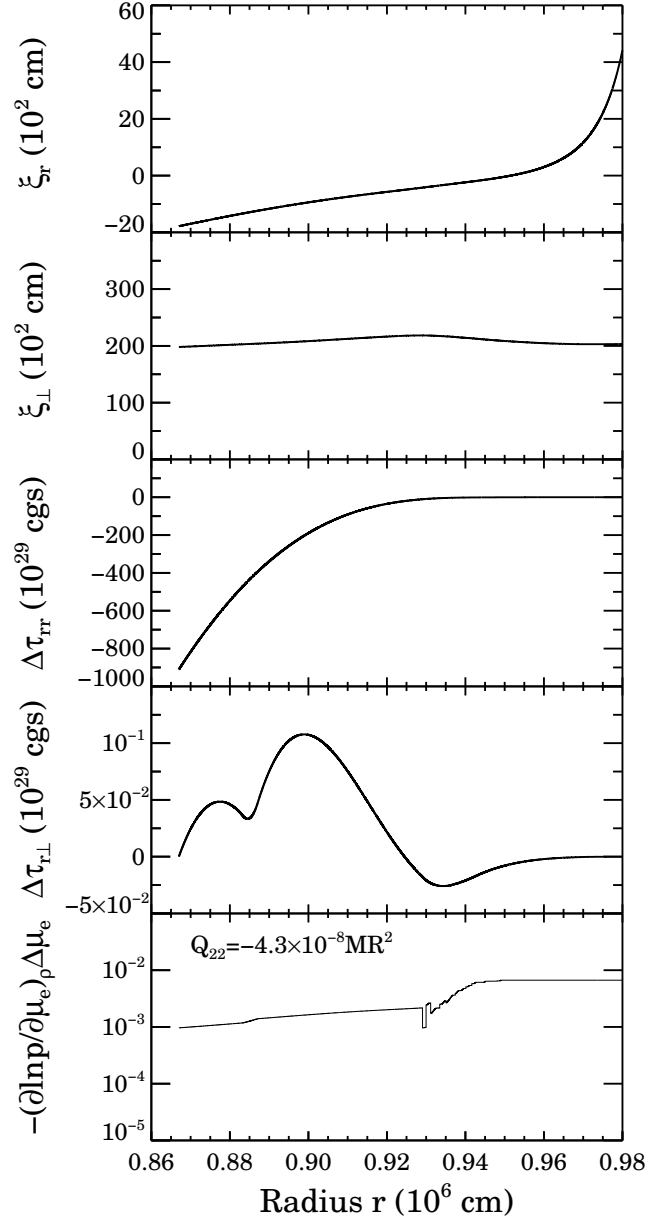}
\end{center}
\caption{\label{fig:smooth-displ} 
Response of the crust to a smooth composition variation,
$\Delta\mue/\mue=5\times10^{-3}$, throughout the crust.  Top two
panels show vertical and horizontal Lagrangian displacements, $\xi_r$
and $\xi_\perp$, in meters, the two middle panels show the stresses,
$\Delta\tau_{rr}$ and $\Delta\tau_{r\perp}$, in $10^{29} \ergcc$. The
bottom panel shows the negative of the source term, $-(\partial\ln
p/\partial\mue)_{\rho} \Delta\mue$ (see
Eq.~[\ref{eq:mue-source-term}]).}
\end{figure}

We now describe the solutions to our crustal perturbation equations,
both for the case where the capture layers are deformed due to local
temperature variations $\delta T$, and for the case where the
perturbations are sourced by a radially uniform composition
perturbation $\Delta\mue/\mue$.

Figures~\ref{fig:shallow-displ}, \ref{fig:medium-displ},
and~\ref{fig:deep-displ} show the response of the crust to a
temperature perturbation $\delta T(r) {\it Re}\{Y_{22}\}$ acting on
shallow (threshold energy $Q=23$~MeV), medium ($Q=42.4$~MeV) and deep
($Q=95$~MeV) capture layers, respectively.  The background model has a
normal core, and the accretion rate is $0.5\dot{M}_{\rm Edd}$.  The
background thermal structure is shown in
Figure~\ref{fig:thermal-struct}, and $\delta T(r)$ is plotted in
Figure~\ref{fig:thermal-pert}.  The top two panels show the vertical
($\xi_r$) and horizontal ($\xi_\perp$) Lagrangian displacements of the
fluid elements in the crust, as defined in
Eq.~(\ref{eq:xi-components}). The next two panels show the Lagrangian
components of the perturbed stress tensor, $\Delta\tau_{rr}$ and
$\Delta\tau_{r\perp}$, defined in Eq.~(\ref{eq:dt-rr-and-rperp}),
while the last panel displays the negative of the source term, $-(d\ln
p/d\ln T)_{\rho}\delta T$, defined in
Eq.~(\ref{eq:thermal-source-term}).  The response of the crust is
qualitatively different in these cases, as we discuss below.

For concreteness about signs, let us focus attention on an angular
location in the star where $Re\{Y_{22}(\theta, \phi)\}$ is positive,
and consider a perturbation with positive $\delta T(r)$ at the capture
layer.  (Figures~\ref{fig:shallow-displ}--\ref{fig:deep-displ} are
drawn to reflect this choice.)  We imagine drawing, at this $(\theta,
\phi)$, a cylindrical tube that extends through the crust and into the
fluid on both sides.  Since we take $\delta T$ to be positive at the
capture layer, in our tube the e$^-$ captures occur at
lower-than-average $\Ef$, which tends to make the capture layer
relatively ``underpressured'' (i.e., the source term $(d\ln
p/dT)_{\rho}\delta T$ is negative, see the bottom panels of
Figures~\ref{fig:shallow-displ}~--~\ref{fig:deep-displ}).

In Figures.~\ref{fig:shallow-displ}~--~\ref{fig:deep-displ} the
location of the capture layer, $r_{\rm lay}$, is identified by a sharp
``kink'' in $\xi_r$ (for the deep capture layer shown in
Figure~\ref{fig:deep-displ}, the kink is quite small, but its location
is coincident with the extremum of the source term).  If the capture
layer were infinitely thin, then there would be a discontinuity in
$\xi_r$ at that location. This is because we allow changes of
composition (i.e., relabeling of fluid elements) at the capture
boundary, and so $\xi_r(r_{\rm lay})$ does not indicate the perturbed
location of the capture layer.

The kink in $\xi_r$ is easy to understand: $d\xi_r/dr$ is large and
negative there because the crustal matter is vertically compressed
around the capture boundary.  We can estimate the jump in $\xi_r$ at
the capture layer by integrating Eq.~(\ref{eq:perts}a) in a thin
region around $r_{\rm lay}$, where the delta-function-like source term
$\Delta S$ is dominant and balances $dz_1/d\ln r$.  With this
approximation, we get
\begin{equation}
\Delta\xi_r\bigg|_{r_{\rm lay}}\approx \Upsilon h
	\frac{\Delta\rho}{\rho}
	\frac{\kB\delta T}{Q}
	\frac{d\ln p}{d\ln\Ef} = \dzd \frac{\Delta\rho}{\rho},
\end{equation}
where $h$ is the scale height at the capture layer with threshold
energy $Q$, and $\Delta\rho/\rho$ is the density jump across the
capture layer.

Compression of the capture boundary occurs in all solutions,
regardless of the capture threshold $Q$. Most other qualitative
features of the solution depend on whether the capture boundary occurs
at relatively low density, $Q \lesssim 25$~MeV, or high density, $Q
\gtrsim 40$~MeV.

Consider first the case of a deep capture layer, shown in
Figure~\ref{fig:deep-displ}. This is the most interesting case, since
it generates the largest $Q_{22}$. To understand the solution, we find
it helpful to picture the displacements as built up in a series of
``steps.''  (Of course, the implied time-ordering is fictitious; the
``steps'' are simply an imaginative device to guide intuition.)  Due
to the ``extra'' e$^-$ captures in our imaginary cylindrical tube, the
capture layer starts out ``underpressured.''  Step~I is vertical
adjustment of the crust to compress the capture layer and de-compress
the crust ($d(r^2\xi_r)/dr > 0$) above and below it. In effect, the
crust adjusts vertically to ``share'' the ``underpressure.'' After
step~I, the entire crust in the tube is relatively
underpressured. Hence, in step~II, the crustal matter gets ``pushed in
from the sides'' at all depths. One can see this behavior in our
solution: in Figure~\ref{fig:deep-displ}, $\xi_\perp$ is positive
everywhere, so the divergence of $\xi_\perp(r) \nabla^a Y_{22}$ is
negative (see Eq.~[\ref{eq:delta-rho}]).  This convergence of solid
matter in from the sides causes the crust to be ``thicker than
average'' in the tube; the bottom of the crust is pushed down and the
top pushed up.  Finally, step~III: the bulging out of the crustal
boundaries displaces the fluid in the ocean and the core, pushing it
out of the tube.

In summary, for deep captures, solid (elastic) matter is pushed into
the tube from the sides, while liquid in the ocean and the core is
pushed out. The center of mass of the tube also sinks somewhat.
Moreover, it turns out that {\it more fluid goes out of the tube than
solid comes in}.  Both of these effects~--~the net loss of mass from the
tube and the fact that the tube's center of mass sinks~--~contribute
to $Q_{22}$ with a negative sign.  Therefore, not only is the
resulting $Q_{22}$ smaller than the fiducial estimate, $Q_{\rm fid}$,
but it has the opposite sign!  (Recall that in the picture behind the
fiducial estimate, matter simply comes in from the sides in response
to the ``underpressure'' of the capture layer. But the actual response
of the crust is much more complicated, and numerical solutions were
crucial to reforming our intuition.)

In our deep-capture solution, $\xi_r$ is of order $\sim 10$~m and
$\xi_\perp \sim 100$~m (for accretion rate of $0.5\dot{M}_{\rm Edd}$),
so the displacements are rather large. This is clearly due to the
smallness of the crust's shear modulus relative to pressure.  Note
also that the shear stress term $\delta\tau_{r\perp}$ is sizeable over
a region roughly $\sim 1$~km thick. Thus, despite the fact that the
capture boundaries are rather thin (several meters), the induced
stresses are `shared' by a sizeable fraction of the crust.  This is
important for the consistency of our approach. If we had found that
all the shear stresses were concentrated near the capture layer, then
they would certainly be large enough there to crack the crust,
invalidating our solution.  Even given that the stresses are shared by
a large fraction of the crust, it is still a serious question whether
the NS can sustain them; we address this question in detail in
\S~\ref{sec:max_intro}.

We consider next solutions sourced by a uniform $\Delta\mue/\mue$,
since they are extremely similar to deep-capture solutions, as seen
evident from comparing Figures~\ref{fig:deep-displ}
and~\ref{fig:smooth-displ}.  Again, to fix signs, take
$\Delta\mue/\mue$ to be positive inside our imaginary tube.  The
perturbation corresponds to the crust inside the tube being ``a little
underpressured everywhere.''  Compare this to the deep-capture layer
case, where the crust starts out ``very underpressured in a thin
layer'', but, in step~I (described above), adjusts vertically to
``share the underpressure.'' That is, in the deep capture layer case
the crust adjusts vertically to resemble the uniform $\Delta\mue/\mue$
case.  Of course, one can also imagine building up a smooth source
term from a sum of delta-function-like sources.  Since the deep
capture solutions all have a similar profile, and since our equations
are linear, the solution sourced by a uniform $\Delta\mue/\mue$ must
share that profile.

We see from Figure~\ref{fig:shallow-displ} that the behavior for
shallow capture layers is qualitatively quite different.  In that case
the crust above the capture layer sinks and compresses (the opposite
of what we find for deep capture layers). The crust below the capture
layer also sinks and compresses, but less so. Crust and fluid enter
the tube above the capture layer and get pushed out of the tube below
the capture layer. We find that the net effect is a positive
contribution to $Q_{22}$, though a much smaller one than the fiducial
estimate.

We see that for fixed $\delta T$, the contribution of a capture layer
to $Q_{22}$ changes sign as one goes from shallow to deep layers.
There is an intermediate regime, $Q\sim 25-35$~MeV, where the
displacements produce almost no contribution to the star's quadrupole
moment.  This does not mean that the effects from shallow capture
layers in the neutron star will mostly cancel the effects from the
deep ones; contributions to $Q_{22}$ from deep capture layers are
roughly two orders of magnitude larger than those from shallow layers,
so the deep ones dominate.  A complete description of how $Q_{22}$
varies with the capture depth and other variables is presented in
\S~\ref{sec:quadrupole-scalings}.

\section{General Properties of Quadrupoles from Capture Boundaries}
\label{sec:quadrupole-scalings}

In this section, we show our results for the NS quadrupole moment
$Q_{22}$, and explain how $Q_{22}$ scales with physical parameters,
such as $\dot M$.  Our results basically validate the original
mechanism put forward by \citet{Bildsten98:GWs}: small temperature
perturbations in the deep crust can easily generate the $Q_{22}$
needed for GWs to balance accretion torque.  We defer to
\S~\ref{sec:max_intro} the question of whether the crust can actually
sustain such large stresses without cracking or yielding.

\begin{figure}
\begin{center}
\epsfig{file=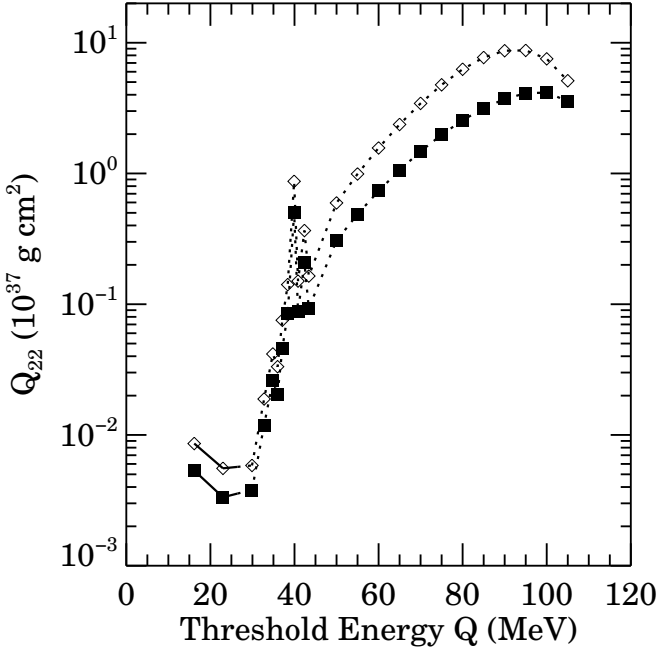}
\end{center}
\caption{\label{fig:quadrup-dT} 
Quadrupole values for a single capture layer, sourced by the
temperature perturbations computed as described in
\S~\ref{sec:thermal-pert}.  The horizontal axis gives the threshold
energy of the reaction.  Filled squares correspond to model~S
(superfluid core), while open diamonds correspond to model~N (normal
core).  The accretion rate is $0.5\dot{M}_{\rm Edd}$ in all cases. The
absolute value of $Q_{22}$ is plotted, however, the sign of it is
positive where the lines are solid, and negative where they are
dotted.}
\end{figure}

\begin{figure}
\begin{center}
\epsfig{file=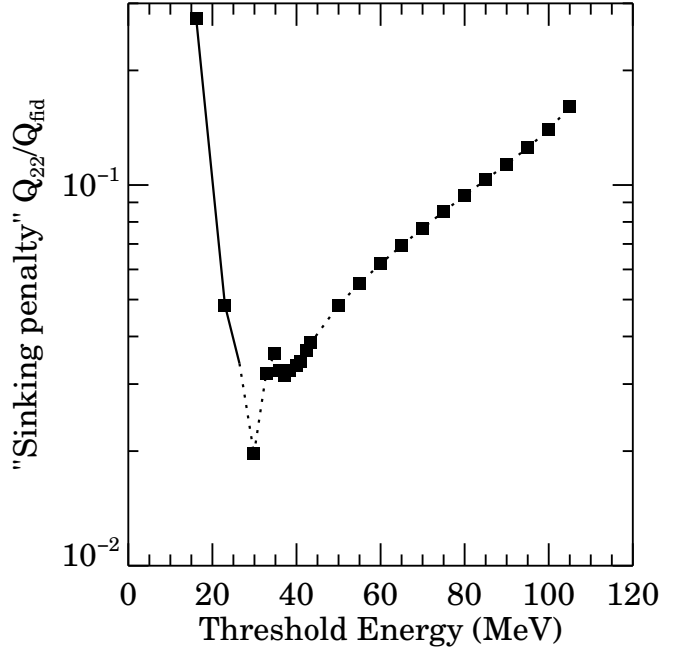}
\end{center}
\caption{\label{fig:quadrup-sink} 
The ``sinking penalty'' for the thermal perturbations at
$0.5\dot{M}_{\rm Edd}$. The absolute value is plotted, however, the
sign of it is positive where the lines are solid, and negative where
they are dotted.}
\end{figure}

\begin{figure}
\begin{center}
\epsfig{file=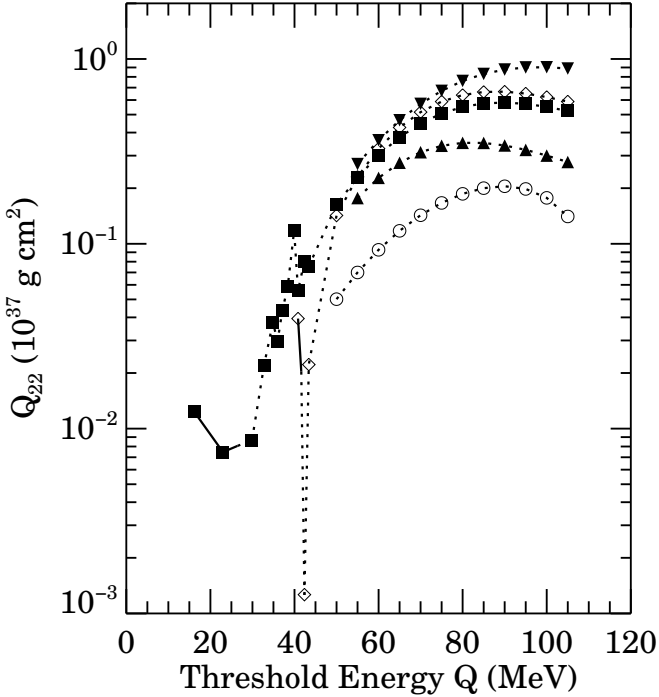}
\end{center}
\caption{\label{fig:quadrup-dz}
Quadrupole values for a single capture layer, with $\dzd=100\ {\rm
cm}$ plotted as function of the threshold energy of the reaction.  The
lines marked with filled squares correspond to the full crust, while
the open diamonds represent the case where the crust is melted around
neutron drip. The line marked with upward-pointing triangles
corresponds to the crust with the shear modulus $\mu$ artificially
decreased by a factor of $2$ from the value given by
Eq.~(\ref{eq:shearmodulus}), while the line marked with
downward-pointing triangles corresponds to the case where $\mu$ is $2$
times larger than the value from Eq.~(\ref{eq:shearmodulus}).
Finally, the line marked with open circles corresponds to the case
where there is a $5\%$ density jump at the crust-core boundary
(i.e. $\rho_l/\rho_s=1.05$). The absolute value of $Q_{22}$ is
plotted, however, the sign of it is positive where the lines are
solid, and negative where they are dotted.}
\end{figure}

\subsection{The Dependence of $Q_{22}$ on the Neutron Star Parameters}
\label{sec:quadrupole-vs-depth}

Figure \ref{fig:quadrup-dT} shows the $Q_{22}$ resulting from a single
capture layer (at e$^-$ capture threshold energy $Q$). The thermal
perturbation is due to a local nuclear heating rate that is assumed to
vary laterally by $10\%$ (i.e., $\fnuc=0.1$). Filled squares are for
the background model with a superfluid core (gap energy
$\Delta=1$~MeV) and open diamonds for the model with a normal core.
The two models have the same background hydrostatic structure and
accretion rate ($\dot{M}=0.5\dot{M}_{\rm Edd}$).

Clearly, deep capture layers generate a much larger $Q_{22}$ than
shallow ones.  We understand this as follows.  If we ignore the
vertical readjustment of the crust (as was done by
\citealt{Bildsten98:GWs}), then a local increase in temperature,
$\delta T$, increases the local e$^-$ capture rate and causes the
location of the capture layer to shift upward by a distance $\dzd$,
given by Eq.~(\ref{eq:dzd-vs-deltaT}) (see also
Figure~\ref{fig:dzd-vs-deltaT} and
Eq.~(\ref{eq:dzd-vs-deltaT-precise}) in
Appendix~\ref{sec:source-terms}).  In particular, in the outer crust,
where the pressure is supplied by degenerate relativistic electrons,
\begin{equation}
\left.\dzd\right|_{\rm outer\ crust} \approx
\Upsilon\frac{\kB T}{\mue m_{\rm p} g}\frac{\delta T}{T}
        =30\ {\rm cm}\left(\frac{\delta T}{10^7\ {\rm K}}\right)
        \left(\frac{2}{\mue}\right),
\end{equation}
where $\Upsilon\sim10-20$ is introduced in Eq.~(\ref{eq:beta-Efermi})
and defined precisely in Appendix~\ref{sec:source-terms}.  When the
elastic readjustment of the crust is ignored, the quadrupole is just
$Q_{\rm fid}=\Delta\rho\dzd R^4$
(\citealt{Bildsten98:GWs} and see also Figure~\ref{fig:sink-diagram}
in \S~\ref{sec:asymm}).  In the outer crust, where in each capture
layer only 2 electrons are consumed and no neutrons are emitted, the
density jump is just $\Delta\rho/\rho=\Delta\mue/\mue=2/Z$, so
\begin{equation}
Q_{\rm fid}\approx
	1.3\times10^{37}\ {\rm g\ cm}^2\ R_6^4
	\left(\frac{\delta T}{10^7\ {\rm K}}\right)
	\left(\frac{Q}{30\ {\rm MeV}}\right)^3.
\end{equation}
In the inner crust, the capture layer is rather thick, so rather than
talk of a ``jump'' in density across the boundary, we define the
fiducial quadrupole moment as
\begin{eqnarray}
Q_{\rm fid} &\equiv& 
        \int{\left.\frac{\partial\ln \rho}{\partial\ln T}\right|_p
         \frac{\delta T}{T} \rho r^{l+2} dr}  
\label{eq:1qlm-fiducial-definition}\\
&=& \Delta z_d \int{\frac{\Gamma_{\mue}}{\Gamma_{\rho}}
\frac{d\ln \mue}{d r}
\rho r^{l+2} dr}\; ,
\label{eq:2qlm-fiducial-definition} 
\end{eqnarray}
\noindent
where $\Gammarho$ and $\Gammamue$ are defined in
Appendix~\ref{sec:source-terms}; i.e., we integrate the Lagrangian
density perturbation $\Delta\rho$ under the condition of $\Delta p=0$
and $\xi_r=0$ (no elastic readjustment of the crust, see
Eq.~[\ref{eq:delta-T-source-dp}]).  In going from
Eq.~(\ref{eq:1qlm-fiducial-definition}) to
Eq.~(\ref{eq:2qlm-fiducial-definition}) we used Eqs.~(A10), (A14), and
(A18) from Appendix~\ref{sec:source-terms}.

In the inner crust the dependence of $\Delta z_d$ and $Q_{\rm fid}$ on
depth is more complicated than in the outer crust because of the
variation in number of electrons captured and neutrons emitted in each
capture layer (see Fig.~\ref{fig:dzd-vs-deltaT}).  Also, while $Q_{\rm
fid}$ given by Eq.~(\ref{eq:1qlm-fiducial-definition}) does increase
with depth for a fixed $\delta T$, because of the radial structure of
the temperature perturbations for fixed $\fnuc$ or $\fcomp$, $Q_{\rm
fid}$ must decrease near the very bottom of the crust.  The high
thermal conductivity of the core forces $\delta T\approx 0$ at the
crust-core boundary (see Figure~\ref{fig:thermal-pert}). This explains
the decrease in the induced $Q_{22}$ very close to this boundary seen
in Fig.~\ref{fig:quadrup-dT}.

Figure~\ref{fig:quadrup-dT} shows that a {\it single} capture layer
near the bottom of the crust generates $Q_{22}\approx
6\times10^{37}$~g~cm$^2$~$(f/0.1)$, where $f$ is either $\fnuc$ or
$\fcomp$. There will be more than one capture layer near the bottom of
the crust, and so the full $Q_{22}$ for the NS is proportionately
higher.  The numbers we quote will always be for a single capture
layer, so this multiplicative factor must be kept in mind.  

   Though the scaling of the fiducial estimate, $Q_{\rm fid}$, is
helpful for our understanding, it neglects an essential piece of
physics.  As described in \S~\ref{sec:nature-solutions}, the crust
prefers to sink in response to the shift in capture layers, and this
reduces the actual quadrupole moment significantly below the fiducial
estimate.  In Figure~\ref{fig:quadrup-sink} we plot this ``sinking
penalty'', $Q_{22}/Q_{\rm fid}$, for the same model as in
Figure~\ref{fig:quadrup-dT}.  As one can see, the penalty is quite
large ($\sim20-50$) for the shallow capture layers
($Q\lesssim40$~MeV), but is only $\sim5-10$ for the capture layers in
the deep crust.  The fact that the sinking penalty decreases with
greater depth, while $Q_{\rm fid}$ increases, means that deep capture
layers are the dominant contributors to $Q_{22}$. We find that
transverse temperature contrasts of $10^6-10^7$~K are sufficient to
generate $Q_{22} \sim 10^{37}-10^{38}$~g~cm$^2$. This is much smaller
than the $\sim10^8$~K contrast originally required by
\citet{Bildsten98:GWs} for the shallow capture layers.

In Figure~\ref{fig:quadrup-dz} we explore the effects of the magnitude
of the shear modulus $\mu$, the physical extent of the crust, and the
possible density discontinuity at the crust-core interface.  In order
to simplify the discussion, we computed the quadrupole moments
resulting from a fixed shift $\dzd=100$~cm of the capture layers,
rather than from a temperature perturbation $\delta T$ that has a
non-trivial radial dependence (see Appendix~\ref{sec:source-terms} for
a detailed discussion).  Filled squares show $Q_{22}$ for the standard
model used throughout this section. The $Q_{22}$ values are different
from those in Figure~\ref{fig:quadrup-dT} since in Fig.~\ref{fig:quadrup-dz}
the perturbations are
normalized to yield the same vertical shift $\Delta z_d$ of the capture
layers regardless of its position in the star.

As shown by \citet{Brown99}, at high $\dot M$ the energy input from
crustal nuclear reactions can melt the crust near neutron drip. In
this case there will be a liquid layer in the middle of the crust, and
our outer elastic boundary condition needs to be applied there.  The
line marked with open diamonds in Fig.~\ref{fig:quadrup-dz} 
shows $Q_{22}$ as function of capture layer depth for this case. 
The quadrupole moment due to deep
capture layers is virtually unaffected (differs by $\sim10\%$) by the
presence of a melted layer near neutron drip.

The lines marked with triangles survey the dependence of $Q_{22}$ on
the crust shear modulus $\mu$.  Doubling $\mu$ compared to the
canonical value roughly doubles $Q_{22}$ (the line marked with
downward-pointing triangles), and halving $\mu$ (line marked with
upward-pointing triangles) roughly halves $Q_{22}$. This is consistent
with the general dependence of the quadrupole moment on the shear
modulus, as derived in \S~\ref{sec:max_intro}.

Throughout this paper we have assumed that the density is continuous
across the crust-core boundary.  However, this is not certain (see
\citealt{PethickRavenhall95} for a review), as many models for the
transition contain density jumps. Though gravitationally stable, this
density discontinuity adds to the restoring force when the crust
pushes down into the core.  The line marked with open circles shows
$Q_{22}$ for a NS model where the liquid just inside the core is $5\%$
denser than the crust immediately above it (i.e.,
$\rho_l/\rho_s=1.05$).  In this case, the rippled surface of the
crust-core boundary provides some extra restoring force that reduces
$Q_{22}$.

\subsection{Dependence of $Q_{22}$ on Accretion Rate, and the 
Overall Picture} 
\label{sec:quadrupole-vs-mdot}

  We now turn our attention to the scaling of $Q_{22}$ with $\dot
M$. This behavior is important for comparing our results to the
observed distribution of NS spin frequencies in LMXB's, since in our
picture the close similarity in spin frequencies (all $\sim 300$ Hz)
over a large range in $\dot M$ implies that $Q_{22}$ scales (roughly)
like $\dot M^{1/2}$.  In Figure~\ref{fig:quadrup-mdot}, we show the
$Q_{22}$ induced by $10\%$ perturbation in nuclear heating
($\fnuc=0.1$, lines marked with filled triangles) and by $10\%$
perturbations in $Z^2/A$ ($\fcomp=0.1$, lines marked with open
circles), as a function of $\dot M$. Solid lines correspond to models
with a superfluid core, while dashed lines indicate models with a
normal core. The dotted line displays, as a function of $\dot{M}$, the
quadrupole required for spin equilibrium at $\nu_s=300$~Hz (see
Eq.~[\ref{eq:qneed}]).

 It should be clear from this figure that the $Q_{22}$ generated by a
single capture layer can easily account for the spins of the accreting
neutron stars.  Given that there are several capture layers in the
deep crust, the required asymmetry in either the heat sources or
compositions is thus proportionately lower than $10\%$.  Hence,
despite the reduction in $Q_{22}$ (below $Q_{\rm fid}$) due to the
vertical readjustment of the crust, the basic mechanism of electron
capture induced density jumps can indeed account for the spin rates.
It is also evident from Figure~\ref{fig:quadrup-mdot} that the case of
superfluid core with $\delta T$ sourced by lateral composition
variations $\fcomp$ (solid line marked with circles) can be eliminated
as a viable source of the quadrupole moment, especially at low
accretion rates, as $\fcomp$ would need to be close to unity.

\begin{figure}
\begin{center}
\epsfig{file=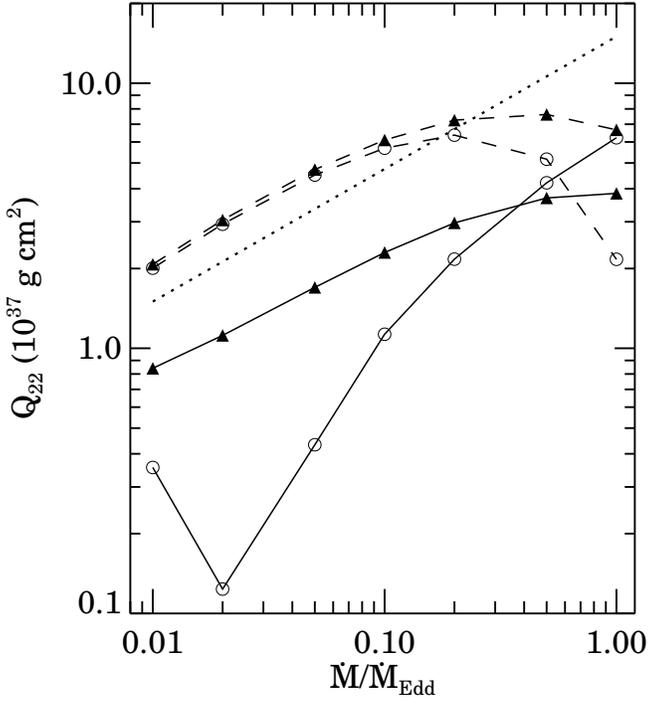}
\end{center}
\caption{\label{fig:quadrup-mdot} 
The quadrupole moment $Q_{22}$ due to a single capture layer with
threshold energy $Q=95$~MeV. Solid lines denote the results for the
model with a superfluid core (gap energy $\Delta=1$~MeV), while dashed
lines denote the values for the model with a normal core ($\Delta=0$).
Lines marked with open circles are for the quadrupole moment sourced
by the composition perturbations, $\fcomp=0.1$, while lines marked
with filled triangles denote the quadrupoles sourced by the variation
in the local heating of the crust, $\fnuc=0.1$.  Finally, the dotted
line is the relation given by Eq.~(\ref{eq:qneed}), i.e., the
quadrupole moment necessary for spin equilibrium at $\nu_s=300$~Hz as
a function of $\dot{M}$.}
\end{figure}

 Remarkably, for accretion rates $\dot{M}\lesssim0.2\dot{M}_{\rm
Edd}$, $Q_{22}$ for a constant $\fnuc$ or $\fcomp$ scales as
$\dot{M}^{1/2}$, i.e., in the same way as the required quadrupole
moment, Eq.~(\ref{eq:qneed}).  At high accretion rates, the
temperature profile in the inner crust is set by the balance of the
heat input due to compression-induced nuclear reactions near neutron
drip and neutrino emission from the crust and the core
\citep{Brown99}.  On the other hand, at low accretion rates, the
effect of heating near neutron drip and crustal bremsstrahlung is
diminished, which leads to a more direct dependence of the temperature
profile on the accretion rate and the outer boundary temperature.

As we said before, the NS's total $Q_{22}$ depends on the magnitude of
the asymmetry (i.e., $\fnuc$ or $\fcomp$) and the number of capture
layers, i.e., the curves showing $Q_{22}$ for different models in
Figure~\ref{fig:quadrup-mdot} can be shifted up or down by overall
factors.  Despite this uncertainty in the prefactor of
$Q_{22}(\dot{M})$, our calculation has no uncertain parameters that
would alter the calculated {\it scaling} of $Q_{22}$ with $\dot{M}$
for an $\dot{M}$-independent $\fnuc$ or $\fcomp$.

As discussed in \S~\ref{sec:resulting-temperature-variations}, the
thermal perturbation calculations also predict a modulation in the
X-ray flux exiting the NS crust.  The temperature perturbations in the
deep crust that displace the capture layers also give rise to a lateral
variation of the flux at the neutron star surface.  These hot and cold
spots, when moving in and out of the view of the observer due to the
rotation of the neutron star, generate a modulation in the persistent
emission.

The perturbed flux $\delta F_r$ arising from either $\fnuc$ or
$\fcomp$ equal to $10\%$ was shown in Figure~\ref{fig:flux-dT-pert} of
\S~\ref{sec:resulting-temperature-variations}.  However, as we see in
Figure \ref{fig:quadrup-mdot}, a $10\%$ $\fnuc$ or $\fcomp$
perturbation in a star with a normal core produces $Q_{22}$ from a
single capture layer that is too large at low $\dot M$ and too small
at high $\dot M$.  If the equilibrium spin frequency is {\it exactly}
independent of $\dot{M}$, then for our deformed capture layer model to
accord with these observations, $\fnuc$ or $\fcomp$ must vary with
$\dot M$.

\begin{figure}
\begin{center}
\epsfig{file=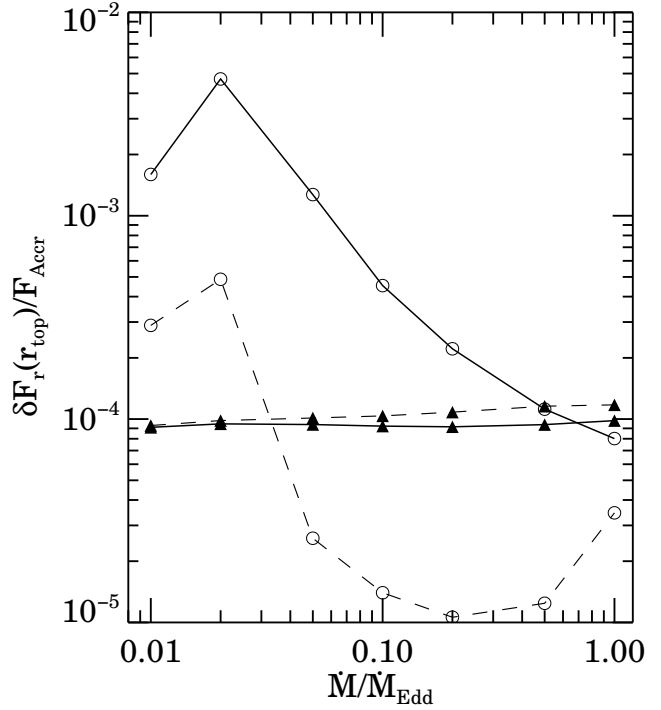}
\end{center}
\caption{\label{fig:flux-need} 
The amplitude of flux perturbations $\delta F_r$ at the top of the
crust that corresponds to a $Q_{22}$ due to a {\it single} capture
layer given by equation~[\ref{eq:qneed}], i.e. a quadrupole moment
that ensures equilibrium $\nu_s=300$~Hz at a given accretion rate.
The legend is same as in Figure~\ref{fig:quadrup-mdot}.  $\delta F_r$
is proportional to the number of capture layers contributing to
$Q_{22}$, so with many capture layers in the deep crust the required
$\delta F_r$ is reduced accordingly.  According to
Eq.~(\ref{eq:qneed}), $\delta F_r\propto\nu_s^{-5/2}$, i.e., for
equilibrium spin of 600~Hz the estimated $\delta F_r$ is reduced by a
factor of 5.}
\end{figure}

In Figure~\ref{fig:flux-need} we show $\delta F_r$, normalized by the
accretion flux $F_A\approx200\ {\rm MeV}/4\pi R^2/$accreted baryon,
under the assumption that the $Q_{22}$ due to a single capture layer
at $Q=95$~MeV is exactly that needed to set $\nu_s=300$~Hz
(Eq.~[\ref{eq:qneed}]). In other words, we adjust $\fnuc$ and $\fcomp$
as a function of $\dot{M}$ to get the required $Q_{22}$.  The actual
flux asymmetry in the NS is of course lower than that shown in
Figure~\ref{fig:flux-need} by the factor of (roughly) the number of
deep capture layers, as additional capture layers in the deep crust
will increase the quadrupole moment for the same $\delta T/T$.

The ratio $\delta F_r/F_A$ shown in Fig.~\ref{fig:flux-need} presumes
that accretion is steady.  However, there are systems, such as neutron
star transients, that accrete only episodically, with time-averaged
accretion rates $\bigl<\dot{M}\bigr>\lesssim10^{-11}
M_\odot$~yr$^{-1}$.  When accretion halts in these systems, the
thermal emission from the surface should be visible directly
\citep{brown98:transients}. Hence, in quiescence, the modulation is
$\delta F_r/F(\rtop)\sim10^{-2}(f/10\%)$, rather than $\delta F_r/F_A$.

Therefore, our thermal calculations lead us to predict a certain level
of modulation in the X-ray luminosity of LMXB's.  Can this modulation
be detected? Possibly.  We have not considered how the ocean would
respond to such lateral flux asymmetry emerging from the crust. One
could imagine transverse flows being generated from the resulting
transverse temperature gradients. Presuming that the flux at the NS
photosphere maintains the asymmetry of the flux at the top of the
crust, it can provide a critical window onto processes in the deep
crust. The bright source Sco~X-1 would be a good place to start, as it
has the highest count rate. The best current limit on a coherent pulse
from Sco~X-1 is $\lesssim1\%$ \citep{vaughan94}. Our capture-layer
mechanism for quadrupole moment generation might then be tested by
observations that are accessible today, without having to wait $5-10$
years for gravitational wave detectors to reach the requisite
sensitivity.  In addition to the excitement of finding a coherent
pulse in the X-rays, such discovery would {\it substantially} reduce
the parameter space that would need to be searched for the
gravitational wave detection.

\section{Shear strength of the Crust and the Maximum Possible Crustal 
Quadrupole}
\label{sec:max_intro}

In the previous sections we described how lateral pressure
gradients due to composition asymmetries can deform a neutron
star crust and computed the resulting quadrupole moments.  However,
there is a maximum degree of deformation that a crust can sustain, 
which is set by its yield or breaking strain
$\bar\sigma_{\rm max}$.  The yield strain places
a fundamental limit on $Q_{22}$ of a neutron star.  Unfortunately,
yield strains of even terrestrial materials, let alone neutron star
matter, are poorly understood.  In
\S~\ref{sec:maximum-strain-level} we summarize the current folklore
regarding $\bar\sigma_{\rm max}$.

Wavy electron capture layers are a very special way of straining the
crust.  However, absent strong magnetic fields, a (static) neutron
star's quadrupole moment must be generated by shear stresses in the
crust, no matter how the stresses arise. Imagine somehow turning off
the crust's shear modulus. Then the matter equations are simply the
Euler equations for a perfect fluid, and then $Q_{22}$ must be zero in
static equilibrium. These considerations suggest that there should be
an expression for $Q_{22}$ that involves the shear-stress forces only.
In \S~\ref{sec:maxint} we derive such an expression,
Eq.~(\ref{eq:qform}).

Eq.~(\ref{eq:qform}) allows us to estimate the quadrupole moment that
results from a ``typical'' strain amplitude. It also gives us an upper
limit on $Q_{22}$, set by the yield stress (derived in
\S~\ref{sec:maxQ}).  Given the tremendous uncertainty in
$\bar{\sigma}_{\rm max}$, this formula allows us to lump all our
ignorance into a single parameter. Note that our expression
(\ref{eq:qform}) for $Q_{22}$, and the resulting upper limit
(\ref{eq:qmax}), do {\it not} depend on several other assumptions made
in the rest of this paper: that the ``non-stressed'' state of the crust
is spherical, that the stress-strain relation is in the linear regime,
or that the perturbation is sourced by temperature or composition
gradients.  It applies regardless of the crust's detailed evolution,
which could have undergone an arbitrary amount of creep or cracking.
Of course, our formula does {\it not} apply when $Q_{22}$ is generated
by some force other than shear stresses, such as magnetic fields.

In \S~\ref{sec:wavy-capture-strain}, we use the formalism developed
earlier to evaluate the strain induced in the crust by the wavy e$^-$
capture layers or uniform composition gradients. We compare the
strains to the crust's maximum strain level and discuss the implications.

Finally, since the analysis in this paper ignores the self-gravity of
the perturbations, in \S~\ref{sec:self-gravity} we estimate the size of
the correction when self-gravity is included.

\subsection{Maximum Strain Level for a Neutron Star Crust}
\label{sec:maximum-strain-level}

Throughout this paper we have made a simplifying assumption that the
response of the crust to the density and pressure perturbations is
purely elastic. However, the response of solids to applied stresses is
more complicated than that.  There are two related issues here. First,
solid materials behave elastically only up to some maximum strain
$\bar\sigma_{\rm max}$, beyond which they usually either crack or
deform plastically.  Upon relieving the stress, the solid does not
return to its initial shape.  Second, even at strains well below the
yield strain, a solid that has been strained for a very long time
tends to ``forget'' its former equilibrium shape; i.e., the
equilibrium shape undergoes irreversible relaxation or ``creep.''
This behavior is called viscoelastic: solids respond elastically on
short timescales, while on very long time scales they behave more like
very viscous fluids. As argued in \S~\ref{sec:crust-deformation}, we
expect that viscoelastic relaxation, while perhaps important to the
detailed picture of ``NS mountain-building,''will not drastically
change our conclusions about the likely magnitudes of NS crustal
deformations.  However, the question of the maximum strain that a
neutron star crust can sustain {\it is} a crucial one for our work, as
is the question of what happens when this strain is exceeded.

Yield strains $\bar\sigma_{\rm max}$ of even ordinary materials, let alone
neutron star crusts, are very hard to predict theoretically.  For perfect
one-component crystals, simple theoretical considerations lead to
$\bar\sigma_{\rm max}\sim 10^{-2}-10^{-1}$ \citep{Kittel56}.  However the
maximum strain of most solids is determined by the motion of
dislocations and other defects.
Early discussions by
\citet{Smoluch70} placed $\bar\sigma_{\rm max}$ of neutron star crusts in
the range of $10^{-2}$ to $10^{-5}$ by analogy with a variety of
terrestrial materials with chemical and lattice imperfections.
\citet{Ruderman91} puts the maximum strain in the range of $10^{-4}-
5\times 10^{-3}$, which implies a large number of cracking events in
the lifetime of a spinning down young pulsar.
 
Recent breakthroughs in the understanding of the soft-gamma ray
repeaters as magnetars may shed some light on our problem as
well. \citet{Thompson95} argued that the energy release that powers
these events is magnetic in nature, but that the events are triggered
by ``cracking'' of the crust, which accumulates strain as the
magnetic field evolves. The maximum energy of such events implies (in
a model-dependent way) that the maximum strain is in the $10^{-3}$
range. Phenomenological support for the idea that crust cracking is
the origin of the bursts comes from the rather striking similarity of
the power-law distribution of burst energy with that of
terrestrial earthquakes \citep{Cheng96}.

However, cracking is just one way in which the crust can relieve the
applied stress.  \citet{SmoluchWelch70} argued that hot pulsar crusts
might undergo large amounts of plastic or ductile deformation and not
crack nearly as often, an effect more recently argued by
\citet{Link98} as arising from high pressures (namely, $p\gg \mu$).
Indeed, terrestrial materials undergo a brittle to ductile transition
when pressures become much greater than their shear modulus
\citep{Turcotte82}. Thus, there is a large range of options, none of
which we are confident to exclude.

\subsection{Integral Expression for $Q_{22}$ in Terms of Shear Stresses}
\label{sec:maxint}

We write the stress-energy tensor of the solid as 
\begin{equation}\label{eq:tauab}
\tau_{ab} = -p \, g_{ab} + t_{ab} \ , 
\end{equation}
where $t_{ab}$ is the (trace-free) shear stress tensor of the crust.
Consider the ``deformed'' star as a spherical star plus a small
perturbation, and consider $t_{ab}$ to be a first-order quantity, so
that
\begin{equation}\label{eq:max-dtab}
\delta\tau_{ab} = -\delta p \, g_{ab} + t_{ab}.
\end{equation}
Equilibrium between gravity and hydro-elastic forces implies
\begin{equation}\label{eq:gradtab}
\nabla^a \delta\tau_{ab} = \delta\rho \, g(r) \, \hat r_b,
\end{equation}
\noindent
where we have neglected the influence of the perturbation on the
star's gravitational field (the Cowling approximation).  We expand the
perturbation in spherical harmonics:
\begin{eqnarray}\label{eq:max-tab}
t_{ab} &=& t_{rr}(r) Y_{lm} (\hat r_a \hat r_b -\frac{1}{2}  e_{ab})
\\ \nonumber
&+& t_{r\perp}(r) f_{ab} 
+ t_{\Lambda}(r) (\Lambda_{ab} + \frac{1}{2} Y_{lm} e_{ab}) ,
\end{eqnarray}
\noindent
where $e_{ab}, f_{ab}$ and $\Lambda_{ab}$ were defined in
Eq.~(\ref{eq:dtab-term-abbreviations}). The above expansion is
automatically trace-free. We have left out terms in $t_{ab}$
proportional to $(\hat r_a \, \epsilon_{bcd} + \hat r_b \,
\epsilon_{acd})\, r^d \nabla^c Y_{lm} $ or $(\nabla_aY_{lm} \,
\epsilon_{bcd} + \nabla_bY_{lm} \, \epsilon_{acd})\, r^d \nabla^c
Y_{lm} $ because, having opposite parity, they decouple from the other
shear stress terms and cannot generate a quadrupole moment.

Projecting Eq.~(\ref{eq:gradtab}) along $\hat r^b$, we obtain:
\begin{equation}\label{eq:drho}
\delta \rho = \frac{1}{g(r)}\biggl[-\frac{d\,\delta p}{dr} + 
\frac{d\,t_{rr}}{dr} + \frac{3}{r}t_{rr} - 
\frac{\beta}{r}t_{r\perp}\biggr] \;.
\end{equation}
We now replace the $\delta p$ term on the right side of
Eq.~(\ref{eq:drho}) in favor of shear stress terms by projecting
Eq.~(\ref{eq:gradtab}) along $\nabla^b Y_{lm}$, which yields
(specializing to $l=2$):
\begin{equation}
\delta p(r) = -\frac{1}{2}t_{rr} 
+ \frac{3}{\beta}t_{r\perp}
-\frac{1}{3}t_{\Lambda} + 
\frac{r}{\beta} \frac{dt_{r\perp}}{dr} \;.
\end{equation}
Using $Q_{22} \equiv \int{\delta\rho \, r^4 dr}$, and integrating by
parts (using the fact that the shear stress vanishes above and below
the crust (see \S~\ref{sec:boundary-conditions}), we obtain:
\begin{eqnarray}\label{eq:qform}
Q_{22} = &-&\int\frac{r^3}{g}\biggl[\, 
        \frac{3}{2}\left(4 - \tilde U\right) t_{rr}  
+\frac{1}{3} \left(6 - \tilde U\right) t_{\Lambda} \\ \nonumber
&+&\sqrt{\frac{3}{2}} \left(8 - 3\tilde U + \frac{1}{3} \tilde U^2 
-\frac{1}{3} r\, \frac{d\tilde U}{dr}\right)t_{r\perp}\, \biggr]\,dr,
\end{eqnarray}
\noindent
where $\tilde{U}$ is defined in Eq.~(\ref{eq:defs}).  This expression
gives the quadrupole moment of the crust, so long as it is in
hydro-elastic balance.  Since this expression involves only shear
stresses, it also makes clear that (static) perfect fluid stars cannot have a
quadrupole moment.

\subsection{Maximum $Q_{22}$ Set by Crustal Yield Strain}
\label{sec:maxQ}

We now reinterpret Eq.~(\ref{eq:qform}) in terms of strains, rather
than stresses, by defining $\sigma_{ab} = t_{ab}/\mu$, where $\mu$ is
the shear modulus.  We define $\bar \sigma$ by $\bar \sigma^2 =
\frac{1}{2} \sigma_{ab}\sigma^{ab}$, and we assume that the crust will
yield when $\bar\sigma>\bar\sigma_{\rm max}$.  This criterion for
yielding is called the von Mises criterion
\citep{Turcotte82}. Different empirical criteria are sometimes adopted
(such as the Tresca criterion, which depends on the difference between
the maximum and minimum eigenvalues of $\sigma_{ab}$), but for our
purpose they would all give similar answers, and the von
Mises criterion is easier to use.

Note that $\tilde U = 4\pi r^3\rho/M_r\ll 1$ in the crust, so all the
coefficients of the stress components in Eq.~(\ref{eq:qform}) are
positive.  Therefore, we can compute the upper bound on $Q_{22}$ by
making the stress components $t_{rr}$, $t_{r\perp}$, and $t_{\Lambda}$
as large as possible, subject to $\sigma_{ab}\sigma^{ab} <
2\bar\sigma^2_{max}$.  Of course, the crustal yield strain could vary
with density, but given that its value is so uncertain, we shall
simply take it to be some constant, characteristic of the entire
crust.

Using our expansion
(\ref{eq:max-tab}) of $t_{ab}$ , we write%
\footnote{Recall that we defined all our variables to be real.}
\begin{eqnarray}
\sigma_{ab}\sigma^{ab} &=& \frac{3}{2} \sigma^2_{rr} [Re(Y_{lm})]^2 +
\sigma^2_{r\perp} [Re(f_{ab})]^2 \\ \nonumber 
&+& \sigma^2_{\Lambda}
[Re(\Lambda_{ab} + \frac{1}{2} Y_{lm} e_{ab})]^2 \ .
\end{eqnarray}
For $l=m=2$, the following identity holds among our basis tensors:
\begin{equation}\label{eq:ident}
\frac{3}{4}[Re(Y_{lm})]^2 + \frac{3}{4}[Re(f_{ab})]^2 +
\frac{9}{2}[Re(\Lambda_{ab} + \frac{1}{2}Y_{lm} e_{ab})]^2 = \frac{15}{32\pi}.
\end{equation}
Therefore $\bar \sigma$ will attain its maximum value, $\bar
\sigma_{\rm max}$, at {\it every} point in the crust $-$ i.e., the
crust will be everywhere stressed to the maximum $-$ if we set
\begin{mathletters}\label{eq:sigmax}
\begin{eqnarray}
\sigma_{rr} &=& \left(\frac{32\pi}{15}\right)^{1/2}\bar \sigma_{\rm max}, \\
\sigma_{r\perp} &=& \left(\frac{16\pi}{5}\right)^{1/2}\bar \sigma_{\rm max},\\ 
\sigma_{\Lambda} &=& \left(\frac{96\pi}{5}\right)^{1/2}\bar \sigma_{\rm max}.
\end{eqnarray}
\end{mathletters}
With this substitution, all the terms in the integrand in
Eq.~(\ref{eq:qform}) share a common prefactor, $\mu r^3/g$. Moreover,
since $\tilde U\ll 1$, the integral in Eq.~(\ref{eq:qform}) can be
expressed as $Q_{22} = \gamma \bar \sigma_{\rm max}I$, where the $I$ is the
integral
\begin{equation}\label{eq:qmax-integral-scaling}
I\equiv \int\frac{\mu r^3 dr}{g} \ ,
\end{equation}
which depends strongly on $M$, $R$, and the location of the
crust-core boundary, while $\gamma$ is a numerical prefactor that depends
very weakly on those parameters. $I$ is approximately given by $I
\approx (2/7) \left<\mu/p\right>(p_{b}/g)R^3\Delta R$, where $p_b$ is
the pressure at the crust-core interface, $\Delta R$ is the crust
thickness, and $\left<\mu/p\right>$ is a (suitably) weighted average
of the shear modulus in the crust.  At densities well above neutron
drip, the shear modulus is $\mu/p\sim 10^{-3} (Z/20)^2 (88/A)^{4/3}
\rho_{14}^{-1/3}$, assuming the neutron fraction $\Xn=0.8$.  The
pressure is approximately ideal degenerate neutron pressure,
$p_n\propto\rho^{5/3}$, and $\Delta R$ is roughly $5/2$ times scale
height at $p_b$, i.e., $\Delta R\sim (5/2) p_b/\rho_b g$.  Hence, we
expect the integral $I$ to scale as $I\propto \rho_b^2 R^7 M^{-2}$. By
fitting to our detailed numerical calculations, we get the more
precise scaling $I=3.61\times10^{37}$~g~cm$^2$~$\rho_{14}^{2.07}
M_{1.4}^{-1.2} R_{6}^{6.26}$. Collecting terms, we find
\begin{eqnarray}\label{eq:qmax}
Q_{\rm max}&=&1.2 \times 10^{38}\mathrm{g\ cm}^2
        \left(\frac{\bar\sigma_{\rm max}}{10^{-2}}\right)
        \frac{R_{6}^{6.26}}{M_{1.4}^{1.2}} 
        \left(\frac{Z}{20}\right)^2
        \left(\frac{88}{A}\right)^{4/3}
        \nonumber \\
        &\times&
        \left(\frac{1-\Xn}{0.2}\right)^{4/3}
        \left(\frac{\Xn}{0.8}\right)^{5/3} 
        \left(\frac{\rho_b}{2.1\times10^{14}\mathrm{g\ cm}^{-3}}
                                \right)^{2.07}, \nonumber \\
\end{eqnarray}
where the values of $Z$, $A$, and $\Xn$ are understood to be averaged
over the crust, with a weighting prefactor $pr^3/g$, which is heavily
biased to select the values near the bottom of the crust.  In
Eq.~(\ref{eq:qmax}) the prefactor and the scalings for $\rho_b$, $M$,
and $R$ are from numerical calculations, while the terms in $Z$, $A$,
and $X_n$ simply come from our analytical scaling formula for the
shear modulus.

Eq.~(\ref{eq:qmax}) is very powerful.  It shows that, so long as only
elastic forces are important, the maximum quadrupole moment attainable
for a NS crust is $Q_{22} \approx 10^{38} {\rm g\ cm}^2
(\bar\sigma_{\rm max}/10^{-2})$, no matter how the strains arise. The
exact upper limit does depend on the NS radius, however. Note that for
a given NS EOS and crust composition, the upper bound on $Q_{22}$ is
smaller for heavier NS's (which have smaller radii).  We also
emphasize that even to approach this upper limit, almost all the
strain must be in the $Y_{22}$ spherical harmonic; strain in other
harmonics pushes the crust closer to the yield point without
contributing to $Q_{22}$.

Also note that, besides providing an upper limit, our formula
(\ref{eq:qform}) provides an estimate of the $Q_{22}$ that can result
for a given level of strain in the crust: $Q_{22}\approx 10^{38} {\rm
g\ cm}^2 \left(\left<\sigma_{22}\right>/10^{-2}\right)$, where
$\left<\sigma_{22}\right>$ is some (appropriately) weighted average of
the ``22-piece'' of the crustal strain.

Note that our formula, suitably interpreted, also holds if the crust's
stress-strain relation is in the non-linear regime. In that case, we
just define an ``effective'' strain $\sigma^{\rm eff}_{ab}$ by
$\sigma^{\rm eff}_{ab} = t_{ab}/\mu$, where $\mu$ is the shear modulus
valid in the linear regime. Then Eq.~(\ref{eq:qmax}) continues to hold
if we replace $\bar \sigma_{\rm max}$ by $\bar \sigma^{\rm
eff}_{max}$, the maximum value of the effective strain.

\begin{figure}
\begin{center}
\epsfig{file=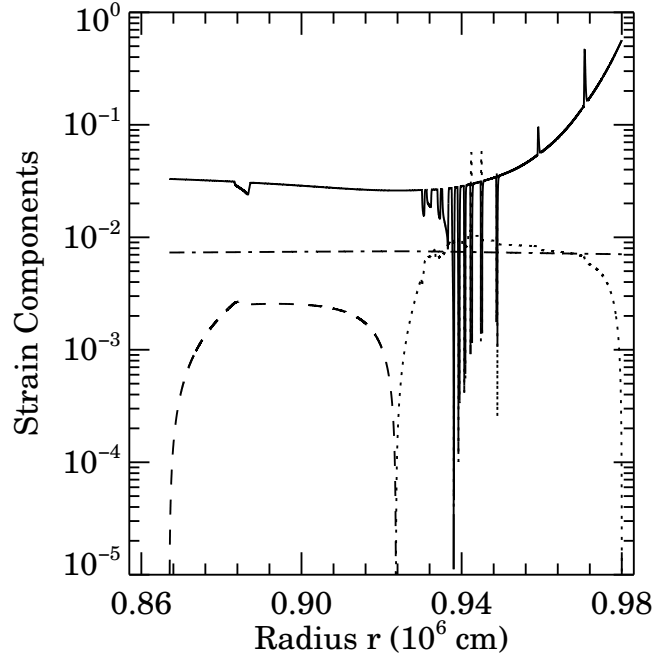}
\end{center}
\caption{\label{fig:thermal-strain} 
Angular maximum values of the shear strain components
$\max_{(\theta,\phi)}\{\sigma_{ab}\}$ as a function of the position in
the crust due to the motion of a deep capture layer ($Q=95$~MeV) in
response to an $\fnuc=0.1$ thermal perturbation. The background model
has a normal core, the accretion rate is $0.5\ \dot{M}_{\rm Edd}$,
and the quadrupole moment is $Q_{22}=8.8\times10^{37}$~g~cm$^2$. Solid
line denotes $(15/32\pi)^{1/2}\sigma_{rr}(r)$, dashed line indicates
$(5/8\pi)^{1/2}\sigma_{r\perp}(r)$, and dash-dotted line shows
$(5/48\pi)^{1/2}\sigma_{\Lambda}(r)$, where $\sigma_{ab}(r)$ are as
defined in Eq.~(\ref{eq:sigma-ab}).}
\end{figure}

\begin{figure}
\begin{center}
\epsfig{file=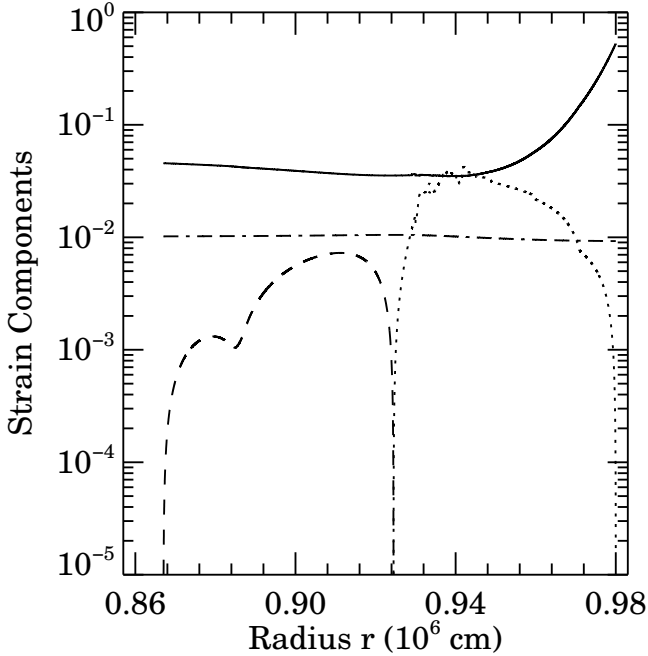}
\end{center}
\caption{\label{fig:smooth-strain} 
Same as Figure~\ref{fig:thermal-strain}, but for displacements sourced
by a smooth composition gradient with
$\Delta\mue/\mue=5\times10^{-3}$.  The quadrupole moment in this case
is $Q_{22}=1.2\times10^{38}$~g~cm$^2$.}
\end{figure}

\subsection{Crustal Strain Due to Electron Captures}
\label{sec:wavy-capture-strain}

What strains are induced by wavy e$^{-}$ capture layers or a uniform
composition gradient?  Expanding the shear tensor as functions of $r$
times (same-parity) basis tensors built from $Y_{lm}$, as in
Eq.~(\ref{eq:dtab-terms}), and then expressing the radial functions in
terms of our variables $z_i$, we have
\begin{mathletters}\label{eq:sigma-ab}
\begin{eqnarray}
\sigma_{rr}(r)&=&\frac{4}{3}\left(\frac{\partial\xi_r}{\partial r}
                        -\frac{\xi_r}{r}\right) +
        \frac{2}{3}\frac{\beta\xi_\perp}{r} \\ \nonumber
         &=& \frac{4}{3}\frac{dz_1}{d\ln r}+\frac{2}{3}\beta^2 z_3 \\
\sigma_{r\perp}(r) &=& 
        r\frac{\partial}{\partial r}\frac{\xi_\perp}{r}
        +\frac{\beta\xi_\perp}{r}
                =\frac{\beta^2 z_4}{\mu} \\
\sigma_{\Lambda}(r)&=& 2\frac{\beta\xi_\perp}{r} = 2 \beta^2 z_3.
\end{eqnarray}
\end{mathletters}
\noindent
The rr-component of strain at any point in the crust is then
$\sigma_{rr}(r,\theta,\phi)=\sigma_{rr}(r) Re\{Y_{lm}(\theta,\phi)\}$,
and similarly for the other components.  In particular, for $l=m=2$,
the maximum value of the angular factor is $(15/32\pi)^{1/2}=0.39$ for
$\sigma_{rr}$, $(5/8\pi)^{1/2}=0.45$ for $\sigma_{r\perp}$, and
$(5/48\pi)^{1/2}=0.18$ for $\sigma_{\Lambda}$.  In other words,
$\max\{\sigma_\Lambda^2[Re(\Lambda_{ab}+(1/2)Y_{lm}e_{ab})]^2\}=5/48\pi$,
and similarly for $\sigma_{rr}$ and $\sigma_{r\perp}$.

Figures \ref{fig:thermal-strain} and~\ref{fig:smooth-strain} show the
components of shear strain induced by the wavy capture layers and
smooth composition gradients, respectively. In
Figure~\ref{fig:thermal-strain} the temperature perturbation comes
from a thermal model with $\dot{M}=0.5\dot{M}_{\rm Edd}$, and
$\fnuc=0.1$; the resulting temperature perturbation at the $Q = 95$
MeV capture layer is $\delta
T=9\times10^6$~K ($\delta T/T\approx2\%$, see
Figure~\ref{fig:thermal-pert}) , and
the deformation of this layer leads to
$Q_{22}=8.8\times10^{37}$~g~cm$^2$.  The magnitude of the composition
gradient in Figure~\ref{fig:smooth-strain} is
$\Delta\mue/\mue=5\times10^{-3}$, which results in
$Q_{22}=1.2\times10^{38}$~g~cm$^2$.  Note that for these fiducial
perturbations, the resulting $Q_{22}$ values are basically what is
needed for gravitational waves to balance accretion torque in LMXB's,
with $\dot M \approx\dot M_{\rm Edd}$.  Of course both the strains and the
quadrupole moment scale linearly with $\fnuc$ or $\Delta\mue/\mue$.

From these figures it is evident that $\sigma_{rr}$ is much bigger
than $\sigma_{r\perp}$ and $\sigma_{\Lambda}$. Our mechanism is thus
not optimally efficient at producing $Q_{22}$, since the crust ends up
being much more strained at the equator than at the poles.  (I.e., a
solution that is near the yield strain at the equator is still
relatively unstrained near the poles, and so the polar regions could
be ``doing more'' to hold up the quadrupole.)  In fact, if
$\sigma_{rr}$ is the dominant stress component, then
Eq.~(\ref{eq:qmax}) becomes $Q_{22} =
3\times10^{37}$~g~cm$^2$~$\left(\left<\sigma_{rr}\right>/10^{-2}\right)$.
The strain $\sigma_{rr}$ near the bottom of the crust in these two
models is $\sim4-5\times10^{-2}$, which shows that this estimate
accords nicely with our numerical solutions.

The second thing to note about Figures~\ref{fig:thermal-strain}
and~\ref{fig:smooth-strain} is that $\sigma_{rr}>10^{-1}$ near the top
of the crust.  Thus our linear elastic model is inconsistent there, in
the sense that our solutions must certainly exceed the yield
strain. To us, this suggests that our assumed level of composition
inhomogeneity would drive a continual plastic deformation of the crust
there.
Although we are therefore
using the ``wrong physics'' to describe the top of the crust (i.e., we
should be using equations that somehow incorporate plastic
deformation), we believe this will not lead to significant errors in
our main results--the values of $Q_{22}$.  This is because, for the
solutions shown in Figures~\ref{fig:thermal-strain}
and~\ref{fig:smooth-strain}, the top of the crust contributes only a
small fraction to the total $Q_{22}$.  For example, if we repeat these
calculations, but with top of the crust set at around neutron drip
($r=9.2\times10^5$~cm), we find that $Q_{22}$ differs by $\sim10\%$
from the value obtained with a full crust (see
Figure~\ref{fig:quadrup-dz}).  As long as most of the crustal mass is
strained below the yield level, our results for $Q_{22}$ are
reasonably accurate.

The fact that a smooth composition gradient induces a
$Q_{22}=1.2\times10^{38}$~g~cm$^2$~$((\Delta\mue/\mue)/0.5\%)$ has
interesting astrophysical implications. As is evident from
Figure~\ref{fig:smooth-strain}, even such small composition gradients
induce sizeable strains in the crust. Moreover, unlike the quadrupole
moments produced by deformed capture layers, which would be wiped out
in the time it takes to accrete the mass in a capture layer
\citep{Bildsten98:GWs}, if the thermal gradient is turned off, the
uniform composition gradient can only be eliminated by replacing the
entire crust; i.e., accreting $\sim0.05M_\odot$ of material.  Hence,
even when accretion ceases, the crust of an LMXB neutron star is
likely to have a remnant quadrupole moment.  In transiently accreting
systems, this remnant quadrupole moment may set an upper limit on the
spin frequency.

In general, it is clear that at high accretion rates ($\gtrsim
0.5\dot{M}_{\rm Edd}$), the quadrupole moment needed to balance
accretion requires the crustal strain to be $\gtrsim 10^{-2}$.  This
is probably higher than the yield strain, so if such equilibrium
prevails, it seems the entire crust must be in a state of continual
plastic flow.  Assuming that accretion continually deforms the entire
crust by the above mechanism, then the stresses are likely to stay
near the yield value.  This can provide a natural explanation for the
similarity of spin frequencies in near-Eddington accretion rate
systems. We have not attempted to model the resulting plastic flow,
but that may be worth pursuing.  We caution, however, that this
picture of inhomogeneity-driven plastic flow is based solely on the
folklore regarding yield strains of materials, as no definitive
calculations of $\bar\sigma_{\rm max}$ exist, and estimates are
typically based upon extrapolating experimental results for ordinary
terrestrial materials by $>10$ orders of magnitude.  Our
approximations {\it have\/} allowed us to quantify quite clearly how
large the crustal stresses must be for gravitational wave emission to
be appreciable in accreting neutron stars. How the crust responds to
such high stresses is a problem for future research.

Since the quadrupole moment scales linearly with the crustal strain,
in lower accretion rate systems ($\lesssim0.5\dot{M}_{\rm Edd}$), the
required strains are correspondingly lower
($\sim10^{-3}-10^{-2}$). Moreover, we showed in
\S~\ref{sec:quadrupole-scalings} (see Figure~\ref{fig:quadrup-mdot}),
that for a fixed composition asymmetry ($\fnuc$ or $\fcomp$) $Q_{22}$
has exactly the right scaling with $\dot{M}$ to balance the accretion
torque by mass quadrupole gravitational radiation at a fixed spin
frequency, independent of the accretion rate. On the other hand, for
high accretion rate systems ($\gtrsim0.5\dot{M}_{\rm Edd}$), in order
to explain the spin frequency clustering at {\it exactly} $300$~Hz,
our mechanism requires that the asymmetry in the crust ($\fnuc$ or
$\fcomp$) correlate with $\dot{M}$ in a well-defined way (see
Figure~\ref{fig:quadrup-mdot}).  Alternatively, if $\fcomp$ and
$\fnuc$ are the same as in lower accretion rate systems, then we would
expect brighter LMXB's to have higher spin frequencies.  Given the
uncertainty in the spin frequency measurements for these sources,
such a possibility cannot be ruled out at present.

\subsection{Correction for Self-Gravity of the Perturbations}
\label{sec:self-gravity}

In deriving the maximum quadrupole formula, Eq.~(\ref{eq:qform})
and~(\ref{eq:qmax}), as well as in the rest of this paper, we have
neglected the changes in the gravitational force due to the
perturbation itself $-$ i.e., we have neglected the deformation's
self-gravity (the Cowling approximation). Very roughly, one would
expect these self-gravity corrections to increase $Q_{22}$ by a
fractional amount proportional to $I_{\rm crust}/I_{\rm NS} \sim
0.1$. However, including the gravitational potential perturbation,
$\delta\Phi$, allows non-zero $\delta\rho$ in the interior of the
stellar core, something that was forbidden within the Cowling
approximation. For perturbations with low $l$, the self-gravity effect
turns out to be much larger than ${\cal O}(I_{\rm crust}/I_{\rm NS})$.
We now derive a formula for $Q_{22}$ similar to Eq.~(\ref{eq:qform}),
but including the self-gravity of the deformation. However, while
Eq.~(\ref{eq:qform}) was exact (within the Cowling approximation), our
improved version will include the effects of self-gravity only in an
approximate way.

Including Newtonian self-gravity, our expressions (\ref{eq:tauab}),
(\ref{eq:max-dtab}), and (\ref{eq:max-tab}) for $\tau_{ab}$,
$\delta\tau_{ab}$, and $t_{ab}$ remain valid, but now the equilibrium
$\delta\tau_{ab}$ satisfies
\begin{equation}\label{eq:2gradtab}
\nabla^a \delta\tau_{ab} = \delta\rho \, g(r) \, \hat r_b \ + \ \rho \nabla_b 
\delta\Phi.
\end{equation}
Projecting Eq.~(\ref{eq:2gradtab}) along $\hat r^b$, we obtain:
\begin{equation}\label{eq:2drho}
\delta \rho = \frac{1}{g(r)}\biggl[-\frac{d\,\delta p}{dr} + 
\frac{d\,t_{rr}}{dr} + \frac{3}{r}t_{rr} - 
\frac{\beta}{r}t_{r\perp}\biggr]  - \frac{\rho}{g} \frac{d\delta\Phi}{dr}   \;.
\end{equation}
Replacing $\delta p$ in (\ref{eq:2gradtab}) by projecting
Eq.~(\ref{eq:2gradtab}) along $\nabla^b Y_{lm}$, and integrating by
parts to eliminate radial derivatives of the shear terms, we obtain

\begin{eqnarray}\label{eq:2qform}
Q_{22} = {\rm rhs \ of \ }(\ref{eq:qform}) + 
\int_0^R{\frac{r^4}{g}\frac{d\rho}{dr}\delta\Phi} dr
\end{eqnarray}
where the ``extra'' term ' in Eq.~(\ref{eq:2qform}) is an integral
over the entire star, not just the crust.  Deformations in the crust
change the potential throughout the star, so now $\delta\rho$ is
nonzero everywhere.

Equation~(\ref{eq:2qform}) is exactly true, but not very useful
without knowing $\delta\Phi(r)$.  We introduce an approximation that
allows us to obtain a closed-form expression for $Q_{22}$. We expect
that for large quadrupole moments, most of the density perturbation
lies near the bottom of the crust.  As an approximation then, we use
the $\delta\Phi(r)$ appropriate to a thin deformed shell at radius
$\rbot$, the location of the crust-core boundary:
\begin{equation}\label{eq:dphi}  
\delta\Phi(r) = -\frac{4\pi}{5}Q_{22}
\left\{ \begin{array}{ll} 
          r^2/r^5_{\rm bot}   
           & \mbox{ $r < \rbot$} \nonumber \\
          1/r^3   
           & \mbox{ $r > \rbot$.} \\
                \end{array}
        \right.
\end{equation}
Plugging Eq.~(\ref{eq:dphi}) into~(\ref{eq:2qform}) we obtain
\begin{eqnarray}
Q_{22} \approx &-&(1-{\mathcal F})^{-1}\int\frac{r^3}{g}\biggl[\, 
\frac{3}{2}\left(4 - \tilde U\right) t_{rr} 
+\frac{1}{3} \left(6 - \tilde U\right) t_{\Lambda}
\nonumber \\
&+&\sqrt{\frac{3}{2}} \left(8 - 3\tilde U + \frac{1}{3} \tilde U^2
-\frac{1}{3} r\, \frac{d\tilde U}{dr}\right)t_{r\perp}\, \biggr]\,dr,
\label{eq:3qform}
\end{eqnarray}
where
\begin{equation}\label{eq:F}
{\mathcal F} \equiv -\frac{4\pi}{5}\biggl[r^{-5}_{\rm bot} 
	\int_0^{\rbot}\frac{r^8}{m(r)}
{\frac{d\rho}{dr} dr \ + \ \int_{\rbot}^R}{\frac{r^3}{m(r)}
\frac{d\rho}{dr} dr }\biggr] \; .
\end{equation}
Note that ${\mathcal F}$ is manifestly positive (since $d\rho/dr < 0$),
so the factor $(1-{\mathcal F})^{-1}$ coming from self-gravity always
leads to an enhancement in $Q_{22}$ over the value given by our
formula (\ref{eq:qform}). Typical values of $\mathcal{F}$ are
$0.2-0.5$, depending on the exact core model. Hence, we expect that
including self-gravity in the full calculation will enhance the
resulting $Q_{22}$ by a factor of $25\%-200\%$.

\section{Conclusions} 
\label{sec:conclusions}

We have investigated whether the accretion-driven spin-up of neutron
stars in low-mass X-ray binaries (LMXBs) can be halted by
gravitational wave emission due to mass quadrupole moments generated
in their crusts.  The quadrupole moment needed to reach this spin
equilibrium is $Q_{22}\approx 10^{37}-10^{38}$~g~cm$^2$ (see
Eq.~[\ref{eq:qneed}]) for the relevant accretion rates in LMXBs,
$10^{-10}-2\times10^{-8} M_\odot$~yr$^{-1}$. How to form and sustain a
quadrupole this large is the main problem we addressed.  We have
undertaken a series of calculations that substantially extend the
original idea of \citet{Bildsten98:GWs} that electron capture
reactions can deform the crust by large amounts.  The major results of
our work are as follows:

\begin{enumerate}
\item By self-consistently solving the elastic equilibrium equations
(\S~\ref{sec:crust-deformation} and~\ref{sec:quadrupole-scalings}) we
have found that the predominant response of the crust to a lateral
density perturbation is to sink, rather than move sideways. For this
reason, the quadrupole moments due to temperature-sensitive e$^-$
captures in the {\it outer} crust (i.e., the case considered by
\citealt{Bildsten98:GWs}) are actually much too small to buffer the
accretion torque. However, a single e$^-$ capture layer in the deep
{\it inner} crust can easily generate an adequate mass
quadrupole. This requires lateral temperature variations in the deep
crust of order $\lesssim5\%$, and the realistic case of multiple
electron capture layers requires a proportionately smaller temperature
variation.  Because of the much larger mass involved in generating
$Q_{22}$ in the inner crust, the temperature contrasts required are
only $\sim10^6-10^7$~K, rather than $\sim10^8$~K originally envisioned
by \citet{Bildsten98:GWs}.  Alternatively, a 0.5\% lateral variation
in the charge-to-mass ratio can generate a $Q_{22}$ sufficient to
balance the accretion torque even in the absence of a temperature
gradient.

\item Our thermal perturbation calculations show that the temperature
variations required to induce a quadrupole moment this large can
easily be maintained if the compressionally-induced nuclear reactions
around neutron drip inject heat with about $\sim10\%$ lateral
variations.  Lateral variations of the same magnitude in the average
$Z^2/A$ of nuclei in the crust can also maintain a similar temperature
asymmetry. This is despite the strong thermal contact with the
isothermal core of the neutron star (see \S~\ref{sec:thermal-pert}).
However, if accretion halts or slows considerably, then these
temperature variations will be wiped out in a thermal time, i.e., a
few years at the crust-core interface (Eq.~[\ref{eq:thermal-time}]).
In this case the e$^-$~capture boundary deformations will be smoothed
out in the time it takes to accrete the mass in a capture layer,
$\sim2.5\times10^6$~yr~$(\Mdot/10^{-9}\ M_\odot\ {\rm yr}^{-1})$ near
the crust-core boundary.

\item While it is not possible to estimate the size of the
compositional or nuclear heating asymmetries a priori, the temperature
variations in the deep crust lead to lateral variations in the
persistent thermal flux emerging from the neutron star.  Since the
neutron star is spinning, a certain $Q_{22}$ implies an amplitude of
the modulation of the persistent X-ray flux (see
Figures~\ref{fig:flux-dT-pert} and \ref{fig:flux-need}). Though a
small effect, these periodic variations can be searched for
observationally.  Detection of such modulation would help tremendously
in the search for gravitational wave emission.

\item If the size of the heating or composition asymmetry is a
constant fixed fraction, then we showed that for
$\dot{M}\lesssim0.2\dot{M}_{\rm Edd}$ the scaling of $Q_{22}$ with
$\dot M$ is just that needed for all of these low accretion rate NS's
to have the same spin frequency (see Figure~\ref{fig:quadrup-mdot}).

\item Quadrupoles this large require a strain in the crust of order
$10^{-3}-10^{-2}$ (depending on the accretion rate, with strains
exceeding $10^{-2}$ for $\dot{M}\gtrsim0.5\dot{M}_{\rm Edd}$, see
discussion in \S~\ref{sec:wavy-capture-strain}), {\it regardless} of
the detailed mechanism for generating the strain. 

\item We have derived a general relation between the maximum
quadrupole moment $Q_{\rm max}$ that a crust can support via elastic
deformation and its breaking strain (see \S~\ref{sec:max_intro}).  Our
relation (Eqs.~[\ref{eq:qform}] and [\ref{eq:qmax}]) is more complete
than previous work and is widely applicable.  In addition to
determining $Q_{\rm max}$, this relation allows one to robustly
estimate the quadrupole moment for a given level of strain, even in
the nonlinear regime, regardless of the amount of plastic flow or
relaxation that the crust has undergone.
\end{enumerate}

Our work has thus clarified many of the important outstanding
questions for the hypothesis that gravitational wave emission due to a
crustal quadrupole moment can buffer the accretion torque. We now
discuss the implications of our results for the millisecond radio
pulsars and for the possibility of gravitational waves from wobbling
NS's.

 Ever since the discovery of the first millisecond radio pulsar
\citep{backer82}, observers have been looking for rapidly rotating
neutron stars near the breakup limit of $\nu_{\rm b} \approx
1476$~Hz~$M_{1.4}^{1/2} R_6^{-3/2}$ (see \citealt{Cook94} for an
exhaustive survey of the breakup limits for different nuclear
equations of state). However, few have been found. Even today, it is
still the case that the majority of millisecond radio pulsars are
spinning at 300 {\rm Hz} or less. By considering the characteristic
ages of those pulsars that reside in binaries,
\citet{backer98:_neutr_star} concluded that most of the initial spin
frequencies of the radio pulsars are near 300~Hz, consistent with the
later inferences from the accreting population. In this paper, we have
shown that accretion-induced lateral density variations in neutron
star crusts can indeed account for limiting the initial spins of
millisecond pulsars.

However, there are a few millisecond pulsars that do spin much faster
than 300~Hz (e.g., B1937+21 and B1957+20, which have periods of
roughly 1.6~ms).  For these fast pulsars, a possible explanation is
that their crusts are very uniform, so that the composition asymmetry
is $\ll10\%$ as required to generate the appropriate $Q_{22}$.  A much
more likely hypothesis is that these pulsars were spun up in
transiently accreting systems with recurrence times greater than a few
years, or accreting at a very slow rate. Lateral temperature
variations in this regime would not persist, and hence the quadrupole
moment resisting accretion would be much smaller.
 
We finally turn to the possibility of gravitational waves from
wobbling NS's.  Our calculation of the maximum $Q_{22}$ sustainable by
the crust can also be used to place a strong limit on the strength of
such wobble radiation.  A NS will ``wobble'' if its angular momentum
$J^a$ is slightly misaligned from some principal axis of the crust.
This phenomenon occurs in the Earth, and is called the Chandler
wobble. It is a somewhat more complicated version of the torque-free
precession of rigid bodies that is treated in undergraduate mechanics
texts.  For the sake of brevity, we will simply quote many of the
relevant results from this area, and refer the reader to \citet{mm60},
\citet{ps72}, \citet{cutlerjones}, and \citet{jones} for more details.

Consider first a rotating (and hence oblate) NS, which is spinning
along some principal axis of the crust.  If the relaxed state of the
star is spherical, then, no matter how the NS is kicked, it will not
precess.  However, if the crust has relaxed to the oblate state (i.e.,
the zero-stress state is oblate), then kicking the NS will cause
precession.

Let the star's angular momentum $J^a$ be along $z^a$, which is
displaced (by some kick) from the principal axis of the crust $n^a$ by
angle $\theta_w$.  To a sufficient approximation, the inertia tensor
of the wobbling body will be
\begin{equation}\label{eq:I_Pines2}
I^{ab} = I_0 e^{ab} + \Delta I_{\Omega}(\hat z^a \hat z^b - {1\over 3}g^{ab}) + \Delta I_d(\hat n^a \hat n^b - {1\over 3}g^{ab}) \; .
\end{equation} 
and this ``figure'' will rotate rigidly around $z^a$ with precession
frequency $\omega_p\approx\Omega$.  Thus $\Delta I_d$ is the piece of
the inertia tensor that ``follows'' the principal axis of the crust.
Defining $\hat x^a$ by $\hat n^a \approx \hat z^a + \theta_w \hat
x^a$, we see that $I_{ab}$ for the wobbling star can be re-written as
\begin{equation}\label{eq:I_Pines3}
I^{ab} = I_0 e^{ab} + (\Delta I_{\Omega}+ \Delta I_d )(\hat z^a \hat z^b - {1\over 3}g^{ab}) + \theta_w\Delta I_d(\hat z^a \hat x^b + \hat x^a \hat z^b) \; .
\end{equation}
Here the piece oriented along $\hat z^a$ is ``held up'' by centrifugal force,
while the piece proportional to $\theta_w\Delta I_d$ is ``held up''
by crustal shear stresses and is responsible for the star's $Q_{21}$ 
moment. A little algebra shows that 
\begin{equation}\label{eq:Q21}
Q_{21} = \sqrt{{15\over{2\pi}}} \, \theta_w \Delta I_d   \; .
\end{equation}
The dominant gravitational wave emission is at the wobble frequency, so 
$\omega_{\rm gw} \approx \Omega$, and the gravitational wave 
luminosity is 
\begin{equation}
L_{\rm gw}^{\rm wobble} = {2\over 5} (\theta_w \Delta I_d)^2\Omega^6
= {{4\pi}\over {75}} Q_{21}^2\Omega^6 \; ,
\end{equation}
Comparing with Eq.~(3), we see that for a fixed spin frequency
$\Omega$, $L_{\rm gw}^{l=m=2}/L_{\rm gw}^{\rm wobble} = 64
(Q_{22}/Q_{21})^2$.  The same ratio holds for the backreaction
torques.  Thus, if the accretion torque is balanced by the wobble
radiation, rather than $l=m=2$ radiation, resulting in $\nu_s\approx
300$~Hz, then the required quadrupole moment is $8$ times larger for
the wobble case. This would require an average strain $\bar \sigma
\approx 2.3 \times 10^{-2}$ for $\dot M = 10^{-9} M_{\odot}/$yr and
our fiducial NS mass and radius.  Alternatively, if one fixes the
gravitational wave frequency (instead of the spin frequency), and
asks, what is the maximum gravitational wave luminosity achievable by
either $l=m=2$ radiation or wobble radiation, we see that the limit
set by crust cracking is exactly the same for the two cases (up to
small corrections of order $(\Omega/\Omega_{max})^2$ and $Q/MR^2$)--a
rather pretty result!

  We thank Andrew Cumming, Chris McKee, and Kip Thorne for many
initial discussions. Ed Brown provided much assistance with the
construction and cross-checking of our thermal models. Patrick Brady
shared his notes on normalizations and detectability of pulsar GW
signals.  D.~I. Jones pointed out an error in our treatment of NS
wobble, and suggested the outlines of the correct answer.
G.U. acknowledges the Fannie and John Hertz foundation for fellowship
support.  This research was supported by NASA via grants NAG5-4093,
NAG5-8658, and NAGW-4517, and by the National Science Foundation under
Grant No.~PHY94-07194.  L.B. is a Cottrell Scholar of the Research
Corporation.

\appendix

\section{Coefficients and Source Terms of the Elastic Perturbation Equations}
\label{sec:source-terms}

  In this section we describe how to compute the coefficients and
source terms of Eqs.~(\ref{eq:perts}) from the background models
developed in \S~\ref{sec:crust-structure} and~\ref{sec:thermal-pert}.

We use the shear modulus as computed by \citet{Strohmayer91} by
Monte-Carlo simulations of both bcc crystals and quenched
solids. Their results can be conveniently rewritten in terms of
the pressure of degenerate relativistic electrons,
\begin{equation}\label{eq:shearmodulus}
\frac{\mu}{p_e}=\frac{6\times10^{-3}}{1+0.595(173/\Gamma_{\rm Coul})^2}
	\left(\frac{Z}{8}\right)^{2/3},
\end{equation}
where $\Gamma_{\rm Coul}$ is defined in Eq.~(\ref{eq:gamma-coul}).  We
neglect the slight dependence of the shear modulus on
temperature. Because of the uncertainty of the charge of nuclei at the
bottom of the crust, we varied the numerator of
Eq.~(\ref{eq:shearmodulus}), as described in
\S~\ref{sec:quadrupole-vs-depth}.  At densities higher than neutron
drip, we have $\alpha_1=(\mu/p_e)(p_e/p)$.  The run of $\mu$ with
pressure in the crust is shown in the top panel of
Figure~\ref{fig:gamma-rho-and-mue}. 

\begin{figure}
\begin{center}
\epsfig{file=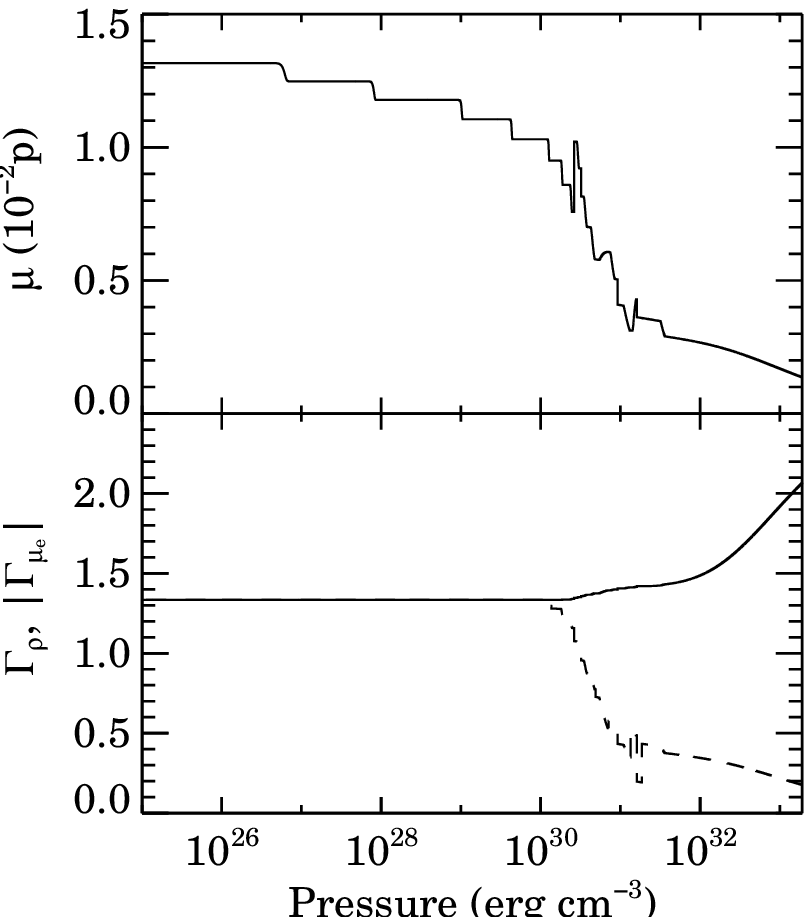}
\end{center}
\caption{\label{fig:gamma-rho-and-mue} 
Top panel: The run of the shear modulus $\mu/p$ with depth.  Bottom
panel: The run of $\Gammarho$ (solid line) and $|\Gammamue|$ (dashed
line) with pressure in our model.  Note that $\Gammamue$ is negative,
so its absolute value is plotted. For pressures $p\lesssim 10^{30}\
\ergcc$ (i.e. for densities less than neutron drip)
$\Gammarho=|\Gammamue|$.}
\end{figure}

In addition to the shear modulus, we require a few thermodynamic
derivatives.  Let us first consider the case of the perturbations
generated by a smooth composition variation in the crust,
i.e., $\mue(r,\theta,\phi)=\mue(r)+\Delta\mue(r)
Y_{lm}(\theta,\phi)$.  In our computations we took
$\Delta\mue(r)/\mue(r)={\rm const}$, but the formalism applies for
arbitrary variation of $\Delta\mue$ with depth.

In this case, we write $p=p(\rho,\mue)$ and $\Delta p$ using
Eq.~(\ref{eq:delta-mue-source-dp}). The index $\Gamma$ in this case is
computed at constant composition,
\begin{equation}
\Gamma=\Gammarho\equiv
	\left.\frac{\partial\ln p}{\partial\ln\rho}\right|_{\mue}
	=\frac{4}{3}\frac{p_e}{p}+
	\left.\frac{\partial\ln p_n}{\partial\ln\rho}\right|_{\Xn}
	\frac{p_n}{p},
\end{equation}
where $p_e\propto(\rho/\mue)^{4/3}$ is the relativistic electron
degeneracy pressure, and $p_n(\rho,\Xn)=p_n(\Xn\rho/\mb)$ is the
neutron pressure. If the neutrons were completely degenerate,
non-relativistic and non-interacting, the prefactor of $p_n/p$ would
be $5/3$. The actual run of $\Gammarho$ with pressure is shown by the
solid line in Figure~\ref{fig:gamma-rho-and-mue}.  

Following Eq.~(\ref{eq:mue-source-term}), we write the source term as
$\Delta S=\Gammamue(\Delta\mue/\mue)$, where
\begin{equation}
\Gammamue\equiv
  \left.\frac{\partial\ln p}{\partial\ln\mue}\right|_{\rho}=
	-\frac{4}{3}\frac{p_e}{p}+
	\left.\frac{\partial\ln p_n}{\partial\ln\mue}\right|_\rho
	\frac{p_n}{p}.
\end{equation}
Using the chain rule, we express the prefactor of the $p_n/p$ term in
terms of the prefactor of the similar term in $\Gammarho$ as
\begin{equation}
\left.\frac{\partial\ln p_n}{\partial\ln\mue}\right|_\rho=
\frac{d\ln\Xn}{d\ln\mue}
\left.\frac{\partial\ln p_n}{\partial\ln\rho}\right|_\Xn,
\end{equation}
where
\begin{eqnarray}
\frac{d\ln\Xn}{d\ln\mue}&=&-\frac{1}{\Xn\mue}\frac{X_{n1}-X_{n2}}{1-X_{n1}}
\left[
	\frac{Z_1}{A_1}-\frac{Z_2}{A_2}\left(\frac{1-X_{n2}}{1-X_{n1}}\right)
\right]^{-1} \nonumber \\
&\approx&\frac{1-\Xn}{\Xn},
\end{eqnarray}
which follows from Eqs.~(\ref{eq:Xn}) and~(\ref{eq:mue}).  Using the
above formalism, one can compute the source term $\Delta S$ and the
index $\Gamma$ of Eq.~(\ref{eq:mue-source-term}).  The source term
$\Delta S$ for this case is shown in the bottom panel of
Figure~\ref{fig:smooth-displ}. 

Most of the discussion in this paper was concerned with the quadrupole
moments generated by shifts in capture layers.  For most of the
calculation we have adopted the physical picture described by
Eq.~(\ref{eq:thermal-source-term}), i.e., that a temperature variation
$\delta T$ ``undoes'' some electron captures on the ``cold'' side of
the star and induces more captures on the ``hot'' side, and hence
moves the capture layer vertically.  This picture is physically
consistent, and the formalism for deriving the source terms is
described below.  However, it is somewhat easier to first understand
the Lagrangian $\dzd$ picture introduced in
\S~\ref{sec:origin-of-crustal-Q}.

In this picture, we first imagine that the crust is infinitely rigid,
and hence does not allow either vertical or horizontal elastic
adjustment. We then go in and ``by hand'' undo some captures on one
side of the star, to a height $\dzd$ above the original capture layer,
and induce some captures to the same depth on the other side of the
star. The perturbed crust then looks like the sketch in
Figure~\ref{fig:sink-diagram}.  After this perturbation, we allow the
crust to respond elastically.  This perturbation is Lagrangian in the
sense that we know a priori which fluid elements have had their $\mue$
perturbed. 

We now formulate the above picture mathematically.  The pressure at
which the capture reaction occurs varies laterally.  In a spherically
symmetric star, the capture layer occurs at $r=r_c$ (i.e., $(A_1,Z_1)$
element is, say, $50\%$ depleted at $r=r_c$).  When we put in the
$\mue$ perturbation, the depletion occurs at $r=r_c+\dzd Y_{lm}$. It
is an excellent approximation to assume that the shape of the capture
layer does not change, only its position in the star.  In this case,
we write $\mue(r,\theta,\phi)\approx\tilde{\mu}_{\rm
e}\left(r-r_c-\dzd Y_{lm}(\theta,\phi)\right)$ and Taylor-expand to
find $\Delta\mue/\mue\approx-(d\ln\mue/dr)\dzd Y_{lm}$. The source
terms are then
\begin{equation}\label{eq:source-approx-dzd}
\Delta S \approx - \Gammamue\frac{d\ln\mue}{dr}\dzd,
\end{equation}
where $\mue(r)$ is the run of electron mean molecular weight in the
equilibrium model. Since $\mue(r)$ is constant outside the capture
layers, this source term is non-zero only in the capture layers.  A
plot of $d\ln\mue/d\ln r$ is shown in Figure~\ref{fig:dlnmue-dlnr}.
Since $\mue(r)$ resembles a step function with steps at the locations
of the capture layers, $d\ln\mue/d\ln r$ resembles a delta function at
the locations of the capture layers.

\begin{figure*}
\begin{center}
\epsfig{file=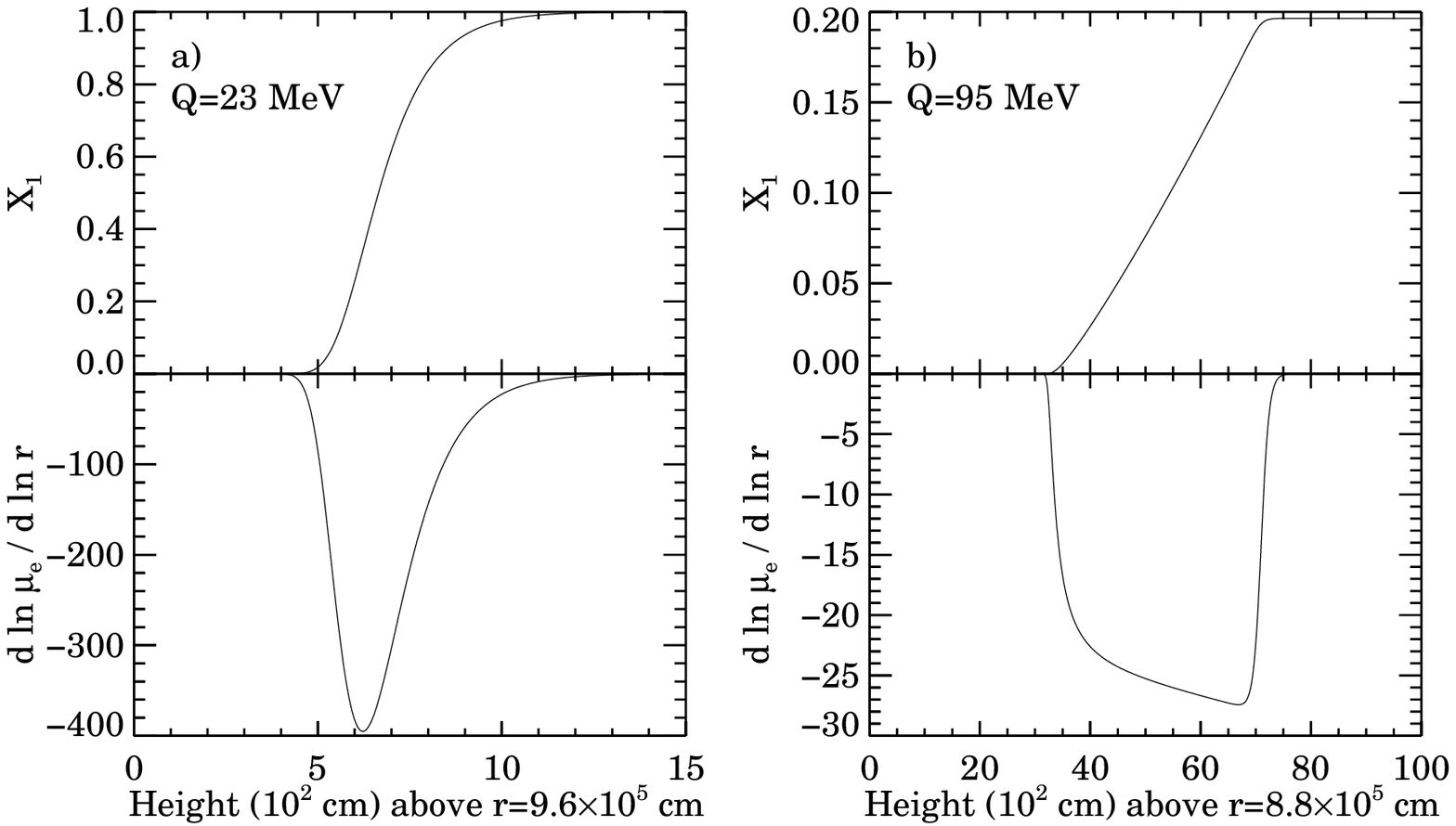}
\end{center}
\caption{\label{fig:dlnmue-dlnr}
Left (a): The run of the mass fraction $X_1$ (top panel) and
$d\ln\mue/d\ln r$ with radius for a shallow capture layer at threshold
energy $Q=23$~MeV.
Right (b):  Same as (a), but for a deep capture layer at $Q=95$~MeV.
}
\end{figure*}

Now consider the case where the capture layers are moved around by
lateral temperature variations, i.e. the case of $\delta T$-sourced
perturbations.  In our equation of state, the dependence on
temperature only comes in through $\mue$,
i.e. $\rho=\rho(p,\mue(p,T))$.  Therefore,
\begin{equation}\label{eq:dlnrho-dlnp-const-T}
\left.\frac{\partial\ln\rho}{\partial\ln p}\right|_{T} =
\frac{1}{\Gammarho}+
\left.\frac{\partial\ln\rho}{\partial\ln\mue}\right|_{p}
\left.\frac{\partial\ln\mue}{\partial\ln p}\right|_{T}.
\end{equation}
The derivative at constant $p$ in the second term above is displayed
more conveniently by changing back to $(\rho,\mue)$ as independent
variables,
\begin{equation}\label{eq:dlnrho-dlnmue-const-p}
\left.\frac{\partial\ln\rho}{\partial\ln\mue}\right|_{p}=
-\frac{\Gammamue}{\Gammarho},
\end{equation}
so in case of $\delta T$ perturbations,
\begin{equation}\label{eq:gamma-dT-perturbations}
\Gamma=\Gammarho\left[1-\Gammamue
\left.\frac{\partial\ln\mue}{\partial\ln p}\right|_{T}\right]^{-1}.
\end{equation}
The term in square brackets is equal to $1$ outside the capture
layers, and hence $\Gamma$ reduces to the same value as in the case of
the perturbations caused by a smooth composition gradient (see
Figure~\ref{fig:gamma-rho-and-mue}).  However, the term in the square
brackets, which we compute below, is significantly smaller than $1$ in
the capture layers.

Using the chain rule as in Eq.~(\ref{eq:dlnrho-dlnp-const-T}) and
using Eq.~(\ref{eq:dlnrho-dlnmue-const-p}), we find that
\begin{equation}
\left.\frac{\partial\ln\rho}{\partial\ln T}\right|_p=
-\frac{\Gammamue}{\Gammarho}
\left.\frac{\partial\ln\mue}{\partial\ln T}\right|_{p}.
\end{equation}
We can now evaluate the prefactor of the source term by switching
independent variables from $(\rho,T)$ to $(p,T)$ to obtain
\begin{equation}\label{eq:dlnp-dlnT-const-rho}
\left.\frac{\partial\ln p}{\partial\ln T}\right|_\rho=
\Gammamue\left.\frac{\partial\ln\mue}{\partial\ln T}\right|_{p}
\left[1-\Gammamue
\left.\frac{\partial\ln\mue}{\partial\ln p}\right|_{T}\right]^{-1}.
\end{equation}
Eqs.~(\ref{eq:gamma-dT-perturbations})
and~(\ref{eq:dlnp-dlnT-const-rho}) allow us to evaluate both the
source term $\Delta S$ and the index $\Gamma$ (see
Eq.~[\ref{eq:thermal-source-term}]), if we can find the partial
derivatives of $\mue(p,T)$ with respect to $p$ and $T$.  We now
describe how to find these derivatives.

While constructing an equilibrium stellar model, we computed the
dependence of $\mue$ on $r$ for each capture layer.  However, at least
in principle, we could compute $\mue(p,T)$ using the following
approach. We combine the rate equation (\ref{eq:rate-spherical}) and
the equation of hydrostatic balance (\ref{eq:hydro-mass}) to obtain
\begin{equation}\label{eq:xofp}
\frac{d\ln X_1}{d\ln p}=-\frac{p}{\dot{m}g}\Rec(p,X,T),
\end{equation}
where $\dot{m}=\dot{M}/4\pi r^2$ is the local accretion rate.
Approximating $r^2$ and $g$ as constant over the capture region, we
then solve Eq.~(\ref{eq:xofp}) to obtain $X_1(p,T)$, which, from
Eq.~(\ref{eq:mue}), immediately gives us\footnote{
To be more precise, $\mue$ is a function of $p$ and a functional of
$T$, since it depends not just on the $T$ at the point but on $T$ in
the whole capture region. However, \citet{BC98} showed, and we confirm
for the case of deep capture layers that include free neutrons, that
the variation in $T$ over the capture region is quite small.  Thus we
treat $T(r)$ as constant over the capture region, in which case there
is no practical distinction between a function and a functional. }
$\mue(p,T)$.  Eq.~(\ref{eq:xofp}) was derived assuming the neutron
star is in hydrostatic balance, but we shall use it to determine
$\mue(p,T)$ even in the perturbed star. This is consistent, because
corrections to $\mue(p,T)$ are due to the nonspherical perturbation
itself, and so would enter our perturbation Eqs.~(\ref{eq:perts}) only
as higher-order corrections.

In practice, rather than to first compute $\mue(p,T)$ and then
differentiate it, it is much more convenient to use a simple
approximation to evaluate the derivatives of $\mue$.  This
approximation is essentially the same as the one used to evaluate the
source term in the case of $\dzd$ perturbations,
Eq.~(\ref{eq:source-approx-dzd}).  We presume that the shape of the
capture layer does not change as we perturb the local temperature by
$\delta T$, but only its location shifts.  Mathematically speaking, we
write $\mue(p,T) \approx \tilde\mu_{\rm e}(p-p_c(T))$, where $p_c$ is
the (temperature-dependent) pressure at the ``center'' of the capture
region, which we define more precisely below. We use the tilde on
$\tilde\mu_{\rm e}$ to distinguish this ``approximate version'' from
the actual function $\mue$.

With this approximation,
\begin{equation}\label{eq:dlnmue-dlnp-const-T}
\left.\frac{\partial\ln\mue}{\partial\ln p}\right|_{T}=
\frac{d\ln\tilde\mu_{\rm e}}{d\ln p}=-\frac{1}{\tilde{V}}
\frac{d\ln\mue}{d\ln r},
\end{equation}
where $\mue(r)$ is just the run of $\mue$ in the background model (see
Figure~\ref{fig:dlnmue-dlnr}). $\tilde{V}$ is defined in
Eq.~(\ref{eq:defs}) and is equal to $r/h$, where $h$ is the local
pressure scale height. In the crust $h/r\lesssim0.1$ ($\ll0.1$ in the
outer crust). It is then evident from Figure~\ref{fig:dlnmue-dlnr}
that the term in square brackets in
Eq.~(\ref{eq:gamma-dT-perturbations}) is indeed much smaller than $1$
inside capture layers.

Similarly, we have
\begin{equation}\label{eq:dlnmue-dlnT-const-p}
\left.\frac{\partial\ln\mue}{\partial\ln T}\right|_{p}=
-\frac{d\ln\tilde\mu_{\rm e}}{d\ln p}\frac{d\ln p_c}{d\ln T}
=\frac{1}{\tilde{V}}\frac{d\ln\mue}{d\ln r}\frac{d\ln p_c}{d\ln T}.
\end{equation}
In the above equation, $p_c$ is a function of $T$ for a given capture
region, and is not a function $r$ or $p$.

Eq.~(\ref{eq:xofp}) exhibits the competition between the local
compression timescale $t_{\rm comp}=p/\dot{m}g$ and the electron
capture timescale $t_{\rm ec}=1/\Rec$.  Most electron captures happen
when $t_{\rm comp}\sim t_{\rm ec}$.  We use this to {\it define} the
center of the capture layer, $p_c$:
\begin{equation}\label{eq:pc-definition}
\frac{p_c}{\dot{m}g}\Rec(\Ef(p_c),T)\equiv 1.
\end{equation}
Differentiating the above equation with respect to $T$ and
using the capture rate (\ref{eq:Rec}) we find 
\begin{equation}\label{eq:dlnpc-dlnT}
\frac{d\ln p_c}{d\ln T}=
	\left[\frac{\Ef-Q}{\kB T}-3\right]
	\left[1+\frac{\Ef}{\kB T}\frac{d\ln\Ef}{d\ln p}\right]^{-1}
	\bigg|_{p=p_c}. 
\end{equation}
All the terms on the right-hand side of the above equation are
evaluated at $p=p_c$. But to find $p_c$ we would have to solve the
transcendental Eq.~(\ref{eq:pc-definition}). However, we note that the
above expression enters in Eq.~(\ref{eq:dlnmue-dlnT-const-p})
multiplied by $d\ln\mue/d\ln r$, which is a sharply peaked function of
$r$.  Thus we can treat the right hand side of~(\ref{eq:dlnpc-dlnT})
as a function of $r$ and let the delta-function like shape of
$d\ln\mue/d\ln r$ pick out the correct value of $p\approx p_c$.

Finally, to evaluate Eq.~(\ref{eq:dlnpc-dlnT}) we need
\begin{equation}\label{eq:dlnEf-dlnp}
\frac{d\ln\Ef}{d\ln p}=\frac{1}{3}\left[\frac{1}{\Gammarho}+
\frac{1}{\tilde{V}}\left(\frac{\Gammamue}{\Gammarho}+1\right)
\frac{d\ln\mue}{d\ln r}\right],
\end{equation}
which follows from $\Ef\propto(\rho/\mue)^{1/3}$ and a modest amount
of algebra.  We can now evaluate both $\Gamma$ and the source term
$\Delta S$ using the properties of the background model:  use
Eqs.~(\ref{eq:gamma-dT-perturbations})
and~(\ref{eq:dlnmue-dlnp-const-T}) to compute $\Gamma$, and
Eqs.~(\ref{eq:thermal-source-term}), (\ref{eq:dlnp-dlnT-const-rho}),
(\ref{eq:dlnmue-dlnp-const-T}), (\ref{eq:dlnmue-dlnT-const-p}), and
(\ref{eq:dlnpc-dlnT}), as well as the local $\delta T$, to compute
$\Delta S$.  The source term $\Delta S$ is plotted in the bottom
panels of
Figures~\ref{fig:nthermal-shallow-displ}--\ref{fig:nthermal-deep-displ}. 

Using Eq.~(\ref{eq:dlnpc-dlnT}) we can establish a precise relation
between the vertical shift $\dzd$ of the ``center'' of the e$^-$
capture layer (in absence of elastic readjustment) and the local
temperature perturbation $\delta T$,
\begin{equation}\label{eq:dzd-vs-deltaT-precise}
\frac{\dzd}{r}=-\frac{1}{\tilde{V}}\frac{d\ln p_c}{d\ln T}
	\frac{\delta T}{T}.
\end{equation}
This relation is plotted in the bottom  panel of
Figure~\ref{fig:dzd-vs-deltaT} in \S~\ref{sec:el-cap-rates}.
Similarly, the change in electron Fermi energy at which most captures
occur in response to a local temperature perturbation is just
\begin{equation}
\Delta\Ef=\frac{d\ln p_c}{d\ln T}\frac{d\ln\Ef}{d\ln p}
	\frac{\Ef}{\kB T}\kB\delta T,
\end{equation}
or $\Delta\Ef=-\Upsilon(\kB\delta T)$, where 
\begin{equation}\label{eq:beta-vs-deltaT}
\Upsilon=-\frac{d\ln p_c}{d\ln T}\frac{d\ln\Ef}{d\ln p}
	\frac{\Ef}{\kB T}.
\end{equation}
The function $\Upsilon$ is plotted in the top panel of
Figure~\ref{fig:dzd-vs-deltaT} in \S~\ref{sec:el-cap-rates}. The
complicated dependence of $\dzd$ and $\Upsilon$ on the capture layer
depth stems from the differences in the number of electrons captured
and number of neutrons emitted in each capture layer.

\newpage
\begin{table}
\caption{List of Variables}
\begin{tabular}{lp{5.5cm}c}
Name		&  Description 			& Section \\ \\
$\Mdot$		&  Mass accretion rate		&\ref{sec:introduction}\\
$\nu_s$		&  Spin frequency		&\ref{sec:introduction}\\
$B$		&  Magnetic field		&\ref{sec:introduction}\\
$Q_{22}$	&  Mass quadrupole moment	&\ref{sec:introduction}\\
$\Mdot_{\rm Edd}$&  Eddington accretion rate	&\ref{sec:introduction}\\
$N_a$		& Accretion torque		&\ref{sec:grav}\\
$\delta\rho$	& Eulerian density perturbation &\ref{sec:grav}\\
$Q_{lm}$	& Mass multipole moment		&\ref{sec:grav}\\
$\nu_{gw}$	& Gravitational wave frequency	&\ref{sec:grav}\\
$N_{gw}$	& Gravitational wave torque	&\ref{sec:grav}\\
$M$		& NS mass			&\ref{sec:grav}\\
$R$		& NS radius			&\ref{sec:grav}\\
$Q_{eq}$	& Mass quadrupole moment needed for equilibrium 
		between accretion and GW emission&\ref{sec:grav}\\
$M_{1.4}$	& NS mass in units of $1.4\ M_\odot$ &\ref{sec:grav}\\
$R_6$		& NS radius in units of $10$~km &\ref{sec:grav}\\
$I_{ab}$	& Mass quadrupole tensor	&\ref{sec:grav}\\
$\Omega$	& Spin frequency 		&\ref{sec:grav}\\
$\bar\sigma$	& Average shear strain		&\ref{sec:grav}\\
$\nu_{\rm s, eq}$& Equilibrium spin frequency	&\ref{sec:grav}\\
$d$		& Source distance		&\ref{sec:grav}\\
$\dot{E}_{gw}$	& Energy loss due to GWs	&\ref{sec:grav}\\
$h_a$		& GW strain			&\ref{sec:grav}\\
$F_x$		& X-ray flux			&\ref{sec:grav}\\
$A$		& Nuclear mass number &\ref{sec:origin-of-crustal-Q}\\ 
$Z$		& Nuclear charge	&\ref{sec:origin-of-crustal-Q}\\ 
$\delta T$	& Lateral temperature perturbation&\ref{sec:origin-of-crustal-Q}\\ 
$\dzd$		& Vertical shift of a capture layer&\ref{sec:origin-of-crustal-Q}\\ 
$\Delta\rho$	& Lagrangian density perturbation&\ref{sec:origin-of-crustal-Q}\\ 
$\Upsilon$	& Sensitivity of e$^-$ capture layer 
		location to temperature&\ref{sec:origin-of-crustal-Q}\\ 
$\mu$		& Shear modulus&\ref{sec:origin-of-crustal-Q}\\ 
$\kB$		& Boltzmann constant&\ref{sec:origin-of-crustal-Q}\\ 
$\rhond$	& Neutron drip density &\ref{sec:crust-structure}\\
$K$		& Conductivity&\ref{sec:crust-structure}\\
$t_{\rm comp}$ & Compression timescale &\ref{sec:el-cap-rates}\\
$\dot{m}$	& Local accretion rate&\ref{sec:el-cap-rates}\\
$g$		& Local gravitational acceleration&\ref{sec:el-cap-rates}\\
$e$		& Electron charge&\ref{sec:el-cap-rates}\\
$\mb$		& Baryon mass&\ref{sec:el-cap-rates}\\
$\Ef$		& Electron Fermi energy&\ref{sec:el-cap-rates}\\
$Q$		& e$^{-}$ capture threshold energy&\ref{sec:el-cap-rates}\\
$ft$		& $ft$ value&\ref{sec:el-cap-rates}\\
$\Rec$		& e$^{-}$ capture rate&\ref{sec:el-cap-rates}\\
$h$		& Scale height&\ref{sec:el-cap-rates}\\
$t_{\rm ec}$	& e$^-$ capture timescale&\ref{sec:el-cap-rates}\\
$X_1$		& Mass fraction of the e$^-$ capture
			element&\ref{sec:Capture-Layers}\\ 
$X_2$		& Mass fraction of the e$^-$ capture
			product&\ref{sec:Capture-Layers}\\  
$X_n$		& Mass fraction of free
			neutrons&\ref{sec:Capture-Layers}\\ 
$n_e$		& Density of electrons&\ref{sec:Capture-Layers}\\ 
$\mue$		& Electron mean molecular
			weight&\ref{sec:Capture-Layers}\\ 
$\vec{v}$	& Compression flow speed&\ref{sec:Capture-Layers}\\ 
$\Delta\Ef$	& Thickness of a capture
			layer&\ref{sec:Capture-Layers}\\ 
$r$		& Radius&\ref{sec:Capture-Layers}\\ 
$p_e$		& Electron pressure&\ref{sec:Hydr-Struct}\\
$\rho_{11}$	& Density in units of $10^{11}\ \grcc$&\ref{sec:Hydr-Struct}\\
$n_n$		& Free neutron density&\ref{sec:Hydr-Struct}\\
$M_r$		& Mass enclosed within radius $r$&\ref{sec:Hydr-Struct}\\
$a$		& Internuclear spacing&\ref{sec:Hydr-Struct}\\
$\Gamma_{\rm coul}$ & Coulomb energy density relative to $\kB T$&\ref{sec:Hydr-Struct}\\
\end{tabular}
\end{table}

\begin{table}
\contcaption{}
\begin{tabular}{lp{5.5cm}c}
Name		&  Description 			& Section \\ \\
$\vec{F}$	& Heat flux 				&\ref{sec:Steady-State-Thermal}\\
$F_r$		& Radial component of the heat flux	&\ref{sec:Steady-State-Thermal}\\
$\enuc$		& Energy generation rate		&\ref{sec:Steady-State-Thermal}\\
$\ebrem$	& Urca neutrino emissivity		&\ref{sec:Steady-State-Thermal}\\
$\epsilon$	& Energy gain/loss rate per gram	&\ref{sec:Steady-State-Thermal}\\
$\Enuc$		& Total energy released in a capture layer&\ref{sec:Steady-State-Thermal}\\
$\Delta$	& Superfluid gap energy			&\ref{sec:Steady-State-Thermal}\\
$\rho_{\rm nuc}$	& Nuclear density		&\ref{sec:Steady-State-Thermal}\\
$L_{\rm core}$	& Energy lost from the NS core		&\ref{sec:Steady-State-Thermal}\\
$T_{\rm burn}$	& H/He burning temperature		&\ref{sec:Steady-State-Thermal}\\
$\fnuc$		& Asymmetry in the nuclear energy release&\ref{sec:possible-dT-sources}\\
$\delta\Enuc$	& Difference total energy released in a capture layer&\ref{sec:possible-dT-sources}\\
$n_k$		& Temperature sensitivity of the conductivity&\ref{sec:possible-dT-sources}\\
$n_e$		& Temperature sensitivity of the neutrino emissivity&\ref{sec:possible-dT-sources}\\
$\fcomp$	& Asymmetry in conductivity&\ref{sec:possible-dT-sources}\\
$\delta(Z^2/A)$  & Difference in charge-to-mass ratio&\ref{sec:possible-dT-sources}\\
$\delta F^a$	& Flux perturbation 			&\ref{sec:thermal-bcs}\\
$\hat{r}^a$	& unit radial vector			&\ref{sec:thermal-bcs}\\
$\nabla^a$	& Derivative operator			&\ref{sec:thermal-bcs}\\
$K_{\rm core}$	& Conductivity of the core 		&\ref{sec:thermal-bcs}\\
$K_{\rm crust}$ & Conductivity of the crust		&\ref{sec:thermal-bcs}\\
$\delta F_{\rm r, core}$ & Radial flux perturbation in the core&\ref{sec:thermal-bcs}\\
$\delta F_{\perp, \rm core}$ &Transverse flux perturbation in the core&\ref{sec:thermal-bcs}\\
$\Delta R$	& Crust thickness&\ref{sec:thermal-bcs}\\
$F_A$		& Accretion flux &\ref{sec:resulting-temperature-variations}\\
$f$		& Composition or nuclear heating perturbation&\ref{sec:resulting-temperature-variations}\\
$t_{\rm th}$	& Thermal time&\ref{sec:resulting-temperature-variations}\\
%
$\xi^a$		& Displacement vector &\ref{sec:crust-deformation}\\
$Q_{lm}^{\rm hist}$ & ``Historical'' multipole moment&\ref{sec:crust-deformation}\\
$Q_{lm}^{\rm pert}$ & Multipole moment due to the ``current''
		perturbation&\ref{sec:crust-deformation}\\
$\Delta\mue$	& Smooth composition gradient&\ref{sec:crust-deformation}\\
$\tau_{ab}$	& Stress tensor &\ref{equations}\\
$g_{ab}$	& Metric tensor &\ref{equations}\\
$\xi_r$		& Radial component of the displacement&\ref{equations}\\
$\xi_\perp$	& Transverse component of the displacement&\ref{equations}\\
$\Phi$		& Gravitational potential&\ref{equations}\\
$\beta$		& $\sqrt{l(l+1)}$&\ref{equations}\\
$\tau_{rr}$, $\tau_{r\perp}$ & Components of the stress tensor&\ref{equations}\\
$e_{ab}$, $f_{ab}$, $\Lambda_{ab}$ & Angular tensors&\ref{equations}\\
$z_1-z_4$	& Eigenfunctions of the elastic perturbation equations&\ref{equations}\\
$\alpha_1-\alpha_4$ & Coefficients of the elastic perturbation
		equations&\ref{equations}\\
$\Gamma$	& Compressibility &\ref{equations}\\
$\Delta S$	& Source term&\ref{equations}\\
$\tilde U$, $\tilde V$ & Polytrope variables&\ref{equations}\\
$\rtop$, $\rbot$ & Top and bottom of the crust &\ref{sec:boundary-conditions}\\
$\rho_s$,$\rho_l$ & Density of the solid and of the liquid at the
			crust-core boundary &\ref{sec:boundary-conditions}\\
$Q_{22}^{\rm pp}$ & Plane-parallel approximation to the quadupole
		moment&\ref{sec:boundary-conditions}\\
$\bar\sigma+{\rm max}$	& Breaking strain of the crust	&\ref{sec:max_intro}\\
$t_{ab}$	& Stress tensor (nonlinear)  &\ref{sec:maxint}\\
$t_{rr}$, $t_{r\perp}$, $t_{\Lambda}$ & Components of $t_{ab}$ &\ref{sec:maxint}\\
\end{tabular}
\end{table}

\begin{table}
\contcaption{}
\begin{tabular}{lp{5.5cm}c}
$\sigma_{ab}$	& Strain tensor  &\ref{sec:maxQ}\\
$\bar\sigma$	& Average strain &\ref{sec:maxQ}\\
$\sigma_{rr}$, $\sigma_{r\perp}$, $\sigma_{\Lambda}$ & Components of $\sigma_{ab}$&\ref{sec:maxQ}\\
$p_b$		& Pressure at the bottom of the crust&\ref{sec:maxQ}\\
$\rho_b$	& Density at the bottom of the crust&\ref{sec:maxQ}\\
$\left<\mu/p\right>$ & Average shear modulus&\ref{sec:maxQ}\\
$\sigma_{ab}^{\rm eff}$ & Effective strain&\ref{sec:maxQ}\\
$I_{\rm crust}$ & Moment of inertia of the crust  &\ref{sec:self-gravity}\\
$I_{\rm NS}$  & Moment of inertia of the neutron star&\ref{sec:self-gravity}\\
$\delta\Phi$	& Gravitational potential perturbation&\ref{sec:self-gravity}\\
${\mathcal F}$  & Correction due to the neglect of self-gravity&\ref{sec:self-gravity}\\
$J^a$ 		& Angular momentum vector&\ref{sec:conclusions}\\
$z^a$		& Direction of the rotation axis &\ref{sec:conclusions}\\
$n^a$		& Pricipal axis direction &\ref{sec:conclusions}\\
$\theta_w$	& Wobble angle&\ref{sec:conclusions}\\
$I^{ab}$	& NS inertia tensor &\ref{sec:conclusions}\\
$I_0$		& Spherically symmetric part of $I^{ab}$&\ref{sec:conclusions}\\
$\Delta I_\Omega$ & Oblate part of $I^{ab}$, for crust whose unstressed state is spherical
&\ref{sec:conclusions}\\
$\Delta I_\Omega+\Delta I_d$ & Oblate part of $I^{ab}$, for a fully relaxed crust &\ref{sec:conclusions}\\
$Q_{21}$	& $m=1$ mass quadrupole moment&\ref{sec:conclusions}\\
$\omega_p$	& Precession frequency&\ref{sec:conclusions}\\
$\omega_{\rm gw}$& GW angular frequency&\ref{sec:conclusions}\\
$L^{\rm wobble}_{\rm gw}$ & GW luminosity due to the wobble&\ref{sec:conclusions}\\
%
$p_c(T)$	& Pressure at the center of the capture layer&\ref{sec:source-terms}\\
$\tilde{\mu}_e$	& Run of $\mue$ with radius&\ref{sec:source-terms}\\
$\Gammarho$	& Logarithmic index of pressure with density&\ref{sec:source-terms}\\
$\Gammamue$	& Logarithmic index of pressure with electron mean
molecular weight&\ref{sec:source-terms}\\
\end{tabular}
\end{table}

\end{document}